\documentclass[%
    amsmath,
    amssymb,
    aps,prx,
    superscriptaddress,
    nofootinbib,
    twocolumn
]{revtex4-2}

\allowdisplaybreaks
\usepackage{bm}
\usepackage[colorlinks=true,allcolors=blue]{hyperref}
\usepackage[capitalize]{cleveref}

\usepackage[ISO]{diffcoeff}[=v4] 
\diffdef{}{long-var-wrap = dv} 
\usepackage{feynmp-auto} 
\usepackage{orcidlink}

\newcommand{\odv}[3][1]{\diff[#1]{#2}{#3}}
\newcommand{\pdv}[3][1]{\diffp[#1]{#2}{#3}}
\newcommand{\fdv}[3][1]{\diff.delta.[#1]{#2}{#3}}
\newcommand{\E}[1]{\left \langle #1 \right \rangle}
\newcommand{\dd}{\mathrm{d}}
\newcommand{\re}{\mathrm{Re}\,}
\newcommand{\im}{\mathrm{Im}\,}
\newcommand{\Oh}{\mathcal{O}}
\newcommand{\tj}{\tilde J}
\renewcommand{\j}{J}
\newcommand{\Je}{\mathcal{J}}
\newcommand{\D}{\mathcal{D}}
\newcommand{\A}{\mathcal{A}}

\newcommand{\Z}{\mathcal{Z}}

\newcommand{\Q}{{\bm q, \omega}}
\newcommand{\Qp}{{\bm q'\!, \omega'}}
\newcommand{\K}{{\bm k, \varpi}}
\newcommand{\Kp}{{\bm k'\!, \varpi'}}
\newcommand{\Kpp}{{\bm k''\!, \varpi''}}
\newcommand{\Km}{{-\bm k, -\varpi}}
\newcommand{\Kpm}{{-\bm k'\!, -\varpi'}}
\newcommand{\Kppm}{{-\bm k''\!, -\varpi''}}
\newcommand{\xiS}{\xi_{\rm s}}
\newcommand{\xiD}{\xi_{\rm d}}
\newcommand{\xic}{\xi_{\times}}

\usepackage{xspace}
\newcommand{\NRpm}{NR$^\pm$\xspace}
\newcommand{\NRp}{NR$^+$\xspace}
\newcommand{\NRm}{NR$^-$\xspace}

\newcommand{\s}{\hspace{-0.5pt}}

\renewcommand{\t}{r}
\renewcommand{\a}{\lambda}
\renewcommand{\b}{\kappa}
\renewcommand{\c}{\mu}
\renewcommand{\l}{\upsilon}

\newcommand{\az}{{{\alpha}_0}}
\renewcommand{\ao}{{{\alpha}_1}}
\newcommand{\bz}{{{\beta}_0}}

\newcommand{\ren}[1]{ \bar{#1}}
\newcommand{\ru}{\ren{u}}
\newcommand{\rr}{\ren{r}}

\newcommand{\raz}{{\ren{\alpha}_0}}
\newcommand{\rao}{{\ren{\alpha}_1}}
\newcommand{\rbz}{{\ren{\beta}_0}}

\newcommand{\ra}{\ren{\a}}
\newcommand{\rb}{\ren{\b}}
\newcommand{\rc}{\ren{\c}}
\newcommand{\rl}{\ren{\l}}
\newcommand{\rt}{\ren{\t}}
\newcommand{\rg}{\ren{g}}

\newcommand{\ded}{{\mathfrak{D}_d}}
\newcommand{\dedc}{{\mathfrak{D}_4}}

\newcommand{\dr}{{\Delta r}}
\newcommand{\daz}{{\Delta \az}}

\newcommand{\es}{{\eta^{\!*}\!}}
\newcommand{\gs}{{\gamma_\t^{\!*}\!}}

\newcommand{\f}{\mathfrak{f}}
\newcommand{\g}{\mathfrak{g}}
\newcommand{\h}{\mathfrak{h}}

\newcommand{\G}{\mathcal{G}}

\usepackage{stmaryrd}
\usepackage{trimclip}

\makeatletter
\DeclareRobustCommand{\shortarrow}{%
  \mathrel{\mathpalette\short@to\relax}%
}
\newcommand{\short@to}[2]{%
  \mkern2mu
  \clipbox{{.5\width} 0 0 0}{$\m@th#1\vphantom{+}{\shortrightarrow}$}%
  }
\makeatother

\newcommand{\rgcond}{|_{(\bm q, \omega) \shortarrow 0 }}

\newcommand{\conj}{btex $*$ etex}

\newcommand{\setval}{
    \fmfset{wiggly_len}{1.5mm}
    \fmfset{arrow_len}{1.5mm}
    \fmfset{arrow_ang}{20}
    \fmfset{dash_len}{2mm}
    \fmfpen{0.25mm}
    \fmfset{dot_size}{1.5thick}
    \fmfcmd{
        style_def Double expr p =
        pickup pencircle scaled 0.25mm;
        draw_double p;
        pickup pencircle scaled 0.15mm;
        enddef;
    }
    \fmfcmd{
        style_def response expr p =
        idraw ("plain", subpath (0,length(p)/2) of p);
        idraw ("wiggly", subpath (length(p)/2,length(p)) of p);
        enddef;
    }
    \fmfcmd{
        style_def plainc expr p =
        draw_plain p;
        pair pos; 
        pos := point 0.5*length(p) of p;
        label(\conj, pos);
        enddef;
    }
    \fmfcmd{
        style_def wigglyc expr p =
        draw_wiggly p;
        pair pos; 
        pos := point 0.5*length(p) of p;
        label(\conj, pos);
        enddef;
    }
    \fmfcmd{
        style_def cc expr p =
        cdraw p;
        pair pos; 
        pos := point 0.7*length(p) of p;
        label(\conj, pos);
        enddef;
    }
    \fmfcmd{
        style_def ccr expr p =
        cdraw p;
        pos := point 0.3*length(p) of p;
        label(\conj, pos);
        enddef;
    }
    \fmfcmd {
        style_def cg expr p =
        idraw ("plain", subpath (0,length(p)/2) of p);
        idraw ("wiggly", subpath (length(p)/2,length(p)) of p);
        pos := point 0.7*length(p) of p;
        label(\conj, pos);
        enddef;
    }
    \fmfcmd {
        style_def cgc expr p =
        idraw ("wiggly", subpath (0,length(p)/2) of p);
        idraw ("plain", subpath (length(p)/2,length(p)) of p);
        pos := point 0.7*length(p) of p;
        label(\conj, pos);
        enddef;
    }
    \fmfcmd {
        style_def cgcr expr p =
        idraw ("plain", subpath (0,length(p)/2) of p);
        idraw ("wiggly", subpath (length(p)/2,length(p)) of p);
        pos := point 0.3*length(p) of p;
        label(\conj, pos);
        enddef;
    }
    \fmfcmd {
        style_def cgr expr p =
        idraw ("wiggly", subpath (0,length(p)/2) of p);
        idraw ("plain", subpath (length(p)/2,length(p)) of p);
        pos := point 0.3*length(p) of p;
        label(\conj, pos);
        enddef;
    }
}

\newcommand{\sigmaoneone}{
    \begin{fmfgraph*}(12,5)
        \setval
        \fmfleft{i,i1}
        \fmfright{o,o1}
        \fmftop{t}
        \fmf{wigglyc}{i,c}
        \fmf{plain}{c,o} 
        \fmffreeze
        \fmf{plain,left}{t,c}
        \fmf{plainc,right}{t,c}
        \vspace{-5mm}
    \end{fmfgraph*}
}

\newcommand{\gammaoneone}{
\parbox{12mm}{
    \begin{fmfgraph*}(12,8)
        \setval
        \fmfleft{i1,i2}
        \fmfright{o1,o2}
        \fmftop{c1}
        \fmfbottom{c2}
        \fmf{wigglyc}{i1,v1}
        \fmf{plainc}{i2,v1}
        \fmf{plain}{v2,o1}
        \fmf{plain}{v2,o2}
        \fmf{cg,left,tension=0.4}{v1,v2}
        \fmf{cc,right,tension=0.4}{v1,v2}
    \end{fmfgraph*}
    }
}

\newcommand{\gammaonetwo}{
    \parbox{12mm}{
    \begin{fmfgraph*}(12,8)
        \setval
        \fmfleft{i1,i2}
        \fmfright{o1,o2}
        \fmftop{c1}
        \fmfbottom{c2}
        \fmf{wigglyc}{i1,v1}
        \fmf{plain}{i2,v1}
        \fmf{plainc}{o1,v2}
        \fmf{plain}{v2,o2}
        \fmf{cg,left,tension=0.4}{v1,v2}
        \fmf{cc,left,tension=0.4}{v2,v1}
    \end{fmfgraph*}
    }
}
\newcommand{\gammaonethree}{
\parbox{12mm}{
    \begin{fmfgraph*}(12,8)
        \setval
        \fmfleft{i1,i2}
        \fmfright{o1,o2}
        \fmftop{c1}
        \fmfbottom{c2}
        \fmf{wigglyc}{i1,v1}
        \fmf{plain}{i2,v1}
        \fmf{plainc}{o1,v2}
        \fmf{plain}{v2,o2}
        \fmf{cgcr,left,tension=0.4}{v1,v2}
        \fmf{cc,right,tension=0.4}{v1,v2}
    \end{fmfgraph*}
    }
}

\newcommand{\sigmatwoone}{
    \parbox{16mm}{
    \begin{fmfgraph*}(16,10)
        \setval
        \fmfleft{i1}
        \fmfright{o1}
        \fmftop{c1}
        \fmfbottom{c2}
        \fmf{wigglyc}{i1,v1}
        \fmf{plain}{v2,o1}
        \fmf{cc,tension=0.3}{v1,v2}
        \fmf{cgcr,left,tension=0.3}{v1,v2}
        \fmf{cc,right,tension=0.3}{v1,v2}
    \end{fmfgraph*}
    }
}

\newcommand{\sigmatwotwo}{
\parbox{16mm}{
    \begin{fmfgraph*}(16,10)
        \setval
        \fmfleft{i1}
        \fmfright{o1}
        \fmftop{c1}
        \fmfbottom{c2}
        \fmf{wigglyc}{i1,v1}
        \fmf{plain}{v2,o1}
        \fmf{cc,tension=0.3}{v2,v1}
        \fmf{cg,left,tension=0.3}{v1,v2}
        \fmf{cc,right,tension=0.3}{v1,v2}
    \end{fmfgraph*}
    }
}

\newcommand{\Sigmaab}{
    \parbox{12mm}{
    \centering
    \begin{fmfgraph*}(12,10)
        \setval
        \fmfleft{i}
        \fmfright{o}
        \fmftop{t}
        \fmf{wiggly}{i,c}
        \fmf{plain}{o,c}
        \fmffreeze
        \fmf{plain,left}{c,t}
        \fmf{plain,left}{t,c}
    \end{fmfgraph*}
    } 
}

\newcommand{\deltag}{
    \parbox{12mm}{
    \begin{fmfgraph*}(12,7)
        \setval
        \fmfleft{i1,i2}
        \fmfright{o1,o2}
        \fmftop{c1}
        \fmfbottom{c2}
        \fmf{wiggly}{i1,v1}
        \fmf{plain}{i2,v1}
        \fmf{plain}{v2,o1}
        \fmf{plain}{v2,o2}
        \fmf{response,left,tension=0.4}{v1,v2}
        \fmf{plain,left,tension=0.4}{v2,v1}
    \end{fmfgraph*}
    }
}

\begin{document}
\begin{fmffile}{feyn}

\title{Scaling behavior in non-reciprocal and odd conserved dynamics near criticality}

\author{Martin Kj{\o}llesdal Johnsrud\,\orcidlink{0000-0001-8460-7149}}
\affiliation{Max Planck Institute for Dynamics and Self-Organization (MPI-DS), D-37077 G\"ottingen, Germany}

\author{Giulia Pisegna\,\orcidlink{0000-0002-5214-0027}}
\affiliation{Max Planck Institute for Dynamics and Self-Organization (MPI-DS), D-37077 G\"ottingen, Germany}

\author{Ramin Golestanian\,\orcidlink{0000-0002-3149-4002}}
\email{ramin.golestanian@ds.mpg.de}
\affiliation{Max Planck Institute for Dynamics and Self-Organization (MPI-DS), D-37077 G\"ottingen, Germany}
\affiliation{Rudolf Peierls Centre for Theoretical Physics, University of Oxford, Oxford OX1 3PU, United Kingdom}

\date{\today}

\begin{abstract}
In recent years, non-reciprocity has been explored as a ubiquitous manifestation of non-equilibrium activity at the microscopic scales for various active matter systems, from mixtures of chemically active enzymes, colloids, and droplets, to engineered light-controlled active colloids and robotic meta-materials. A commonly used minimal model to describe the dynamics of a binary mixture of conserved species with non-reciprocal interactions is the \emph{non-reciprocal} Cahn-Hilliard (NRCH) model. The model is characterized by a temperature-like tuning parameter, which can trigger phase separation, and a non-reciprocal coupling, which can lead to the formation of spatio-temporal patterns, as it represents an intrinsic source of non-equilibrium activity and breaks parity and time-reversal symmetries. Here, we study the scaling behavior of the NRCH model near the critical point using perturbative dynamical renormalization group techniques. We find that structural and dynamical correlations are controlled by different correlation lengths, both of which diverge at the critical point, but governed by different scaling laws. In particular, while the structural correlations are always controlled by temperature and the classical Wilson-Fisher critical exponent, the dynamical correlation length exhibits multiple scaling regimes in which either temperature or non-reciprocal coupling can dominate as the key tuning parameter, with a new critical exponent characterizing the divergence. The critical point corresponds to a conserved equilibrium-like dynamics with odd mobility, which we denote as the \emph{odd} Cahn-Hilliard (OCH) model. Our findings may have important implications on how living systems can control phase separation and spatio-temporal pattern formation using the rates of catalytic reactions, and in general metabolism, as control parameters. 
\end{abstract}

\maketitle


\section{Introduction}
\label{sec: introduction}

\noindent

The hierarchical nature of self-organization in biological systems makes living matter a promising testing ground for applying ideas from the physics of critical phenomena~\cite{kardarStatisticalPhysicsFields2007}, and the associated tools such as the field-theoretic and renormalization group (RG) techniques~\cite{zinn-justinQuantumFieldTheory1989}. Examples of such applications include a proposed mechanism for emergent sensitivity in hearing \cite{rislerUniversalCriticalBehavior2004,rislerUniversalCriticalBehavior2005}, anomalous scaling and hyper-uniformity in chemotaxis and colony growth \cite{Gelimson2015,Thirumalai,Mahdisoltani2021a,BenAliZinati2022,Frey-chemotaxis-RG}, bacterial range-expansion \cite{range-expansion-rg}, motility-induced phase separation \cite{CFLee-MIPS} and active model B+ \cite{caballeroBulkMicrophaseSeparation2018,AMB+RG}, and aspects of brain function \cite{neuron-rg-1}, as well as the emergence of polar order in animal flocks \cite{toner2024,Chen2015,cavagnaDynamicScalingNatural2017,cavagnaNaturalSwarms3992023} and non-reciprocal mixtures \cite{pisegnaEmergentPolarOrder2024}, and anomalous scaling in anisotropic active suspensions \cite{Sriram-aligned-prl,Golestanian2026}. Moreover, real-space RG based on decimation has been used to study scale-dependent properties of DNA \cite{Noy2012}, proteins \cite{Bradde2017}, and neural networks \cite{neuron-decimation}. 

Biological liquid-liquid phase separation has been recently studied as a paradigm for intracellular organization \cite{bananiBiomolecularCondensatesOrganizers2017}, primarily understood to be driven by equilibrium interactions and quasi-statically controlled by tunable parameters, such as composition \cite{brangwynnePolymerPhysicsIntracellular2015,jacobsSelfAssemblyBiomolecularCondensates2021,shrinivasPhaseSeparationFluids2021}. Intracellular environments are, however, non-equilibrium media, and recent proposals have highlighted genuinely non-equilibrium processes that can lead to spatiotemporal structure formation and control of phase-separation via modulating metabolic activity \cite{cotton2022, parkavousiEnhancedStabilityChaotic2025, romanoNonreciprocalInteractionsCondensates2026}. Metabolically active condensates have novel physical characteristics \cite{sweetloveRoleDynamicEnzyme2018, testa2021, Kokkoorakunnel2024}, and have been suggested to play an auto-regulatory role in the cell cycle dynamics \cite{Niebel2019}. Bimolecular condensates are inherently slow in terms of their mechanical response \cite{Jawerth2020}, and controlling their phase behavior by changing temperature or molecular interactions will also engage inherently slow processes familiar from polymer physics \cite{brangwynnePolymerPhysicsIntracellular2015}. It is, therefore, a tantalizing prospect to speculate whether bimolecular condensates can control the agility of the their dynamics or response by tuning their metabolic activity, through available mechanisms in non-equilibrium physics \cite{OuazanReboul2023b}.

A prominent candidate for incorporating non-equilibrium activity at the microscopic scale is non-reciprocal interactions \cite{Soto2014,Cohen2014,ivlevStatisticalMechanicsWhere2015}, which are generically present in mixtures of chemically active enzymes and colloids due to interfacial phoretic transport mechanisms \cite{golestanianPhoreticActiveMatter2022}. These interactions are powerful forces of non-equilibrium self-organization \cite{sahaScalarActiveMixtures2020, youNonreciprocityGenericRoute2020}: they can lead to spontaneous formation of temporally programmed shape-shifting structures \cite{Soto2015,Osat2022} and metabolic cycles with super-exponential coarsening, which may have facilitated the emergence of proto-cells at the origin of life \cite{OuazanReboul2023b,OuazanReboul2023,OuazanReboul2023a}. Active systems with non-reciprocal alignment \cite{Uchida2010a,Uchida2010,Saha2019,fruchartNonreciprocalPhaseTransitions2021,Loos-NR-2023,bhattEmergentHydrodynamicsNonreciprocal2023,duanDynamicalPatternFormation2023,Hickey2023,hanaiNonreciprocalFrustrationTime2024,Popli-NR-2025,rouzaireDynamicsO2Excitations2026,Duan2025,fruchartNonreciprocalManybodyPhysics2026,myinNonreciprocityGenericMechanism2026,akritidisFateIsingUniversality2026,myinBreakdownEmergentChiral2026a} and engineered robotic meta-materials \cite{Souslov2017,Scheibner2020,Shankar2022,guilletMeltingNonreciprocalSolids2025,Veenstra2025,al-izziNonreciprocalBucklingMakes2026} have also revealed a remarkably wide range of emergent features with relatively modest input, opening up promising avenues for engineering and applications, which enhances the already existing momentum towards medical, bioengineering, and environmental applications of active matter \cite{Ju2025,Chen2025}. 
We also note the substantial body of work on open quantum systems~\cite{siebererKeldyshFieldTheory2016,siebererUniversalityDrivenOpen2025}, where external drive and dissipation lead to non-equilibrium effects, associated with non-Hermitian dynamics~\cite{hatanoLocalizationTransitionsNonHermitian1996,heissPhysicsExceptionalPoints2012,hanaiCriticalFluctuationsManybody2020}.
Such systems are therefore promising as platforms for realizing criticality and phase-transitions in non-reciprocal~\cite{siebererDynamicalCriticalPhenomena2013,tauberPerturbativeFieldTheoreticalRenormalization2014,siebererNonequilibriumFunctionalRenormalization2014,foss-feigEmergentEquilibriumManybody2017,youngNonequilibriumFixedPoints2020,youngNonequilibriumUniversalityNonreciprocally2026} or otherwise non-equilibrium systems~\cite{maghrebiNonequilibriumManybodySteady2016,zelleUniversalPhenomenologyCritical2024,davietNonequilibriumCriticalityOnset2024,zelleNonequilibriumOrdersParametrically2026}.

\begin{figure}[t]
    \centering
    \includegraphics[width=\columnwidth]{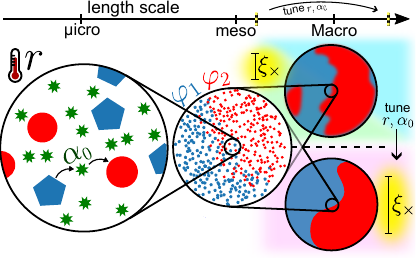}
    \caption{
    Illustration of the hierarchical structure of the system. 
    $\mu$icro: Two species of enzymes or active colloids (red and blue) produce and consume a chemical (green).
    Features such as temperature and non-reciprocal activity set the parameters, such as the tuning parameters $r$ and $\az$.
    meso: The system is described by a density field $\varphi_a$.
    Macro: The crossover length $\xic$ (yellow) depends on the tuning of the microscopic parameters $r$ and $\az$. Adjusting the tuning parameters can induce transitions between different regimes, depending on whether the correlation length is above (green/blue, manifestly Non-Reciprocal regime) or below (magenta, effective-equilibrium regime) the crossover length; see \cref{fig: Illustration} for the scaling properties of these regimes.
    }
    \label{fig: drawing}
\end{figure}

In this paper, we investigate how the large-scale properties of a phase-separating mixture can be controlled in the presence of non-reciprocal interactions that operate alongside the standard tuning parameters inherited from the equilibrium dynamics, such as temperature\footnote{The role of composition as tuning parameter can be readily incorporated in our formalism. Since it will lead to a significant enlargement of the parameter space, we prefer to relegate it to future publications.}, in the context of the non-reciprocal Cahn-Hilliard (NRCH) model \cite{sahaScalarActiveMixtures2020, youNonreciprocityGenericRoute2020}, near its critical point where the correlations are expected to be long-ranged. Our aim is to investigate whether in the vicinity of the critical point, where the system phase separates, the non-reciprocal coupling constant, which is microscopically controlled by catalytic reaction rates, can emerge as a tuning parameter that can control the behavior of the mixture. This can provide a possible physically realizable route to spatio-temporal dynamics in the intracellular environment, controlled by metabolic activity. 

\begin{table}[!b]
    \centering
    \caption{Effective Equilibrium (EE) and manifestly Non-Reciprocal (NR) scaling regimes. We find $\nu = \frac{1}{2} + \frac{\epsilon}{10} + \Oh(\epsilon^2)$, $\nu_{\rm n} = \frac{1}{2}+ \Oh(\epsilon^2)$. The Gaussian (G) scaling behavior, which corresponds to the case where nonlinearities are absent, is presented for a comparison.  
    }
    \begin{tabular*}{.99\columnwidth}{@{\extracolsep{\fill}}l  c c c c c }
     \hline\hline
        Regime & $\xiS$ & $\xiD$ & $\eta$ & $z = 4 - \eta$ & FDR  \\
        \hline
        \hline\\[-4mm]
        G  & $ \sim r^{-\frac{1}{2}}$ & $\sim \xiS$ & $0$ & 4 & Yes \\[1mm]
        EE & $\sim r^{-\nu}$ & $\sim \xiS$ & $\frac{ \epsilon^2 }{ 50 }$ & $4-\frac{ \epsilon^2 }{ 50 }$ & Yes  \\[1mm]
        $\alpha\text{NR}$ & $\sim r^{-\nu}$ & $ \sim |\alpha_0|^{-\nu_{\rm n}}$ & $\frac{ \epsilon^2 }{ 50 }$ & $4-\frac{ \epsilon^2 }{ 50 }$ & No \\[1mm]
        $r\text{(S)NR}$ & $\sim r^{-\nu}$ & $\sim r^{-\nu_{\rm n}}$ & $\frac{ \epsilon^2 }{ 50 }$ & $4-\frac{ \epsilon^2 }{ 50 }$ & No  \\[1mm]
        \hline\hline
    \end{tabular*}
    \label{tab: exponents}
\end{table}

\begin{figure*}[!t]
    \centering
    \includegraphics[width=1\textwidth]{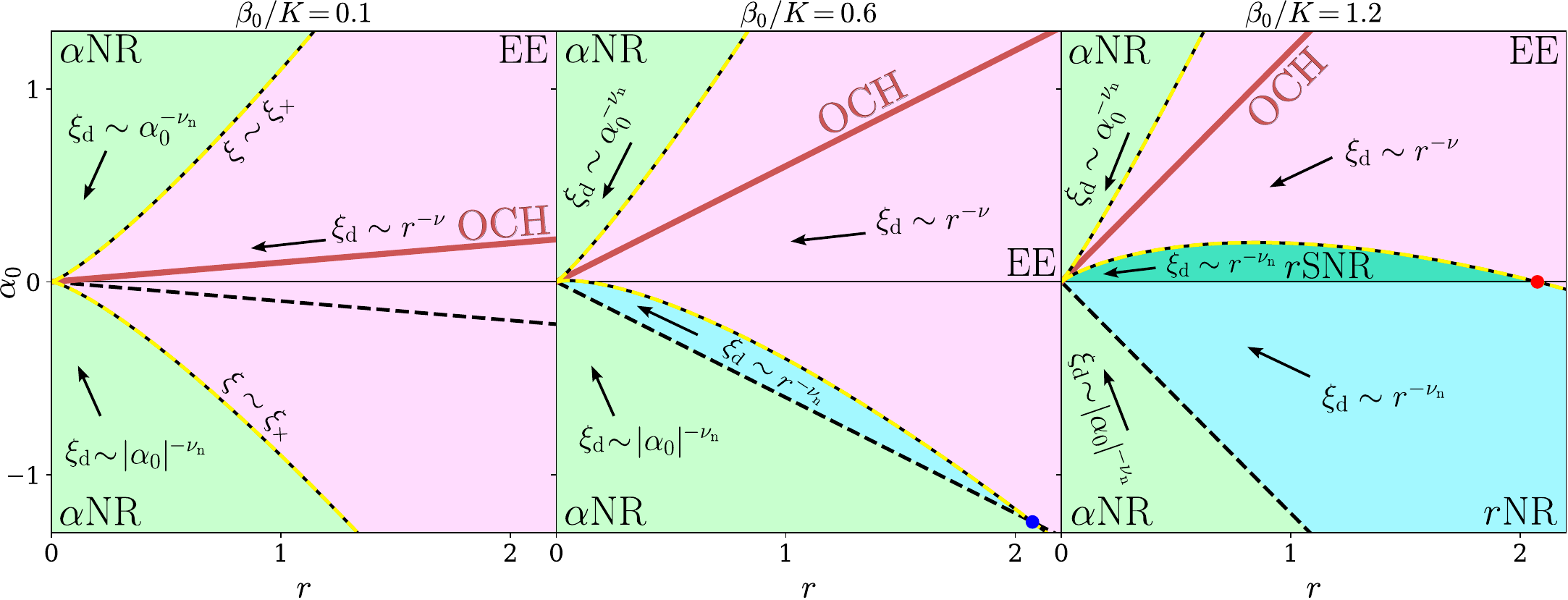}
    \caption{
    The scaling behavior of the NRCH for various values of $\beta_0 > 0$.
    The correlation length $\xiD$ scales with the tuning parameters.
    The scaling regimes are either Effective Equilibrium (EE), or manifestly Non-Reciprocal (NR). In the latter, the scaling is controlled by $r$ in $r$(S)NR and $\az$ in $\alpha$NR.
    In $r$SNR, the non-reciprocity is suppressed due to a competition between $\az$ and $\bz$.
    The values of the exponents are given in \cref{tab: exponents}.
    On the red line, the system is an odd Cahn-Hilliard (OCH) model, and is defined by \cref{eq: NRCH pc}.
    The yellow-and-black line, where $\xiD\sim\xic$, demarcates crossover in scaling exponents.
    The line is defined by $|\sigma|/r^\Delta = 1$, and its intersection with $\az = 0$ is marked with a red dot.
    The dashed black line marks $|\az| = \bz r/K$, where $\az$ becomes the dominant parameter in $\sigma$ over $r$.
    The $r$NR regime terminates where this line crosses the crossover line, marked with a blue dot. \cref{tab: exponents} gives a summary of the values of the exponents and the scaling regimes.
    }
    \label{fig: Illustration}
\end{figure*}

The NRCH model has two tuning parameters, the reduced temperature $r \!\sim\! (T - T_c)/T_c$ near the critical temperature $T_c$, and the linear non-reciprocal coupling $\az$. The former is standard in other critical systems such as the equilibrium model B \cite{hohenbergTheoryDynamicCritical1977} or the Cahn-Hilliard model \cite{brayTheoryPhaseorderingKinetics1994}, while the latter is a consequence of activity at the microscopic level, as illustrated in \cref{fig: drawing}. The presence of these two parameters, together with the conservation law and broken time-reversal symmetry, will give rise to a rich and inherently out-of-equilibrium scaling behavior close to the critical point. In particular, the system features two correlation lengths denoted $\xiS$ and $\xiD$.
The former is defined by structural features, such as the equal-time correlation function, whereas the latter is defined by dynamic features, such as the linear response to perturbations or dynamic auto-correlation functions.
The deviation between $\xiS$ and $\xiD$ is a manifestation of broken time-reversal symmetry and the ensuing departure from the fluctuation dissipation theorem, thus manifesting a signature of inherently active criticality.

We find that the divergence of the structural correlation length is characterized by the equilibrium exponent $\nu$, via $\xiS \sim r^{-\nu}$. We also have equilibrium values for the anomalous scaling $\eta$ and the dynamic exponent $z = 4 - \eta$. However, we find that the divergence of $\xiD$ near the critical point is controlled by a novel non-equilibrium scaling exponent $\nu_{\rm n}$, featuring several scaling regimes, whose properties are summarized in \cref{tab: exponents} and illustrated in \cref{fig: Illustration}.
In the Effective Equilibrium (EE) regime (highlighted in magenta), the scaling behavior of the NRCH model is equivalent to its reciprocal counterpart, model B~\cite{hohenbergTheoryDynamicCritical1977}.
In the green and blue regions, collectively denoted as manifestly Non-Reciprocal (NR), the novel exponent $\nu_{\rm n}$ emerges, the two length-scales $\xiS$ and $\xiD$ separate, and the system features signatures of broken time-reversal symmetry at the scale of $\xiD$.
Within the NR-regime, the scaling of $\xiD$ is either controlled by the $r$-parameter or the $\az$-parameter, denoted correspondingly as $r\text{NR}$ and $\alpha\text{NR}$.
In addition, within the $r$-dominated regime, there is a subset where the non-reciprocity is \emph{suppressed} due to a competition between the two tuning-parameters. We denote the region with $r$-controlled suppressed non-reciprocity as $r\text{SNR}$. The NR regimes signify a departure from criticality as defined in equilibrium systems, where a single correlation length $\xi$ is dominant and control the divergence of all thermodynamic properties as the system approach the critical point.

As the tuning parameters are adjusted, the correlation lengths $\xiS$ and $\xiD$ may pass the crossover length $\xic$, which is defined as the scale at which the system changes from the Effective Equilibrium scaling regime to a manifestly Non-Reciprocal regime. This is illustrated by black-and-yellow lines in \cref{fig: Illustration}. As we will demonstrate further on, this crossover separates the domains for which different RG fixed-point dominate. The black dashed line represents a crossover between $r$-dominated and $\az$-dominated scaling. This happens due to a competition between the tuning parameters.
In addition to the simple linear non-reciprocity $\az$, there is also the non-reciprocal stiffness $\beta_0$, and the non-linear non-reciprocal coupling $\alpha_1$.
These parameters interact with the noise to give rise to an effective, scale dependent non-reciprocity $\alpha_0(\ell)$, where $\ell$ parametrize the scale under consideration.
If $\alpha_0(0)$---the microscopic value of $\alpha_0$---is sufficiently small, $r$ will dominate the scaling of $\xiD$, giving rise to the $r\text{NR}$ regime.
In fact, the effective non-reciprocity can be suppressed and change sign at larger scales.
A similar effect, called effervescence, has been earlier been observed deep in the ordered phase of the NRCH~\cite{sahaEffervescenceBinaryMixture2025}.
The $r\text{SNR}$ region where $\alpha_0(\ell)$ is suppressed and changes sign are highlighted in turquoise in \cref{fig: Illustration}.
The largest value of $r$ for which the effective non-reciprocity changes sign is marked with a red dot in \cref{fig: Illustration}.
The crossover between $r$- and $\alpha_0$-dominated regimes happens at $\beta_0 r /K = |\alpha_0|$, which is illustrated by a black dashed line in \cref{fig: Illustration}.

The rich scaling behavior is a consequence of the RG fixed-points of the model.
There are three non-trivial fixed-points, which we denote odd Cahn-Hilliard (OCH) and manifestly Non-Reciprocal (\NRpm).
The name odd Cahn-Hilliard comes as the equation here takes the form of a system driven by a the gradient flow of a Cahn-Hilliard type free energy, but with odd mobility [see \cref{eq: odd CH}].
As a consequence, it has a Boltzmann-like steady-state distribution and is subject to FDT-like identities~\cite{johnsrudFluctuationDissipationRelations2025,johnsrudFluctuationDissipationRelations2025a}.
The sign in \NRpm denote whether the sign of the effective non-reciprocal parameters $\az(\ell)$ and $\bz(\ell)$ at the fixed point have the same (+) or opposite (-) signs.
The points above the upper crossover line (black-and-yellow) in \cref{fig: Illustration} are dominated by the \NRp fixed-point, while the points below the lower crossover line are dominated by \NRm [see also \cref{fig: flow}].
Our two-loop calculation shows that the non-reciprocal parameters that break time-reversal symmetry do not flow to zero, but rather give rise to large scale signatures of activity.
This is in stark contrast to the equivalent non-conserved model, where the system flows to the equilibrium Wilson-Fisher fixed-point \cite{rislerUniversalCriticalBehavior2004,rislerUniversalCriticalBehavior2005,tauberPerturbativeFieldTheoreticalRenormalization2014,siebererDynamicalCriticalPhenomena2013}.

The rest of this Article is organized as follows. \Cref{sec: NRCH} introduces the NRCH and reports the results in more detail. \Cref{sec: mom shell} contains the one-loop RG calculation using the momentum shell approach, as corroborated with two-loop calculations using the Callan-Symanzik approach presented in \cref{sec: cs approach}. We present concluding remarks in \cref{sec: discussion}. \Cref{app: response field} contains conventions and details about response-field formalism, \cref{app: CS} summarizes the Callan-Symanzik approach to RG, and \cref{app: complex} presents the complex formulation of the NRCH and its Feynman rules. The evaluation of the diagrams using dimensional regularization is detailed in \cref{app: dim reg}.

\section{NRCH and its critical behavior}
\label{sec: NRCH}

The non-reciprocal Cahn-Hilliard model~\cite{sahaScalarActiveMixtures2020,youNonreciprocityGenericRoute2020} consists of scalar fields, $\varphi_a(\bm x, t)$, representing fluctuations of conserved densities governed by over-damped Langevin equations.
We restrict ourselves to two fields, $\varphi_1$ and $\varphi_2$, which are assumed to obey an $\text{SO}(2)$ rotational symmetry in field-space.
The governing equation is
\begin{align}
    \label{eq: EOM}
    \partial_t \varphi_a
    &= 
    -\!\nabla^2\! \left( K_a[\varphi] + W_a[\varphi]\right)
    +
    \sqrt{ 2 D(-\nabla^2) } \xi_{a}.
\end{align}
Here, the square brackets indicate functional dependence.
The indices $a,b\in\{1, 2\}$ run over the components of the field, and we will employ Einstein summation convention throughout the text.
The noise is Gaussian, with $\E{\xi_a(\bm x, t)} = 0$ and
\begin{align}
    \E{ \xi_{a}(\bm x, t)  \xi_{b}(\bm x', t')} =\delta_{ab} \delta^d(\bm x - \bm x') \delta(t - t').
\end{align}
The thermodynamic forces driving the dynamics are divided into a reciprocal term $K_a$ and a non-reciprocal term $W_a$.
The reciprocal term can be written as the functional derivative of a free energy functional ${\cal F}$:
\begin{align}
    \label{eq: eq force}
    K_a[\varphi](\bm x) = - \fdv{ {\cal F}[\varphi] }{ {\varphi_a(\bm x)} }.
\end{align}
We make the symmetry explicit by writing
\begin{align}
    K_a[\varphi] & =  - f_\mathrm{eq}[\varphi]\varphi_a,& 
    W_a[\varphi] & = - f_\mathrm{nr}[\varphi]\varepsilon_{ab} \varphi_b,
\end{align}
where $f_\mathrm{eq}$ and $f_\mathrm{nr}$ are $\mathrm{O}(2)$-invariant functionals, $\varphi \equiv \sqrt{ \varphi_a \varphi_a}$, and $\varepsilon_{ab}$ is the Levi-Civita tensor: $\varepsilon_{11} = \varepsilon_{22} = 0$, $\varepsilon_{12} = - \varepsilon_{21} = 1$.
Throughout the paper we apply the Einstein summation convention.
We expand the functionals to include all RG-relevant terms:
\begin{align}
    f_\mathrm{eq}[\varphi] & = r - K \nabla^2 + u \varphi^2,\\
    f_\mathrm{nr}[\varphi] & = \alpha_0 - \beta_0 \nabla^2 + \alpha_1 \varphi^2.
\end{align}
This corresponds to a free energy
\begin{align}
    {\cal F}[\varphi] = \int_{\bm x} \left[
    \frac{r}{2} \varphi^2 + \frac{K}{2}(\nabla\varphi)^2 + \frac{u}{4}\varphi^4
    \right].
\end{align}
For purposes of renormalization, we formulate this model as within the response-field formalism. 
See \cref{app: response field} for details.

The NRCH was introduced as a minimal model for phase-separation with non-reciprocal interactions \cite{sahaScalarActiveMixtures2020,youNonreciprocityGenericRoute2020}.
A range of variations of the NRCH, with different non-reciprocal couplings, number of components or symmetries, have been applied to obtain a range of exotic non-equilibrium states~\cite{pisegnaEmergentPolarOrder2024,greveCoexistenceUniformOscillatory2025,sahaPhaseCoexistenceNonreciprocal2024,sahaEffervescenceBinaryMixture2025,ranaDefectSolutionsNonreciprocal2024,ranaDefectInteractionsNonreciprocal2024,Rana2026,suchanekEntropyProductionNonreciprocal2023,parkavousiEnhancedStabilityChaotic2025,parkavousiCompositionalDisorderMulticomponent2026,johnsrudStateDiagramNonreciprocal2025}.
It has been shown that the restricted model we consider here, with an $\text{SO}(2)$-symmetry, displays many of the distinctive features of the NRCH~\cite{pisegnaEmergentPolarOrder2024,ranaDefectInteractionsNonreciprocal2024,parkavousiNonreciprocalSurfaceTension2026,sahaEffervescenceBinaryMixture2025}.
The effect of the non-linear non-reciprocity $\ao$ was first investigated in \cite{sahaEffervescenceBinaryMixture2025}, and the non-reciprocal stiffness $\bz$ in \cite{sahaEffervescenceBinaryMixture2025,parkavousiNonreciprocalSurfaceTension2026}.

\subsection{General properties of NRCH and the consequences of the conservation law}

It is well known in the context of equilibrium dynamics that the presence of a conservation law affects certain dynamical features~\cite{hohenbergTheoryDynamicCritical1977}. Although NRCH is not governed by free energy minimization, it is instructive to consider the following class of dynamical models
\begin{align}
    \label{eq: CGLE}
    \partial_t \varphi_a = {\cal M} \big(K_a[\varphi] + W_a[\varphi]\big) + \sqrt{ 2 {\cal D}} \,\xi_a.
\end{align}
where the mobility $\cal M$ and the noise amplitude $\cal D$ might have a number of alternative structures, subject to the constraint ${\cal D}/{\cal M} = D =\mathrm{const}$. In particular, for ${\cal M} = 1$ and ${\cal D}= D$, \cref{eq: CGLE} represents the complex Ginzburg-Landau equation (CGLE) (upon defining the complex field $\varphi = \varphi_1+i \varphi_2$) ~\cite{aransonWorldComplexGinzburgLandau2002,rislerUniversalCriticalBehavior2004,rislerUniversalCriticalBehavior2005,tauberPerturbativeFieldTheoreticalRenormalization2014,siebererDynamicalCriticalPhenomena2013}, whereas for the case of ${\cal M} = -\nabla^2$ and ${\cal D}= D(-\nabla^2)$ it represents the NRCH model \cite{sahaScalarActiveMixtures2020,youNonreciprocityGenericRoute2020,pisegnaEmergentPolarOrder2024,greveCoexistenceUniformOscillatory2025,sahaPhaseCoexistenceNonreciprocal2024,sahaEffervescenceBinaryMixture2025,ranaDefectSolutionsNonreciprocal2024,ranaDefectInteractionsNonreciprocal2024,suchanekEntropyProductionNonreciprocal2023,johnsrudStateDiagramNonreciprocal2025}. These two models are the non-reciprocal extensions of the two-component model A and model B \cite{hohenbergTheoryDynamicCritical1977}, which are recovered for $W_a = 0$. At equilibrium, the conservation law affects the critical behavior by changing the dynamical scaling exponent $z$. These models are robust to changes in the dynamics because they both relax towards the same steady state, as governed by the Boltzmann distribution. This is no longer true when non-reciprocity is introduced, due to the breaking of time-reversal symmetry.

We can get further insight into the effect of the conservation law by considering the functional Fokker-Planck equation
\begin{align}
    \odv{  }{ t } P[\varphi] = - \int_{\bm x} \fdv{ }{ \varphi_a}
    J_a[\varphi],
\end{align}
for the stochastic field theory. Here, the probability current $J_a$ is defined as follows
\begin{align}
     J_a[\varphi] = 
     \left\{{\cal M} \big(  K_a[\varphi] + W_a[\varphi]\big) - {\cal D}\fdv{  }{ {\varphi_a} } \right\}
     P[\varphi].
\end{align}
Within this formulation, one can now make a powerful observation about the statistical properties of the system in steady state if the model satisfies the so-called \emph{potential conditions}~\cite{grahamQuantumStatisticsOptics1973,dekerFluctuationdissipationTheoremsClassical1975,grahamPotentialComplexGinzburgLandau1990}. These conditions require that in the field-space the two forces are orthogonal, i.e.
\begin{align}
    \label{eq: potential condition}
    \int_{\bm x} {\cal M}  W_a K_a = 0,
\end{align}
and that the non-equilibrium force is divergence free, i.e.
\begin{align}
    \label{eq: divergence-free}
\int_{\bm x} \fdv{}{\varphi_a}  {\cal M} W_a= 0.
\end{align}
While the latter is automatically satisfied by the form $W_a = -\varepsilon_{ab} f_\text{nr} \varphi_b$, the validity of the former requires certain conditions to be met; see below. 

When the system satisfies these conditions, the steady state field configurations admit a Boltzmann-like distribution of the form
\begin{align}
 \label{eq: Boltzmann}
    P^{\text{ss}}[\varphi] 
    \propto 
    \exp\left\{- \frac{  {\cal F}[\varphi] }{ D }\right\},
\end{align}
while having a non-vanishing steady-state current
\begin{align}
    \label{eq: ss-current}
    J^\text{ss}_{a}[\varphi] =  {\cal M} W_a[\varphi] P^{\text{ss}}[\varphi].
\end{align}
The current gives rise to a stationary state velocity field in the order parameter space
\begin{align}
    \frac{J^\text{ss}_{a}} {P^\text{ss}} =  {\cal M} W_a = - \varepsilon_{ab}  {\cal M}  f_\text{nr}\varphi_b,
\end{align}
from which we can extract a (pseudo-scalar) \textit{vorticity} as follows
\begin{align}
    \Omega^\text{ss}\equiv
    \varepsilon_{ab} \fdv{}{\varphi_a}  {\cal M} W_b.
\end{align}
The vorticity describes the effective local angular velocity that characterizes the rotation between the two field, and is related to the circulation (or the non-conservative work for ${\cal M}=1$) in a closed loop 
\begin{align}
\oint {\cal M} W_a \dd \varphi_a=\int \fdv{}{\varphi_a}  {\cal M} W_b\, \dd \varphi_a \wedge \dd \varphi_b=\int \Omega^\text{ss} \dd \varphi_1 \dd \varphi_2,
\end{align}
via a functional Stokes theorem \cite{obyrneGeometricTheoryExtended2024,obyrneGeometricTheoryExtended2025,Mahault2022}. This quantity can be used as a (computational of experimental) tool to probe time-reversal symmetry breaking, as has been done in the effervescent phase of nonlinear NRCH with composition as a tuning parameter \cite{sahaEffervescenceBinaryMixture2025}. 

\begin{figure*}
    \centering
    \includegraphics[width=.9\textwidth]{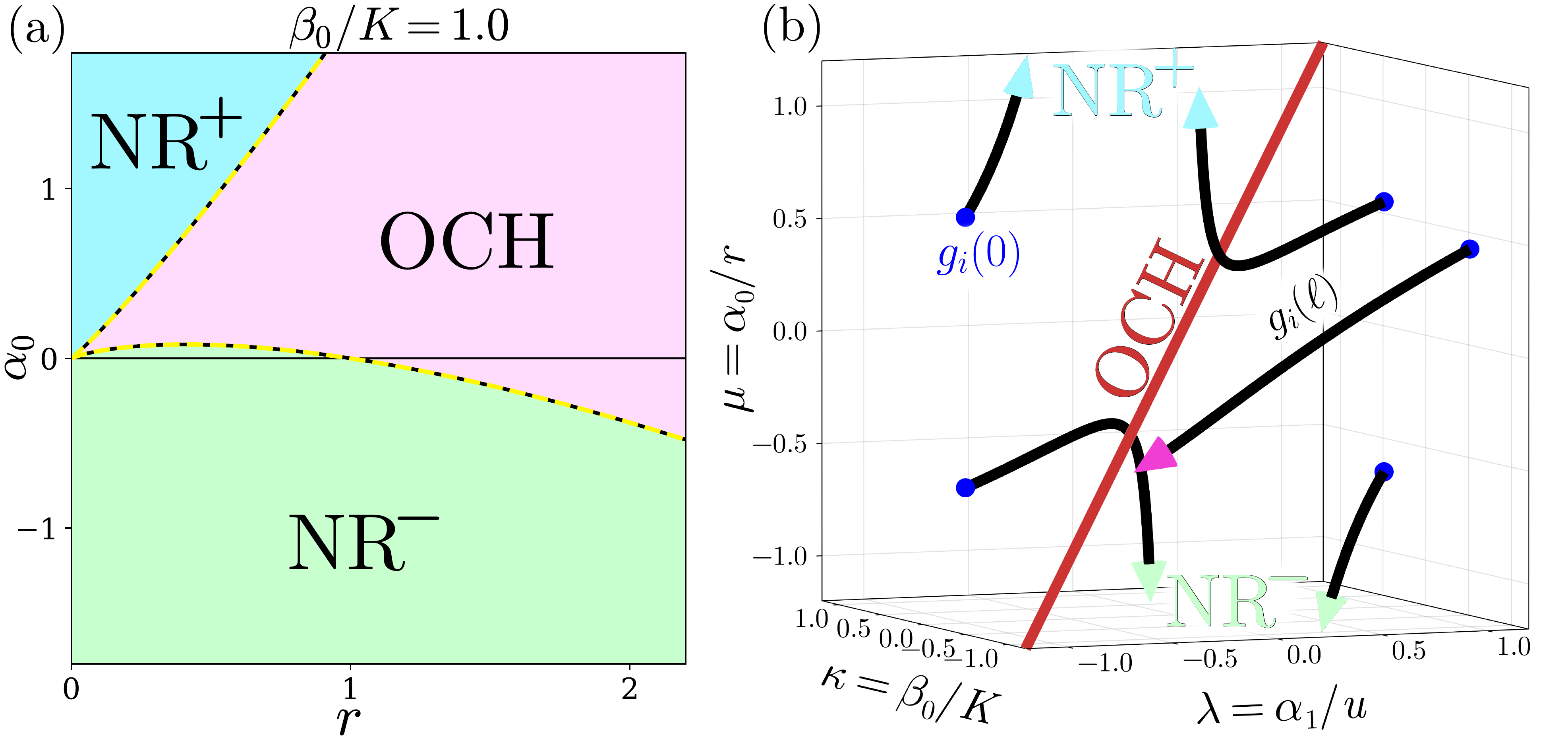}
    \caption{
    (a) The different scaling regimes are connected to different RG fixed-points, OCH and \NRpm.
    (b) The RG flow of the non-reciprocal parameters of the NRCH.
    The OCH fixed-point corresponds to the line $\alpha_1 / u = \beta_0/K  = \alpha_0/r$.
    This is unstable, and so the system will flow to either of the \NRpm fixed-point, at $\alpha_0 / r = \pm \infty$.
    }
    \label{fig: flow}
\end{figure*}

\subsection{The odd Cahn-Hilliard model}

For the CGLE case corresponding to ${\cal M} = 1$, the condition given in \cref{eq: potential condition} can be met if $\frac{\ao}{u} = \frac{\bz}{K}$, for all values of $\az$ and $r$, as we can remove $\az$ from the dynamical equation, \cref{eq: CGLE}, using the global transformation $\varphi_a\rightarrow \varphi_a' = [e^{-\alpha_0 t \varepsilon }]_{ab}\varphi_b$. For the NRCH model, which corresponds to ${\cal M} = -\nabla^2$, the potential condition [\cref{eq: potential condition}] is satisfied only if 
\begin{align}
    \label{eq: NRCH pc general}
    f_\mathrm{eq}\propto f_\mathrm{nr}.
\end{align}
This is a stricter condition because the global transformation mentioned above cannot be used to remove $\az$ due to the presence of the conservation law. If $\alpha_0$, $\beta_0$, and $\alpha_1$ are all non-zero, this condition is satisfied provided we have
\begin{align}
    \label{eq: NRCH pc}
    \frac{\ao}{u} = \frac{\bz}{K} = \frac{\az}{r} = \theta.
\end{align}
Under these conditions, the system can be treated as if it follows a functional gradient dynamics governed by free energy $\cal F$ and an \emph{odd} mobility $\Upsilon_{ab}$, namely
\begin{align}
    K_a + W_a = - \Upsilon_{ab} \fdv{\cal F}{\varphi_b},
\end{align}
where 
\begin{align}
   \Upsilon_{ab} = \delta_{ab} + \theta \,\varepsilon_{ab} \neq \Upsilon_{ba}.
\end{align}
The equation of motion then takes the form
\begin{align}
    \label{eq: odd CH}
    \partial_t  \varphi_a = \nabla^2 \Upsilon_{ab} \fdv{F}{\varphi_b} + \sqrt{2D (-\nabla^2)} \,\xi_a
\end{align}
We denote the above equation the \emph{odd Cahn-Hilliard} (OCH) model. The structure of \cref{eq: odd CH} implies a restricted version of the Fluctuation Dissipation Theorem (FDT)~\cite{johnsrudFluctuationDissipationRelations2025,johnsrudFluctuationDissipationRelations2025a}. This will be discussed in details in \cref{sec: one-loop RG}.

The difference in the requirements for the potential condition is the reason why the critical behavior of NRCH and CGLE differ. Below, we will show that this allows the system to flow far from equilibrium, break the FDT at large scales, and exhibit manifestly non-equilibrium scaling behavior. 

\subsection{Criticality out-of-equilibrium}

At equilibrium, the FDT establishes a strict relation between the equal-time correlation function $C_\text{ET}(\bm x)$ and the total integrated linear response function $\chi_\text{T}(\bm x)$, as follows
\begin{align}
    C_\text{ET}(\bm x) = k_{\rm B} T \chi_\text{T}(\bm x).
\end{align}
The spatial decay of these quantities is controlled by the correlation length $\xi$, defined as the length scale beyond which they are both exponentially suppressed:
\begin{align}
    \chi_{\text T}(\bm x), \, 
    C_{\text{ET}}(\bm x) \sim x^{-\frac{d-1}{2}} e^{-{x}/{\xi}}, &&
    x \gg \xi.
\end{align}
Away from effective equilibrium, where time-reversal symmetry is broken, such a connection between $\chi$ and $C$ does not necessary hold. Therefore, we can define structural and dynamic correlation lengths as follows
\begin{align}
    C_{\text{ET}}(\bm x) &\sim x^{-\frac{d-1}{2}} e^{-{x}/{\xiS}}, &
    x &\gg \xiS , \\
    \chi_{\text T}(\bm x) &\sim x^{-\frac{d-1}{2}} e^{-{x}/{\xiD}}, &
    x &\gg \xiD.
\end{align}
The correlation length $\xiD$ carries information about the non-equilibrium time evolution of the system, and can also be calculated from the dynamic correlation function $C(\bm x,t)$.

In our case, at the linear level of \cref{eq: EOM}, we find (see \cref{app: correlation lengths})
\begin{align}
    \xiS &= \sqrt\frac{K}{r}, &
    \xiD &= \sqrt\frac{K + i\beta_0}{r + i\alpha_0},
\end{align}
where $\xiS$ follows the standard definition of correlation length in equilibrium, while $\xiD$ is complex and depends on the non-reciprocal parameters. Note that $f_\text{eq}\propto f_\text{nr}$ implies $\xiS = \xiD$.
We can then ask how these quantities scale as the critical point is approached. 
The two scaling regimes indicate whether non-equilibrium effects are relevant or irrelevant at the scale of $\xiD$.
In contrast, for the CGLE, it was found that the large scale behavior tends towards equilibrium close to criticality~\cite{rislerUniversalCriticalBehavior2004,rislerUniversalCriticalBehavior2005,tauberPerturbativeFieldTheoreticalRenormalization2014,siebererDynamicalCriticalPhenomena2013}.
To confirm whether this behavior persists beyond the linear analysis and in the presence of a conservation law, we employ perturbative Renormalization Group (RG) techniques.

\subsection{Fundamentals of RG}
\label{sec: fundamentals of RG}

The renormalization group describes how a change of scales affects a model~\cite{kadanoffScalingLawsIsing1966,wilsonRenormalizationGroupCritical1971a,wilsonRenormalizationGroupCritical1971,wilsonRenormalizationGroupExpansion1974}.
A system is characterized by an intrinsic microscopic length scale $a$, which gives rise to a natural cutoff scale $\Lambda \sim 1/a$ in Fourier (wavevector or momentum) space. Upon coarse-graining the microscopic dynamics, all mesoscopic parameters, entering the corresponding field theory, acquire a dependence on this underlying scale. A particular realization of the NRCH, with a given value for its parameters and cutoff scale $\Lambda$, can be described as a point $V_i = (r, \az, u, \ao, K, \beta_0, D, \cdots)$ in the parameter space.
RG describes how changing the length scale from $a$ to $a e^\ell$ affects these parameters. 
The transformations result in a set of flow equations, denoted as \emph{$\beta$-functions}:
\begin{align}
    \label{eq: def beta functions}
    \odv{V_i(\ell)}{\ell} = \beta_i\big(V(\ell)\big).
\end{align}
These describe the \emph{flow} of the parameters in \emph{RG-scale} $\ell$  as we look at the same realization at different scales. The fixed-point structure of \cref{eq: def beta functions} determines the critical behavior of the system, as determined by the natural flow towards the fixed-points along the stable directions and tuning of the parameters along the unstable directions. For NRCH, both $r$ and $\az$ are tuning parameters.

A reduced representation of the RG flow of the NRCH coupling constants is illustrated in \cref{fig: flow} for the ratios between non-reciprocal coupling constants and their reciprocal counterparts, namely, $g_i(\ell) = \{\alpha_1(\ell)/u(\ell), \beta_0(\ell)/K(\ell),\alpha_0(\ell)/r(\ell)\}$. We observe that there is one unstable fixed line, defined by the potential condition \cref{eq: NRCH pc}.
The system may flow towards this for a while, before flowing towards one of the two stable \NRpm fixed-points $\alpha_0(\ell)/r(\ell) = \pm \infty$, which manifest non-reciprocal scaling behavior.

In what follows, we calculate the $\beta$-functions for NRCH first using the momentum shell approach for one-loop calculations, and then complementing it with the Callan-Symanzik scheme for the two-loop order contributions.


\section{Momentum Shell RG}
\label{sec: mom shell}

To investigate the dependence of the physical behavior of a model on length scale, we can exploit the arbitrariness of the ultraviolet cutoff $\Lambda$. The formal program of the RG method consists of three main steps \cite{kardarStatisticalPhysicsFields2007}: i) \emph{Coarse-graining}, in which short-wavelength fluctuations are integrated out by eliminating momentum modes in the shell $\Lambda/b < q < \Lambda$, where $b \gtrsim 1$ is a scaling factor. In the presence of relevant nonlinear interactions, this step is highly nontrivial, since it generates effective couplings and modifies the existing parameters of the theory. ii) \emph{Rescaling}: after coarse-graining, the theory is characterized by a reduced cutoff $\Lambda/b$. However, since the large-scale physics should not depend on the arbitrary choice of microscopic cutoff, one restores the original cutoff $\Lambda$ through appropriate rescaling of space and time. iii) \emph{Renormalization}: the above two steps lead to effective changes in the scale of the fields, which need to be adjusted to achieve self-similarity at the critical point. This transformation maps the theory onto itself while renormalizing the parameters, thereby producing a flow in the parameters space [\cref{eq: def beta functions}], ultimately leading to fixed-points where scale invariance holds. 

In this section, we will perform the one-loop RG calculation of \cref{eq: EOM} at the critical point using the momentum shell approach outlined above.

\subsection{Response-field formalism}
\label{sec: path integral}

We can represent the Langevin equation~\cref{eq: EOM} as a path-integral by introducing an auxiliary field  $\tilde \varphi$, called the response field~\cite{martinStatisticalDynamicsClassical1973,dedominicisFieldtheoreticTechniquesCritical1975,janssenLagrangeanClassicalField1976}. Within this formalism, the expectation value of an operator $\Oh[\tilde\varphi,\varphi]$ is written as an integral over all possible field-configurations in time and space, weighted by the action $\A$:
\begin{align}
    \label{eq: path integral}
    \!\!\
    \E{\Oh[\varphi,\tilde\varphi]} 
    & \propto \int \D \varphi \D \tilde\varphi \,\, \Oh[\varphi,\tilde\varphi]\,\, e^{-\A[\varphi,\tilde\varphi]},\\
    \label{eq: def A real space}
    \!\!\A[\varphi,\tilde\varphi] 
    & =\! \int\limits_{\bm x, t} \! i \tilde \varphi_a\!
    \left[  \zeta\partial_t\varphi_a \!+\! \nabla^2(K_a \!+\! W_a \!+\! Di\tilde\varphi_a) \right]\!,\!\!
\end{align}
where we have introduced a friction coefficient $\zeta$.
For example, the correlation function $C_{ab}$ and the response propagator $G_{ab}$ are defined as
\begin{align}
     \label{eq: correlation function}
     C_{ab}(\Q)\delta_{\bm q + \bm q'} \delta_{\omega+\omega'} 
     & \equiv 
     \E{\varphi_a(\Q)\varphi_b(\Qp)}, \\
    \label{eq: response propagator}
    G_{ab}(\Q)\delta_{\bm q + \bm q'} \delta_{\omega+\omega'} 
    & \equiv 
    \E{\varphi_a(\Q) i \tilde\varphi_b(\Qp)},
\end{align}
in Fourier space. These are collectively denoted as two-point functions. The definition of the Fourier transform and the delta-functions, as well as details on the path-integral, see \cref{app: response field}. 

The starting point to build a perturbative expansion for RG is to define the \emph{free} (linear) theory.
We split the action into free and interacting parts, 
\begin{align}
    \label{eq:intAction}
    \A[\varphi, \tilde\varphi] = \A_0[\varphi, \tilde\varphi] + \A_{\rm I}[\varphi, \tilde\varphi].
\end{align}
Linearizing \cref{eq: EOM} gives a path-integral in terms of the free action $\mathcal{A}_0$, as follows:
\begin{align}
    \A_0 &=\! \int_{\bm x, t}\! i\tilde \varphi_a
    \left[ (\zeta\partial_t + S)\varphi_a + A\varepsilon_{ab}\varphi_b+ \nabla^2 Di\tilde\varphi_b \right ]\!,\\
    \label{eq: realspace S A}
    S &\equiv -\nabla^2 (r - K\nabla^2), \hspace{6mm}
    A \equiv -\nabla^2(\alpha_0 - \beta_0 \nabla^2).
\end{align}
The interacting part can be written as
\begin{align}
    \A_{\rm I}[\varphi, \tilde{\varphi}] 
    &= \int_{\bm x, t} i \tilde \varphi_a(-\nabla^2 U_{abcd} ) \varphi_b \varphi_c \varphi_d,
\end{align}
in terms of a symmetrized interaction tensor, which is defined as
\begin{align}
    \label{eq: def gabcd}
    U_{abcd} &\equiv 
    \frac{1}{3}
    \left( 
        \Xi_{ab}\delta_{cd}
        +\Xi_{ad}\delta_{bc}
        +\Xi_{ac}\delta_{db}
    \right), \\
    \Xi_{ab} & \equiv u\delta_{ab}  + \alpha_1\varepsilon_{ab}.
\end{align}
Expectation values in terms of the free action $\A_0$ are denoted by the $0$ subscript. 
For example, the free two-point functions are
\begin{align}
    \label{eq: free response}
    G_{0,a b}(\bm q, \omega)
    & =  
    [-i \zeta \omega\delta_{ab} + S( q) \delta_{ab} + A( q)\varepsilon_{ab}]^{-1}, \\
    \label{eq: free corr}
    C_{0,ab}(\bm q, \omega) 
    &= 2 D q^2 \left[ G_0(\bm q, \omega) G_0^\dagger(\bm q, \omega) \right]_{ab},
\end{align}
where the Fourier-transformed linear operators defined in \cref{eq: realspace S A} are 
\begin{align}
    S(q)  = q^2(r + K q^2), && A(q)  = q^2 (\az + \bz q^2).
\end{align}

\subsection{One-loop RG}
\label{sec: one-loop RG}

To apply RG transformation on the theory, we rewrite \cref{eq:intAction} in Fourier space, where modes with different wave numbers couple due to the nonlinearities. We decompose the fields as $\varphi_{a}^<(\bm q, \omega)$ corresponding to $q< \Lambda/b$, and $\varphi_{a}^>(\bm q,\omega)$ on the shell $\Lambda/b < q < \Lambda$, together with the corresponding response fields, thus obtaining
\begin{equation}
     \A = \A_{0}^< + \A_{0}^> + \A_{\rm I}^{<,>} . 
\end{equation}
We then integrate out the contributions on the shell, such that the new probability measure depends only on the new cutoff $\Lambda/b$. To achieve this, we apply standard perturbative field-theory techniques using Feynman diagrams.
The graphical building blocks are
\begin{align}
    \parbox{10mm}{
    \begin{fmfgraph*}(10,6)
        \setval
        \fmfleft{i}
        \fmfright{o}
        \fmf{response}{i,o}
        \fmfv{l=$a$,l.d=2pt}{i}
        \fmfv{l=$b$,l.d=2
        pt}{o}
    \end{fmfgraph*}
    }\hspace{2mm}
    & = G_{0,a b}(\bm q_a, \omega_a) \delta_{\bm q_a + \bm q_b} \delta_{\omega_a + \omega_b} ,\\
    \parbox{10mm}{
    \begin{fmfgraph*}(10,6)
        \setval
        \fmfleft{i}
        \fmfright{o}
        \fmf{plain}{i,o}
        \fmfv{l=$a$,l.d=2pt}{i}
        \fmfv{l=$b$,l.d=2pt}{o}
    \end{fmfgraph*}
    }\hspace{2mm}
    & = C_{0,a b}(\bm q_a, \omega_a) \delta_{\bm q_a + \bm q_b} \delta_{\omega_a + \omega_b},\\
    \vspace{2mm}
    \parbox{12mm}{
    \centering
    \begin{fmfgraph*}(12,10)
        \setval
        \fmfleft{i1,i2}
        \fmfright{o1,o2}
        \fmfv{label=$a$,label.angle=180,l.d=2pt}{i1}
        \fmfv{label=$b$,label.angle=180,l.d=2pt}{i2}
        \fmfv{label=$c$,label.angle=0,l.d=2pt}{o1}
        \fmfv{label=$d$,label.angle=0,l.d=2pt}{o2}
        \fmf{wiggly}{i1,v}
        \fmf{plain}{v,o1}
        \fmf{plain}{i2,v,o2}
    \end{fmfgraph*}
    }\hspace{1.3mm}
    &
    = - q_a^2 U_{a b c d} \delta_{\Sigma_i \bm q_i} \delta_{\Sigma_i \omega_i},
\end{align}
which can be used calculate the corrections to the two and four points functions. Due to the conservation law, the one-loop corrections to $D$ are null, while the propagator $G$ obtains non-trivial contributions from the self-energy $\Sigma$, defined as
\begin{equation}
    \tilde{G}_{ab}^{-1}(\bm q, \omega) = G_{0,ab}^{-1}(\bm q, \omega) - \Sigma_{ab}(\bm q, \omega).
\end{equation}
We define this in terms of corrections to $r$ and $\az$:
\begin{align}
    \Sigma_{ab}(\bm q, \omega) 
    \equiv -q^2 (\delta r \delta_{ab} + \delta \az \varepsilon_{ab})
    + \Oh(q^4, \omega).
\end{align}
The leading order contribution to $\Sigma$ is
\begin{align}
    \Sigmaab
    = - 3 q^2 U_{a b c d}\int_{\bm k, \varpi}^{>}\!\! C_{0,c d}(\bm k, \varpi),
\end{align}
where the upper limit symbol $>$ indicates that the integration on $\bm k$ is to be performed over the momentum shell.
The coarse-grained corrections arising from this diagram renormalize the parameters as
\begin{align}
    \tilde r(b) &= r + \delta r(b), \\
    \tilde \alpha_0(b) &= \alpha_0+\delta \alpha_0(b),
\end{align}
while $\zeta,\beta_0,K,D$ are not renormalized. 

The interaction tensor $U_{abcd}$, and thus $u$ and $\ao$, also get diagrammatic corrections.
The one-loop contribution is
\begin{align}
    \!\!\deltag\Big|_\text{sym}
    \equiv -q^2\delta U_{a b c d} + \Oh(q^4, \omega),
\end{align}
where the $\text{sym}$-subscript indicates symmetrization with respect to equivalent external legs.
This contributes to the corrected coupling tensor as
\begin{align}
    \tilde
    U_{abcd} 
    & \equiv U_{abcd} + \delta U_{abcd}(b).
\end{align}
The tensor $\tilde U_{abcd}(b)$ is related to the corrected non-linear couplings $\tilde u(b)$ and $\tilde \ao(b)$ as $U_{abcd}$ is to $u$ and $\ao$, see \cref{eq: def gabcd}.
We calculate the diagram to the leading order in $q^2$:
\begin{align}
    &
    \deltag
    \hspace{-2mm}
    \\
    &= 18 q^2 {\int\limits_{\bm k, \varpi}\hspace{-.7mm}}^{>}\hspace{-1mm}   U_{\s a \s b e \! f} G_{0,e \s g}(\bm k, \varpi) 
    C_{0,f\! h}^T( \s -\bm k, \s -\varpi) k^2  U_{\s g \s h \s c d}\s.
    \label{eq: mom shel diagram 2}
\end{align}
Evaluating these diagrams to integrate out the modes on the shell recasts the effective action into ${\cal A}[\varphi, \tilde \varphi; \tilde{v_i}, \Lambda/b]$, where the parameters are renormalized and the effective cutoff depends on the scaling parameter $b$. 
To restore the cutoff $\Lambda$ we proceed with the rescaling of space and time, and the renormalization of the fields, as follows
\begin{eqnarray}
    && x \to x'=x/b, \qquad t \to t'=t/b^z, \\
    && \varphi \to \varphi' = b^\chi \varphi, \qquad \tilde \varphi \to \tilde \varphi ' = b^{\tilde \chi} \tilde{\varphi},
\end{eqnarray}
defining the dynamical critical exponent $z$, and the field scaling dimensions $\chi,\tilde \chi$. Note that these scaling relations are valid in real space, i.e. performed on \cref{eq: def A real space}. 

The conservation law of the dynamical equation \cref{eq: EOM} produces a relation between the scaling dimensions of the fields. Starting from the transformation
\begin{eqnarray}
      &&  \zeta \int_{\bm x,t} i \tilde \varphi \partial_t \phi  \to \zeta\ b^{d+z- \chi - \tilde \chi -z} \int_{\bm x',t'} i \tilde \varphi \partial_{t'} \varphi,
\end{eqnarray}
and invoking the condition that it remains scale invariant, we obtain the following identity
\begin{align}
    \chi + \tilde \chi & = d.
\end{align}
We can now assign physical dimensions to all the parameters of the theory and define the following recursion relations for the parameters that are renormalized at one-loop
\begin{align}
    \label{eq: dimensional analysis r alpha0}
    r(b) & = b^2 \tilde r(b), 
    & \alpha_0(b) &= b^2 \tilde{\alpha}_0(b),\\ 
    \label{eq: dimensional analysis u alpha1}
    u(b) &= b^{4-d} \tilde{u}(b), & 
    \alpha_1(b) &= b^{4-d} \tilde{\alpha}_1(b) ,
\end{align}
and for those that do not receive corrections and only acquire scaling due to their physical dimensions
\begin{align}
    \label{eq: dimensional analysis K}
    K(b) &= b^{z-4} K, \\
    \label{eq: dimensional analysis beta}
    \beta_0 (b) &= b^{z-4} \beta_0,\\
    \label{eq: dimensional analysis D}
    D(b) &= b^{d+z-2-2 \tilde \chi} D .
\end{align}
The variation of these rescaled quantities with respect to $\ell=\ln b$ measures how the parameters change when performing an infinitesimal RG transformation, and defines the corresponding flow-functions.
The one-loop analysis gives the following RG flow functions 
\\\noindent\null\vspace{-2mm}
\begin{align}
    \label{eq: MS beta r}
    \odv{r(\ell)}{\ell}
    & = 2 r + \frac{ 4 D A_d \Lambda^{d-2} u }{ \zeta(K + r \Lambda\!^{-2}) } ,
    \\
    \label{eq: MS beta a0}
    \odv{\az(\ell)}{\ell}
    & = 2 \alpha_0 + \frac{ 4 D A_d \Lambda^{d-2} \alpha_1 }{ \zeta(K + r \Lambda\!^{-2}) },
\end{align}
\noindent\null\vspace{-8mm}
\begin{widetext}
\noindent\null\vspace{-5mm}
\begin{align} 
\label{eq: MS beta u}
    \odv{u(\ell)}{\ell} 
    & = \epsilon u 
     - \frac{2  D A_d \Lambda^{-\epsilon}}{\zeta (K + r\Lambda\!^{-2})^2}
    \left(5 u^2-\frac{ 
        \left[\ao(K + r\Lambda\!^{-2}) - u (\bz + \az\Lambda\!^{-2})\right]^2}{
            (K + r\Lambda\!^{-2})^2 + (\bz + \az\Lambda\!^{-2})^2 
        }\right), 
    \\
   \label{eq: MS beta a1}
    \odv{\ao(\ell)}{\ell} 
    & =
    \epsilon \ao
    -  \frac{2  D A_d \Lambda^{-\epsilon}}{\zeta (K + r\Lambda\!^{-2})^2}
    \!\left(5 \ao u+ \frac{  \left[\ao(K + r\Lambda\!^{-2})-u (\bz + \az\Lambda\!^{-2})\right]
    \left[\ao (\bz + \az\Lambda\!^{-2}) + u (K + r\Lambda\!^{-2})\right]
                }{
            (K + r\Lambda\!^{-2})^2 + (\bz + \az\Lambda\!^{-2})^2
        }\right)\!
   .
\end{align}
\noindent\null\vspace{-2mm}
\end{widetext}
The geometric factor, defined in \cref{eq: geometric factor}, is $ A_d =  2 / [(4 \pi)^{d/2} \Gamma(d/2)]$, and $\epsilon\equiv4-d$.

Since we are interested in the scaling behavior of the system near the critical point, we can simplify the expressions of the RG flow equation by incorporating $r \ll K \Lambda^2$ and $\az \ll \bz \Lambda^2$. Using the definition
\begin{align}
    \ded \equiv \frac{D A_d}{\zeta K^2}{\Lambda\!}^{-\epsilon},
\end{align}
we write the simplified RG flow equations as follows
\begin{align}
    \odv{r}{\ell} 
    & = 
    2r + 4 \ded u  (K \Lambda^2 - r), \label{eq: MS beta r 2}\\
    \odv{\az}{\ell}
    & = 2\az + 4 \ded \ao (K \Lambda^2 - r), \label{eq: MS beta a0 2}\\
    \odv{u}{\ell}
    & = \epsilon u 
    - 10 \ded  
    \left[u^2-\frac{(\ao K\!-\!u\bz)^2 }{5 (K^2 + \bz^2)}\right], \label{eq: MS beta u 2}\\
    \odv{\ao}{\ell} 
    &= \epsilon \ao 
    - 10 \ded  \bigg[\ao u \!
    +\!
    \frac{\!(\ao K\!-\!u\bz)(\ao\bz\! +\! u K)\!}{5 (K^2 + \bz^2)}
    \bigg].\label{eq: MS beta a1 2}
\end{align}

The fixed-points of the RG flow represent scale invariant points in the parameter space, and are investigated using an $\epsilon$-expansion. In addition to an unstable Gaussian fixed-point for vanishing nonlinearities $u^*, \alpha_1^*=0$, we obtain a non-trivial fixed-point
\begin{align}
\label{eq:fixedpoint1L}
    r^* & = - (K\Lambda^{2}) \frac{\epsilon}{5}  , &
    \alpha^*_0 & = - \frac{\bz}{K} (K\Lambda^{2}) \frac{\epsilon}{5} , \\
    u^* &= \frac{1}{\dedc} \,\frac{\epsilon}{10}, &
    \alpha_1^* &= \frac{\bz}{K} \frac{1}{\dedc} \,\frac{\epsilon}{10}. 
\end{align}
to the leading order.
To understand the physical meaning of this fixed-point, we note the fixed-point values satisfy the following relation
\begin{equation}
    \frac{\alpha_1^*}{u^*} = \frac{\beta_0}{K} = \frac{\alpha_0^*}{r^*}= \theta ,
\end{equation}
where $\theta$ is a free-parameter set by the initial conditions. This relation corresponds to the potential condition of \cref{eq: NRCH pc} that gives rise to the OCH model defined in \cref{eq: odd CH}. 

Note that the fixed-point values explicitly depend on the parameters that are not renormalized at one-loop order, namely $\zeta, \beta_0,K$ and $D$, as well as the cutoff $\Lambda$. For the coefficients $\zeta$ and $D$, as previously discussed, the conservation law imposes that no perturbative corrections are generated at higher orders, so that these parameters retain their bare values throughout the RG flow. The situation is different for $\beta_0$ (see below).
In this case, there is no analogous protection mechanism, and higher-order fluctuations can in principle generate nontrivial renormalizations. Indeed, contributions to their flow are expected to appear already at two-loop order, as observed in closely related models \cite{halperinRenormalizationgroupMethodsCritical1976, rislerUniversalCriticalBehavior2004}. 

For the OCH field theory with $\theta \neq 0$, we find the following \emph{non-perturbative} fluctuation-dissipation relation \cite{johnsrudFluctuationDissipationRelations2025a,johnsrudFluctuationDissipationRelations2025}
\begin{align}\label{eq: FDT1}
    G_{ab}(\bm q, \omega)
    -
    G_{ba}(-\bm q, -\omega)
    = \frac{i\omega}{D} C_{ab}(\bm q, \omega),
\end{align}
to hold, while in general
\begin{align}
    \label{eq: FDT2}
    G_{ab}(\bm q, \omega)
    -
    G_{ab}(-\bm q, -\omega)
    &\neq \frac{i\omega}{D} C_{ab}(\bm q, \omega),\\
    \label{eq: FDT3}
    C_{ab}(\bm q, \omega) - C_{ab}(-\bm q, - \omega) &\neq 0,
\end{align}
in the absence of time-reversal symmetry. Moreover, even \cref{eq: FDT1} may be manifestly violated for the NRCH model away from (but close to) criticality, as we will discuss below.

\begin{figure*}[t]
    \centering \includegraphics[width=.9\textwidth]{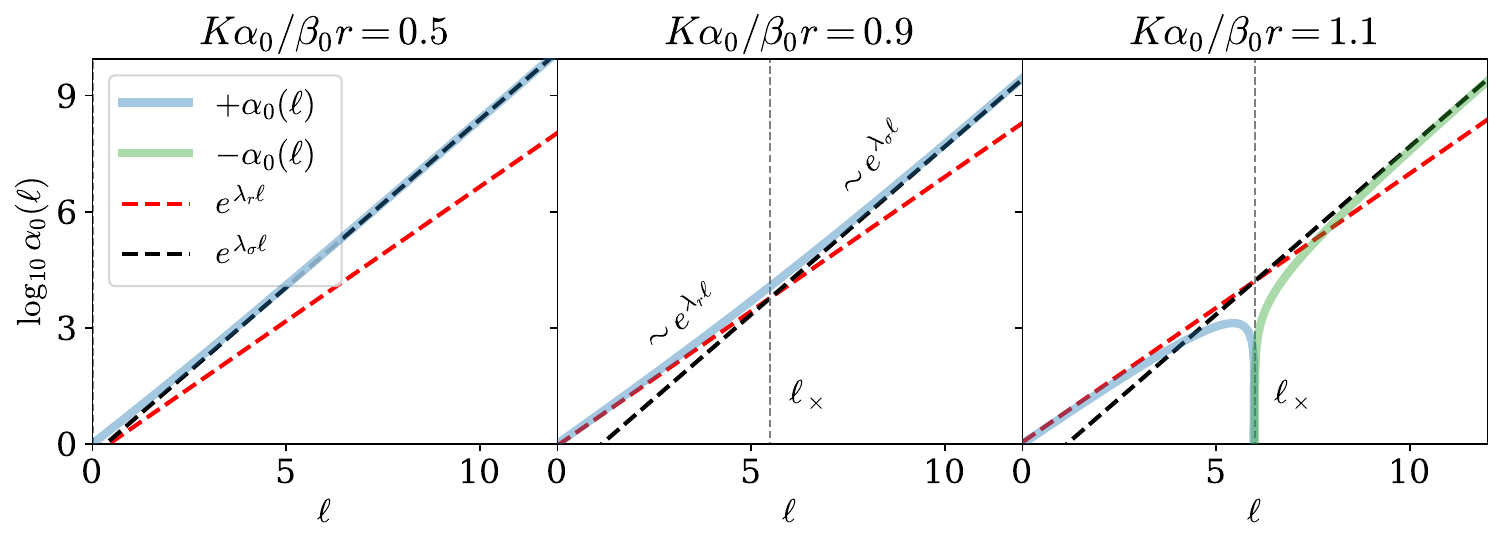}
    \caption{ 
    The flow of the effective tuning-parameter $\az$ according to the $\beta$-functions calculated with the momentum-shell approach, as a function of RG-scale $\ell = \ln b$, for $u(0) = r(0) = 1$.
    Depending on the initial values of $r$, $\az$, $\bz$ and $K$, there is a crossover RG-scale $\ell_\times$, and a corresponding length-scale $\xic$, where the scaling-behavior changes.
    This may be accompanied by a change in sign.
    }
    \label{fig: Crossover}
\end{figure*}

We can now assess the stability of this fixed-point by considering $\theta = \bz / K$ as a finite parameter. We analyze the RG flow in the neighborhood of the fixed-point by linearizing the flow equations around its values as $V_i(\ell) = V_i^* + \Delta V_i(\ell)$, where $V_i = (r, \alpha_0, u, \alpha_1)$. We find
\begin{align} 
    \label{eq:dynpert}
    \odv{}{\ell}\Delta V_i(\ell) = {\cal J}_{ij} \Delta V_j(\ell),
\end{align}
where ${\cal J}$ is the Jacobian matrix of the flow functions evaluated at the fixed-point, namely ${\cal J}_{ij} \equiv \partial_j \beta_i(V^*) $, according to the definition of $\beta_i$ given in \cref{eq: def beta functions}.
It reads
\begin{align}
       &{\cal J}= \nonumber\\* 
       &\!\!
    \begin{bmatrix}
        2 \left(1 - \frac{\epsilon}{5}\right) & 0 
        & 4 \dedc K \Lambda^{2}  \left(1+\frac{\epsilon}{5}\right) & 0 \\
        - 2 \frac{\beta_0}{K} \frac{\epsilon}{5} & 2 & 0 
        & 4 \dedc K \Lambda^{2} \left(1+\frac{\epsilon}{5}\right) \\
        0 & 0 & - \epsilon & 0 \\
        0 & 0 & - 4 \frac{\beta_0}{K} \frac{\epsilon}{5} & - \frac{\epsilon}{5}
    \end{bmatrix}\!. 
     \label{eq:jacobian}
\end{align}
The eigenvalues $\lambda_i$ of this matrix determine the scaling directions of the RG flow and therefore characterize the stability of the fixed-point under infinitesimal perturbations.
The directions with a well-defined scaling forms are the corresponding \emph{left} eigenvectors\footnote{There will be some mixing between the $r,\az$-sector and the $u,\ao$-sector, which can be ignored as the terms are higher order in $\epsilon$ when non-universal quantities are subtracted.}:
\begin{align}
    \label{eq: eiegnvalues momentum shell}
    \lambda_r = 2 - \frac{2\epsilon}{5},&&
    \lambda_\sigma = 2, &&
    \lambda_u = -\epsilon, &&
    \lambda_{w} = - \frac{\epsilon}{5}.
\end{align}
Here, $\sigma \equiv \alpha_0 - \theta r$ and $w = \ao - \theta u$, where $\theta = \bz/K$.
We therefore conclude that the fixed-point is infrared stable with respect to the nonlinear couplings, while exhibiting two relevant directions. The first corresponds to the usual $r$ term, associated with the tuning of temperature across the transition. 
The second is linked to linear non-reciprocity $\az$ and emerges as an additional relevant parameter of the critical theory.
At this point, there is no correction beyond the Gaussian value of the anomalous scaling of the correlation function, and neither to the dynamical exponent, which are connected due to the conservation law, namely
\begin{align}
    \eta = 0 + \Oh(\epsilon^2), && z = 4 - \eta=4+\Oh(\epsilon^2).
\end{align}
Understanding the role of the second tuning parameter requires a more careful analysis.

\subsection{Non-reciprocity as tuning parameter}
\label{subsec: nr as tuning}

What distinguishes a parameter as a \emph{tuning}  parameter of a critical system is that it can be externally controlled to drive the emergence of long-range correlations. This property is associated with the relevance in the RG sense: the corresponding eigenvalue of a perturbation around the fixed-point is positive, as shown in \cref{eq: eiegnvalues momentum shell}. In other words, any nonzero deviation from the fixed-point value grows under coarse-graining, causing the system to flow away from criticality at large scales. Therefore, for the system to exhibit critical behavior, the tuning parameters must be adjusted toward their fixed-point values.

The above analysis shows that both $r$ and the linear non-reciprocal coupling $\alpha_0$ correspond to relevant parameters.
This feature marks a fundamental difference with respect to the CGLE, i.e the non-conserved counterpart of NRCH, in which case the linear non-reciprocal term does not affect the universal critical behavior, as it can be eliminated through a global phase rotation of the order parameter. 

To gain insight into the roles of the two tuning parameters, we integrate \cref{eq:dynpert} by restricting the analysis to the perturbations associated with the two relevant directions. This yields the following expressions for the scale-dependent tuning parameters
\begin{align}
    \label{eq: r(l)}
    \dr(\ell)&= \dr(0) e^{\lambda_r\ell}, \\
    \label{eq: a(l)}
    \hspace{-2mm}
    \daz(\ell)
    &= 
    \frac{\beta_0}{K}
    \dr(0) e^{\lambda_r\ell}
    \! +
    \!\left[ \daz(0) - \frac{\beta_0}{K} \dr(0) \right]\! e^{\lambda_\sigma\ell}\!,
    \!\!
\end{align}
with eigenvalues being $\lambda_r = 2 - 2 \epsilon/5$ and $\lambda_\sigma = 2$. 
The reduced temperature $\Delta r$ evolves independently under the RG flow, whereas the non-reciprocal coupling $\Delta \alpha_0$ is mixed with $\dr$. 
As a consequence, the scale dependence of $\Delta \alpha_0$ contains two competing contributions, associated with the two distinct relevant eigenvalues. 
For simplicity of notation, from now on we will indicate with $r$ and $\alpha_0$ the difference from the corresponding fixed-point values, namely $\dr \to r$ and $\daz \to \alpha_0$, which therefore have null fixed-points, in line with the standard additive renormalization procedure.

The competition between these two behaviors gives rise to a crossover phenomenon: depending on the relative magnitude of the terms in \cref{eq: a(l)}, the scaling behavior of $\az$
can be controlled by its own scaling or by temperature scaling, as illustrated in \cref{fig: Crossover}. The crossover occurs at the RG-scale $\ell_\times$, corresponding to the characteristic length scale
\begin{align}
    \label{eq: crossover momentum shell}
    \xic & = \Lambda^{-1} \left| \frac{ K }{ \beta_0 } \frac{ \alpha_0 }{ r } - 1 \right|^{- \frac{1}{\omega_\c}},\\
    \omega_\c & \equiv \lambda_\sigma - \lambda_r = \frac{2 \epsilon}{5} + \Oh(\epsilon^2).
\end{align}
The competition between the two types of scaling can suppress the effective non-reciprocal coupling under certain conditions, ultimately leading to the emergence of a length scale at which the effective non-reciprocity parameter vanishes. Such an effect has been previously discovered in the context of the emergent phenomenon of \emph{effervescence}, which was observed deep in the ordered phase of the NRCH model~\cite{sahaEffervescenceBinaryMixture2025}.

\subsection{Scaling hypothesis}
\label{subsec: scaling hypo}

\begin{figure*}
    \centering
    \includegraphics[width=\textwidth]{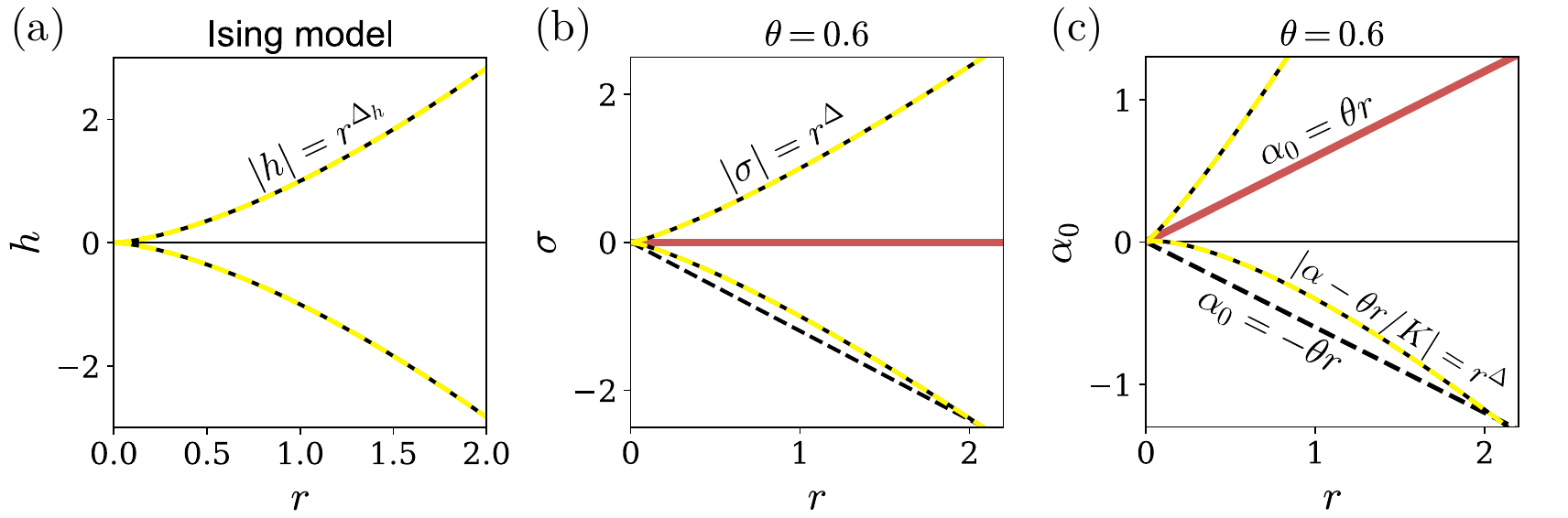}
    \caption{(a) The scaling behavior of the Ising universality class, as a function of the two tuning-parameters $r$ and $h$.
    (b) The scaling behavior of the NRCH, as a function of $\sigma$ and $r$.
    (c) The scaling behavior of the NRCH as a function of $\az$ and $r$, for a given value of $\beta_0$.
    The red line represents the OCH fixed-points.
    The yellow line in all cases illustrates the crossover in the scaling-function, from $|s| \ll 1$ to $|s| \gg 1$.}
    \label{fig: phase diagram}
\end{figure*}

At equilibrium, an analogous scenario is observed when a system can be tuned to criticality by using two different parameters, temperature $r$ and external magnetic field $h$, in the Ising universality class where magnetization is described by a single scalar field $\phi(\bm x)$ governed by the free energy
\begin{align}
    {\cal F}_h[\phi] = \int_{\bm x} \left[ \frac{r}{2}\phi^2 + \frac{K}{2}(\nabla\phi)^2 + \frac{u}{4} \phi^4 - h \phi \right].
\end{align}
Near the critical point, the correlation length diverges as $\xi \sim r^{-\nu}$ when $h=0$, defining the critical exponent $\nu$. On the other hand, when  $r=0$, but $h$ is non-zero, the correlation length scales as $\xi \sim |h|^{- {\nu}/{\Delta_h}}$. These two behaviors are captured by the scaling hypothesis~\cite{kardarStatisticalPhysicsFields2007,goldenfeldLecturesPhaseTransitions2019}
\begin{align}
    \xi(r, h) 
    &=
    r^{-\nu} {\G}_{hr}\left(\frac{h}{r^{\Delta_h}\!\!}\right), \\
    {\G}_{hr}(s) 
    &\sim 
    \begin{cases}
        1, & |s|\ll 1, \\
        |s|^{-{\nu}/{\Delta_h}},  & |s| \gg 1.
    \end{cases}
\end{align}
Here, ${\G}_{hr}$ is the scaling function, and $\Delta_h$ is the gap-exponent that keeps the ratio $h/r^{\Delta_h}$ invariant under RG. We can therefore draw a diagram in the $r-h$ plane, where the line $r = |h|^{\Delta_h}$ distinguishes the two scaling regimes. This is illustrated in \cref{fig: phase diagram}(a).

The NRCH model retains $r$ while another endogenous control parameter $\alpha_0$ emerges naturally without the need for an externally imposed symmetry-breaking field. As previously discussed, $\xiS$ is independent of $\az$ at the linear level, while $\xiD$ is not. We therefore consider $\xi = \xiD$, and propose a scaling hypothesis based on the above observations. \Cref{eq: a(l)} shows that the two tuning-parameters of the NRCH are mixed, as $\az$ does not have scaling behavior alone. This yields a scaling hypothesis involving the standard exponent $\nu$ and a novel scaling exponent $\nu_{\rm n}$, as follows
\begin{align}
    \label{eq: scaling hypothesis}
    \xi(\alpha_0, r)
     &=
    r^{-\nu}
    {\G}_{\sigma r} \left(\frac{ \sigma }{ r^{\Delta} }\right), \\
    {\G}_{\sigma r} (s) & \sim 
    \begin{cases}
        1, & |s| \ll 1, \\
        |s|^{-\nu_{\rm n}}, & |s|\gg 1,
    \end{cases}
\end{align}
\vspace{-5mm}
\begin{align}
    \sigma &\equiv \alpha_0 - \theta r, &
    \Delta \equiv \frac{\nu}{\nu_{\rm n}} &= 1 + \frac{\epsilon}{5} + \Oh(\epsilon^2).
\end{align}
Here, $\theta = \frac{\beta_0}{K}$, and we have defined
\begin{align}
    \nu &\equiv \frac{1}{\lambda_r}= \frac{1}{2} + \frac{\epsilon}{10} + \Oh(\epsilon^2), 
    \\
    \nu_{\rm n} &\equiv \frac{1}{\lambda_\sigma} = \frac{1}{2} + \Oh(\epsilon^2).
\end{align}
With this, we draw a scaling diagram in terms of $r$ and $\sigma$ in \cref{fig: phase diagram}(b).
The scaling behavior is controlled by the relevant asymptotic regime of ${\G}_{\sigma r}(s)$, as determined by where the microscopic parameters are located in relation to the line $|\sigma|=r^\Delta$. The transition between the two regimes coincides with the point where correlation length $\xiD$ equals the crossover length $\xic$, found in \cref{eq: crossover momentum shell}.

We can draw this diagram in terms of the original parameters $r$ and $\az$. We assume that $\bz > 0$, so $\theta > 0$ (without loss of generality), and consider above the critical point, i.e. $r > 0$.
The black-and-yellow lines in \cref{fig: phase diagram}(b) correspond to $|\az - \theta r| = r^\Delta$, and thus they are mapped to $\az = \pm r^\Delta + \theta r$ in \cref{fig: phase diagram}(c).
Between these lines, the scaling is dominated by $r^{-\nu}$.
Outside of these lines, however, the scaling is given by $|\sigma|^{-\nu_{\rm n}}$, but whether the behavior of $\sigma$ is dominated by $r$ or $\az$ depends on their relative strengths.
This is illustrated in \cref{fig: phase diagram}(c), as well as \cref{fig: Illustration}, for more values of $\theta$. We confirm this scaling hypothesis in the Callan-Symanzik approach in \cref{sec: NR scaling}.

\subsection{Universal features}

The predictions obtained so far, including fixed-point values and crossover length scale,  depend explicitly on the parameter $\beta_0/K$. Since the one-loop calculation leaves this ratio unchanged, a conclusive validation of our results requires a calculation at two-loop order, where non-zero perturbative corrections might be obtained. To perform this task, we turn to a Callan–Symanzik RG approach. To this end, we start by presenting our previous results in a way that facilitates the connection between the two approaches.

We start by selecting one relevant operator as the tuning parameter, e.g. $r$, and combining all the others into marginal operators, by using ratios. We then define our coupling constants, collectively denoted as $g_i \equiv (\l, \a, \b,\c)$, as follows
\begin{align}
    \l \equiv \ded u, &&
    \a \equiv \frac{\alpha_1}{u}, &&
    \b \equiv \frac{\beta_0}{K}, &&
    \c \equiv \frac{\alpha_{0}}{ r }.\label{eq: new var def}
\end{align}
Here, we have applied additive renormalization of $2 K \Lambda^2 \l$ and $2K\Lambda^2 \a \l$ to $r$ and $\az$, respectively, to remove non-universal features. At the fixed-point, this reduces to $\dr$ and $\daz$, which we then relabel as $r$ and $\az$, as discussed above.
The flow functions given in \cref{eq: MS beta r 2,eq: MS beta a0 2,eq: MS beta u 2,eq: MS beta a1 2}, now read as follows
\begin{align}
    \label{eq: beta l MS}
    \odv{\l}{\ell} =\beta_\l&= \epsilon \l - 10 \l^2 \left[ 1 - \frac{1}{5} \frac{(\b-\a)^2}{1+\b^2} \right], \\
    \label{eq: beta a MS}
    \odv{\a}{\ell}=\beta_\a&= - 2 \l (\a-\b) \frac{1+\a^2}{ 1+\b^2 }, \\
    \label{eq: beta b MS}
    \odv{\b}{\ell}=\beta_\b
    &= 0 + \Oh(\l^2), \\
    \label{eq: beta c MS}
    \odv{\c}{\ell}=\beta_\c
    &= 4 \l (\c-\a), 
\end{align}
in terms of the new variables defined in \cref{eq: new var def}. We now introduce the second approach to RG.

\section{Callan-Symanzik RG}
\label{sec: cs approach}

In this approach we exploit the arbitrariness of the ``renormalization scale'' $\Lambda$ used in dimensional regularization, which plays a similar role to the cutoff in the momentum shell approach~\cite{kleinertCriticalPropertiesPhi2001,zinn-justinQuantumFieldTheory1989}.
The procedure ensures that the resulting physical observables are independent of the regulator $\Lambda$ by introducing renormalized fields and constants, denoted hereafter by an over bar, namely $\ren \varphi$, $\rr$, $\ru$, etc.
The constants appearing in the equation of motion, \cref{eq: EOM}, called the \emph{bare} constants, are related by renormalization factors, such as $Z$, $Z_r$, $Z_u$, which absorb the dependence on $\Lambda$ and are thus divergent when the regularization is not employed.
Our model is defined in terms of dimensionless constants $g_i = (\l, \a, \b, \c)$ and a tuning parameter (either $r$ or $\az$), and the renormalization factors depend only on $g_i$, as we will see below.
Renormalized quantities are defined in terms of renormalized vertex functions, $ \bar \Gamma^{(\tilde N, N)}$ (see \cref{app: response field}).
In particular, $\ren \Gamma^{(1,1)}_{ab}(\bm q, \omega)$ is the inverse of $\E{\ren\varphi_a(\bm q, \omega) i \ren{\tilde\varphi}_b(\bm q', \omega')}$, and the four-point vertex-function $\ren\Gamma^{(1, 3)}$ defines $\rl$ and $\ra$.

\subsection{Scaling functions}

\begin{figure*}[t]
    \centering
    \includegraphics[width=\textwidth]{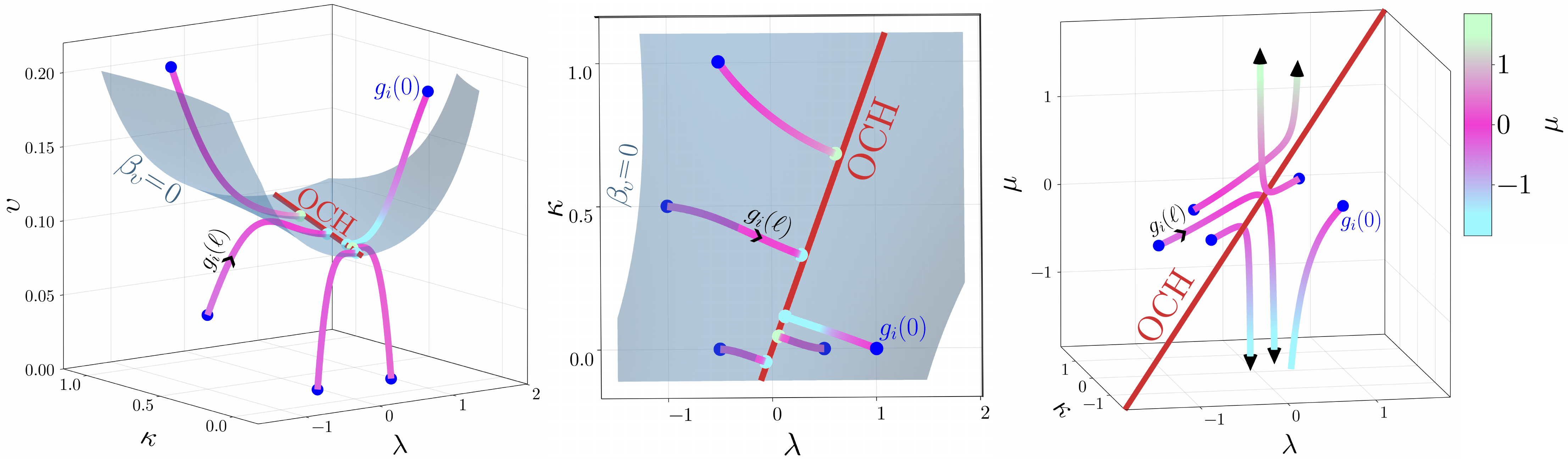}
    \caption{
        The flow of a small sample of points in parameter space.
        The points first flow towards the $\beta_\l = 0$ surface, shown in steel blue, then toward the $\a = \b$ line OCH, shown in red.
        However, $\c$ does not approach a finite value, but rather flows away to $\c = \pm \infty$.
        The color of the line indicates the value of $\c$ along it, according to the color bar.
        The initial point $g_i(0)$ is indicated by a blue dot, and the final point $g_i(\ell_f=16)$ by a dot with the color given by the value of $\c$ at that point, $\c(\ell_f)$.
        The same lines are labeled by $g_i(0)$ and $g_i(\ell)$ in all three plots.
    }
    \label{fig: flow 3D}
\end{figure*}

We now briefly summarize the Callan-Symanzik (CS) approach, with a more in-depth discussion given in \cref{app: CS}.
Bare vertex functions depend on bare constants $\Gamma^{(\tilde N, N)}(\bm q, \omega; g_i, \t)$,
while their renormalized counterparts depend on renormalized quantities and the renormalization scale\footnote{We could introduce $\tilde Z$ to renormalize $\tilde \varphi$, but the conservation law ensures $\tilde Z = 1$.}:
\begin{align}
    \ren\Gamma^{(\tilde N, N)}(\bm q, \omega; \ren g_i, \rt, \Lambda)
    =
    Z^\frac{N}{2}\Gamma^{(\tilde N, N)}.
\end{align}
Given a set of underlying bare parameters, varying $\Lambda$ induces a shift of the renormalized quantities, while the bare vertex function is unchanged.
This observation leads to the RG-equations.
To be quantitative, it implies $\Lambda \odv{}{\Lambda}\Gamma^{(\tilde N, N)}|_{\rm bare} = 0$.
Here, $|_{\rm bare}$ indicate that the derivative is taken with bare parameters fixed, leading to a shift in the renormalized quantities.
Applying the chain-rule, we obtain the CS equation
\begin{align}
    \label{eq: CS}
    \!\!\!\!
    \left[ 
        \Lambda \pdv{  }{ \Lambda } 
        \!-\! \beta_i(\rg) \pdv{  }{ \ren g_i } 
        \!-\! \gamma_\t(\rg) \rt \pdv{  }{ \rt } 
        \!-\! N \eta(\rg) 
    \right] \!
    \ren\Gamma^{(\tilde N , N )} \!\!= \!0,\!\!\!\!
\end{align}
with the $\beta$-functions defined as\footnote{Note that the definition here has the opposite sign as compared to the standard definition of the $\beta$-function in the CS formulation, in order to keep consistency between the two different RG approaches.}
\begin{align}
    \label{eq: CS beta}
    \beta_i(\rg) \equiv -\Lambda \odv{ \rg_i(\Lambda) }{ \Lambda }\Big|_{\rm bare}.
\end{align}
Here, $\gamma_\t$ and $\eta$ are the flow functions for $\t$ and the field, respectively, and are defined in \cref{eq: CS flow}. The equation is solved using the method of characteristics, where the RG-scale $\ell$ is introduced to construct the flow equation for the coupling constants $g_i$ as follows
\begin{align}
    \label{eq: beta CS flow}
    \odv{ g_i(\ell) }{ \ell } & \equiv \beta_i(g(\ell)), && g_j(0) = \rg_j.
\end{align}
Note the difference in sign convention in comparison to the momentum shell definition of $\beta$-functions, given in \cref{eq: def beta functions}. Close to the fixed-points $g_j^*$, defined by $\beta(g^*) = 0$, the CS equation has solutions with a scaling form. 

For the case of $\tilde N = N = 1$, we can use rotational invariance to write (the relevant part of) the vertex function as
\begin{align}
    &
    \bar \Gamma_{ab}^{(1,1)}(\bm q, \omega; \rg_i, \rt, \Lambda) 
    = - i \ren\zeta\omega  \delta_{ab}\\
    & \hspace{8mm} + 
    Kq^2(\Lambda \xi_\t)^{\eta-2}
    [ F_S(q \xi_\t) \delta_{ab} + F_A(q\xi_\t)\varepsilon_{ab} ].
\end{align}
Here, $\eta = \eta(g_i^*)$, and we have introduced the scaling functions
\begin{align}
    F_{S,A}(s)
    & \sim
    \begin{cases}
        1, & s\ll 1, \\
        s^{2-\eta} , & s\gg 1.
    \end{cases}.
\end{align}
The correlation length was found using the matching condition.
We evaluated $\ell = \ell_r$, where $\ell_r$ is defined so that $e^{2 \ell_r}r(\ell_r) = K\Lambda^2$, and identified $\xi_\t\Lambda  \equiv e^{\ell_\t}$.
This gives the scaling relation
\begin{align}
    \xi_\t &= \Lambda^{-1}\left( \frac{r}{K \Lambda^2} \right)^{-\nu},&
    \nu &\equiv \frac{ 1 }{ 2 - \gamma_r^* }.
\end{align}
Away from the fixed-point, the linearized flow is
\begin{align}
    \label{eq: flow of g}
    g_i(b) = g_i^* + \left[e^{J\ell}\right]_{ij}  \delta g_j,
\end{align}
where $J_{ij} = \partial_i\beta_j(g^*)$.
If, for some fixed $i$, $\delta g_i \neq 0$, the scaling exponents is modified $\eta\rightarrow\eta + \omega_i$.
If $\omega_i < 0$, the direction is stable, and so at large scale the system flows toward $g_i^*$.
If $\omega_i > 0$, the direction is unstable, and this analysis holds below a finite crossover length scale, above which the system flows away from $g^*_i$~\cite{zinn-justinQuantumFieldTheory1989,kleinertCriticalPropertiesPhi2001}.


\subsection{Two-loop results}

The input needed to derive the RG equations with the CS formulation are calculated perturbatively, by expanding the vertex-functions in powers of $\l$, using a similar diagrammatic method as for the momentum shell approach. We employ dimensional regularization, where the integrals are analytically continued to non-integer dimensions, and divergences then appear as inverse powers of $\epsilon = 4 - d$. The renormalization factors are calculated by canceling these divergent terms using the minimal-subtraction scheme~\cite{kleinertCriticalPropertiesPhi2001}.

To calculating higher order diagrams, it is more convenient to reformulate the NRCH model in terms of a complex field
\begin{align}
    \varphi = \varphi_1 + i\varphi_2,
\end{align}
and its complex conjugate $\varphi^*$. The details of this derivation and the connection to the vector formulation of \cref{sec: path integral} are presented in \cref{app: complex}.
We have the following propagators and interaction vertices:
\begin{align}
    \parbox{11mm}{
    \centering
    \begin{fmfgraph*}(11,5)
        \setval
        \fmfleft{i}
        \fmfright{o}
        \fmf{cc}{i,o}
    \end{fmfgraph*}
    }, && 
    \parbox{11mm}{
    \centering
    \begin{fmfgraph*}(11,5)
        \setval
        \fmfleft{i}
        \fmfright{o}
        \fmf{cg}{i,o}
    \end{fmfgraph*}
    }, &&
    \parbox{11mm}{
    \centering
    \begin{fmfgraph*}(11,5)
        \setval
        \fmfleft{i}
        \fmfright{o}
        \fmf{cgc}{i,o}
    \end{fmfgraph*}
    }, &&
    \!\!
    \parbox{12mm}{
    \centering
    \begin{fmfgraph*}(12,10)
        \setval
        \fmfleft{i1,i2}
        \fmfright{o1,o2}
        \fmf{wigglyc}{i1,v}
        \fmf{plainc}{i2,v}
        \fmf{plain}{o1,v,o2}
    \end{fmfgraph*}
    },&&
    \!\!\parbox{12mm}{
    \centering
    \begin{fmfgraph*}(12,10)
        \setval
        \fmfleft{i1,i2}
        \fmfright{o1,o2}
        \fmf{wiggly}{i1,v}
        \fmf{plain}{i2,v}
        \fmf{plainc}{o1,v}
        \fmf{plainc}{o2,v}
    \end{fmfgraph*}
    }.
\end{align}
The star represent the complex-conjugate field. With this, we calculate the perturbative corrections to $\Gamma^{(1,1)}$ (self-energy $\Sigma$) and $\Gamma^{(1,3)}$ (denoted $\delta \Gamma$).

\begin{figure}[t]
    \centering
    \includegraphics[width=\columnwidth]{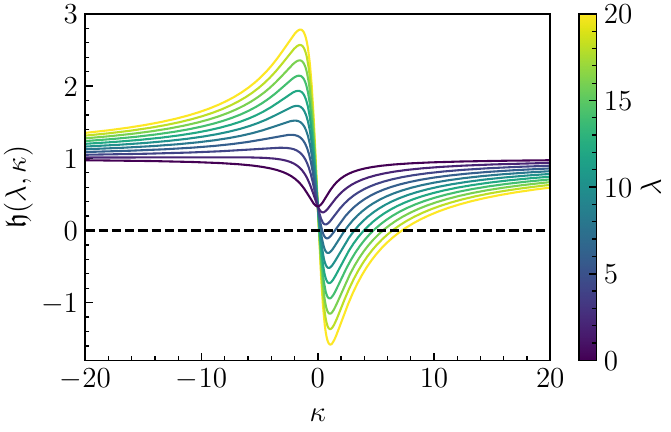}
    \caption{The function $\h(\a,\b) = \h(-\a,-\b)$.}
    \label{fig: h}
\end{figure}

At one loop, the self-energy is
\begin{align}
    \label{eq: c diagrams 1}
    \Sigma_1
    & =
    \sigmaoneone.
    \vspace{-20mm}
\end{align}
The correction to the four-point vertex is
\begin{align}
    \label{eq: c diagrams 2}
    \delta \Gamma_1
    & =
    \gammaoneone
    +
    \gammaonetwo
    +
    \gammaonethree.
\end{align}
To obtain  the leading order corrections to $K$ and $\beta_0$, we must include the following two-loop integrals to the self-energy:
\begin{align}
    \label{eq: c diagrams 3}
    \Sigma_2
    & =
    \sigmatwoone
    +
    \sigmatwotwo.
\end{align}
The calculation of these integrals are left to the Appendices.
There, we obtain the renormalization factors \cref{eq: Z r,eq: Z a0,eq: Z,eq: Z' l,eq: Z' a,eq: Z' b,eq: Z' c}, which give the $\beta$-functions, through \cref{eq: beta from Z}. Combining these results, we find
\begin{align}
    \label{eq: beta l}
    \beta_\l 
    & = 
    \epsilon\l - 10 \l^2
    \left[ 1 - \frac{ 1 }{ 5 } \frac{ (\a - \b)^2 }{ 1 + \b^2 } \right], \\
    \label{eq: beta a}
    \beta_\a & = -2 \l (\a - \b) \frac{ 1 + \a^2 }{ 1 + \b^2 }, \\
    \label{eq: beta b}
    \beta_\b & = -2 \l^2 (\b - \a) \big[ 1 + \h(\a, \b)\big], \phantom{\frac{2}{\b^2}}\\
    \label{eq: beta c}
    \beta_\c & =  4 \l (\c - \a).\phantom{\frac{2}{\b^2}}
\end{align}
The function $\h(\a,\b)$ is defined in \cref{eq: h}, and illustrated in \cref{fig: h}.
These functions depend only on the dimensionless couplings $g_i = (\l, \a, \b, \c)$. These results match the momentum-shell result \cref{eq: beta l MS,eq: beta a MS,eq: beta b MS,eq: beta c MS}.
The couplings have slightly different definitions, see \cref{eq: def g CS}, masking the non-universal regularization scheme dependence.
From \cref{eq: CS flow}, we find the flow functions as
\begin{align}
    \gamma_r = 4 \l, && \gamma_\az = 4 \l \,\frac{\a}{\c},
\end{align}
and
\begin{align}
    \eta = 2 \l^2 \{ 1 + (\b - \a)[\im \g(\b) + \a \re \g(\b)] \}.
\end{align}
The function $\g(\b)$ is defined in \cref{eq: g}, and plotted in \cref{fig: fg}.

The flow induced by the $\beta$-functions, \cref{eq: beta l,eq: beta a,eq: beta b,eq: beta c}, is shown in \cref{fig: flow 3D}.
In addition to the two equilibrium fixed-points---$\a^*=\b^*=\c^* = 0$ and $\l^* = 0$ (Gaussian) or $\l^* = \frac{\epsilon}{10}$ (Wilson-Fisher)---these functions have a non-trivial, non-equilibrium fixed-point 
\begin{align}
    \l^* = \frac{ \epsilon }{ 10 },  && \a^* = \b^* = \c^* \equiv\theta \in \mathbb{R}. && \text{(OCH)}
\end{align}
This is the odd Cahn-Hilliard (OCH) fixed-point, as it fulfills the potential condition \cref{eq: NRCH pc}. The exponents for this fixed-point are 
\begin{align}
    \gamma_r^* =\gamma_{\alpha_0}^* = \frac{ 2 }{ 5 }\epsilon, && \hspace{5mm}
    \eta^* = \frac{ \epsilon^2 }{ 50 }.
    && \hspace{5mm}\text{(OCH)}
\end{align}
The existence of this fixed-point is a deviation from the CGLE, where the alignment of the non-reciprocal parameters is only a slow manifold of the flow, not a fixed-point~\cite{rislerUniversalCriticalBehavior2004,rislerUniversalCriticalBehavior2005,tauberPerturbativeFieldTheoreticalRenormalization2014,siebererDynamicalCriticalPhenomena2013}.

The flow close to a fixed-point is governed by the Jacobian $J_{ij} \equiv \partial_i\beta_j(g^*)$, which in this case is
\begin{align}
    J_{ij}
    & =
    \begin{bmatrix}
        - \epsilon & 0 & 0 & 0 \\[2mm]
        0 & - \frac{ 1 }{ 10 } \epsilon &  \frac{ 1 }{ 10 } \epsilon & 0 \\[2mm]
        0 & \frac{ \h^* + 1 }{ 25 } \epsilon^2 & - \frac{ \h^* + 1 }{ 25 } \epsilon^2  & 0 \\[2mm]
        0 & - \frac{ 2 }{ 5 } \epsilon & 0 & \frac{ 2 }{ 5 }\epsilon
    \end{bmatrix},
\end{align}
where $\h^* = \h(\a^*, \b^*)$.
The stability of this point is governed by the eigenvalues, which are
\begin{align}
    \omega_\upsilon &= -\epsilon, &
    \omega_{\a+\b+\c} &= 0, \\
    \omega_{\frac{5}{4}\!\a+\c} &= - \frac{\epsilon}{10},  &
    \omega_{\c} &= \frac{2\epsilon}{5},
\end{align}
to the leading order. We write the eigenvalues in terms of their \emph{right} eigen-directions in the parameter space. $\omega_{\upsilon}$ is the $\l$-direction, $\omega_{\a+\b+\c}$ is the degeneracy given by shifting $\a,\b,\c$ by the same constant, $\omega_{\frac{5}{4}\!\a+\c}$ is a stable direction towards the fixed-point and $\omega_{\c} > 0$ is an unstable direction. \Cref{fig: flow} depicts the flow of $g_i$, clearly illustrating the unstable direction.

As there is an unstable direction, any system not initialized exactly on this fixed-point will flow away from it. If we consider $\c^{-1}$ as the variable, the corresponding $\beta$-function is $\beta_{(\c^{-1})} = -4 \l (1 - \a\c^{-1} ) \c^{-1}$.
This shows there is another set of non-critical fixed-points, $g_i(\ell) = g_i'$, where the system is manifestly Non-Reciprocal (\NRpm):
\begin{align}
    \l'\! =\frac{ \epsilon }{ 10 },  && \hspace{2mm}
    \a'\! = \b'\!\equiv\!\theta\! \in \mathbb{R}, && \hspace{2mm}
    \c'\! = \pm \infty.
    && \text{(\NRpm)}
\end{align}
The corresponding exponents at this point are
\begin{align}
    \quad\gamma_r' = \frac{ 2 }{ 5 }\epsilon, && \quad \gamma_{\alpha_0}' =  0, &&\quad \eta' = \frac{ \epsilon^2 }{ 50 },
    && \quad\text{(\NRpm)}\!\!
\end{align}
to the leading non-trivial order. We find that $\partial_{(\c^{-1})}\beta_{(\c^{-1})}(\c=\pm\infty) < 0$, so the \NRpm fixed-points are stable. This means that, as long as the system is not prepared exactly at the set of OCH fixed-points, which represent a measure-zero set in the parameter space, the system will flow towards the \NRpm fixed-point. At $r = \az = 0$, exactly at criticality where $\xiS = \xiD = \infty$, the system obeys the potential conditions \cref{eq: NRCH pc general}, and is thus effective equilibrium. This means the \NRpm fixed-point represent the non-equilibrium behavior of a system with large but finite dynamic correlation length $\xiD$, which is why we denote it as non-critical.
For all the non-trivial fixed-points, the dynamical exponent is
\begin{align}
    z = 4 - \eta = 4 - \frac{\epsilon^2}{50}.
\end{align}

If $r$ and $\az$ are tuned such that $\c(0)$ is close to the OCH fixed-point $ \c^*$, the linear scaling regime is
\begin{align}
    \delta \c(\ell) \sim e^{\omega_\c\ell} \delta \c. 
\end{align}
As the system flows towards large scales, $\ell$ increases and so does $\delta \c$.
The crossover from the OCH fixed-point to the \NRpm is given by $\big|\frac{\delta \c(\ell_\times)}{\c^*} \big| \equiv \big|\frac{\c-\b}{\b} \big| e^{\omega_\c\ell_\times}  \approx 1$, corresponding to the length-scale
\begin{align}
    \xic = \frac{ 1 }{ \Lambda } 
    \left| \frac{ K }{ \beta } \frac{ \alpha_0 }{ r } - 1  \right|^{- \frac{ 1 }{ \omega_\c }},
\end{align}
which matches our previous derivation \cref{eq: crossover momentum shell}, which used the momentum shell RG .

To establish a more direct connection between the two RG approaches, we can write the flow in terms of the original parameters. The comparison requires rescaling by the physical dimension, namely, $r\rightarrow e^{2\ell} r$ etc.; see \cref{sec: cs approach}. This yields
\begin{align}
    \hspace{-2.5mm}
    \odv{r}{\ell}
    & = 2r  - 4 \ded u r, \\
    \hspace{-2.5mm}
    \odv{\az\!}{\ell}\!
    & = 2 \az - 4 \ded \ao r, \\
    \hspace{-2.5mm}
    \odv{\bz\!}{\ell}\!
    & = 2 \ded^2 u ( K\ao-\bz u) \! \left[1 + \h\left(\frac{\ao}{u}, \frac{\bz}{K}\right)\right]\!\!, \\
    \hspace{-2.5mm}
    \odv{u}{\ell}
    & = \epsilon u \!-\! 10 \ded \!\left[u^2 - \frac{(K\ao \!-\! \bz u)^2\!}{5 (K^2 + \bz^2)}\right]\!\!,\\
    \hspace{-2.5mm}
    \odv{\ao\!}{\ell}\!
    & = \epsilon\ao \!-\! 10 \ded\! \left[\ao u + \frac{ (K\ao \!-\! \bz u) (\bz\ao\!+Ku)}{5 (K^2 + \bz^2)} \right]
    \!\!.\!\!
\end{align}
The equations for $u$ and $\alpha_1$ can be directly compared with earlier results for the the CGLE \cite{rislerUniversalCriticalBehavior2004,tauberPerturbativeFieldTheoreticalRenormalization2014}.
The comparison uncovers that the flow of $\az$ and $\bz$ is responsible for the change in critical behavior, as affected by the introduction of the conservation law.


\section{Concluding Remarks}
\label{sec: discussion}

We have studied the scaling behavior of NRCH near its critical point and found multiple scaling regimes that can be tuned independently by the equilibrium and non-equilibrium tuning parameters. The momentum shell and Callan-Symanzik RG calculations have enabled us to characterize the scaling behavior of the NRCH model near the OCH fixed-point. In particular, we find that the equal-time correlation function obeys the scaling form
\begin{align}
    \bar C_{\text{ET}, ab}(\bm q) = \frac{D}{\bar \zeta K} \frac{ \xiS^{2 - \eta} }{\G_\text{s}(q\xiS)} \delta_{ab}, 
\end{align}    
with the structural correlation length scaling as $\xiS \sim r^{-\nu}$. On the other hand, the total response function is
\begin{align}
    \bar \chi_{\text{T}, ab}(\bm q) & 
    = 
    \frac1K \re \frac{ \xiD^{2 - \eta} }{\G_\text{d}(q\xiD)} \delta_{ab} 
    + 
    \frac1K \im  \frac{ \xiD^{2 - \eta}}{\G_\text{d}(q\xiD)} \varepsilon_{ab},
\end{align}
where the dynamical correlation length scales as $\xiD \sim r^{-\nu} $ in the Effective Equilibrium regime, and as $\xiD \sim |\sigma|^{-\nu_\text{n}}$ in the NR regimes, where $\sigma = \alpha_0 - \theta r$. The value of $\theta$ is in general not equal to $\beta_0 / K$, as the two-loop calculation reveals that $\beta_0$ flows, but rather, the fixed-point value of $\a$ and $\b$ on the OCH line, as illustrated in \cref{fig: flow 3D}. The scaling functions $\mathcal{G}_\text{s}(s)$ and $\G_\text{d}$ are constants for $|s|\ll1$, and scale as $|s|^{2-\eta}$ for $|s|\gg 1$.
See \cref{sec: NR scaling} for a detailed derivation.

NRCH features two tuning parameters, $r$ and $\az$, with different biological relevance---$r$ is related to the temperature (as well as the equilibrium interaction parameters), whereas $\alpha_0$ is controlled by metabolic activity. Since the non-reciprocal coupling breaks time-reversal symmetry, the system can feature two correlation lengths, $\xiS$ and $\xiD$, in strong contrast to equilibrium criticality that is controlled by a single, dominant length-scale. 

The critical scaling behavior is divided into \emph{scaling regimes}.
In the EE regime, where the system is close to an odd Cahn-Hilliard model, it features equilibrium scaling $\xiD\sim\xiS\sim r^{-\nu}$. The system enters a manifestly Non-Reciprocal regime at the crossover length $\xic$. In this regime, $\xiD$ scales as $r^{-\nu_\text{n}}$ or $|\az|^{-\nu_\text{n}}$, where $\nu_\text{n}$ is a novel exponent. The regimes are associated with fixed-points of the RG equations. The rich non-equilibrium scaling behavior is a consequence of the following phenomenon: $r$ and $\az$ have identical physical, but they are differently renormalized, and as a consequence adopt different effective scaling dimensions, allowing them to compete via crossovers. More specifically, this is a consequence of $\nu_{\rm n} < \nu$, which in turn renders the OCH fixed-point unstable. We expect this picture to hold true at higher orders. To verify this conjecture, numerical analysis, higher order perturbative calculations, or non-perturbative approaches, could be used \cite{canetNonperturbativeApproachCritical2007,canetNonperturbativeRenormalizationGroup2010,canetNonperturbativeRenormalizationGroup2011,siebererDynamicalCriticalPhenomena2013}.

We find that the conservation law affects the steady state in a significant manner, in contrast to equilibrium dynamics. In particular, the potential conditions, which are a sufficient requirement for a steady-state Boltzmann distribution, are stricter for NRCH than its non-conserved counterpart.
Pronounced dependence on such dynamical characteristics is a common feature of active systems; it appears, for example, in non-equilibrium field theories such as KPZ and its generalizations \cite{kardarDynamicScalingGrowing1986,maireConservationLawsSlow2026,tonerRougheningTwodimensionalInterfaces2023,besseInterfaceRougheningNonequilibrium2023} as well as finite-state Markov chains \cite{maesNonDissipativeEffectsNonequilibrium2018}.

In conserved field theories, broken time-reversal symmetry is predicted to give rise to generic scale invariance~\cite{grinsteinGenericScaleInvariance1991}.
Recent work illustrates this in terms of long-ranged $n$-point functions ($n>2$) or non-local steady states~\cite{metzgerGenericLongrangeCorrelations2026,lucaGenericNonlocalStatistics2026}.
Time-reversal symmetry in dynamical field theories lead to a cancellation of perturbative corrections.
In active systems, this is no longer the case \cite{grinsteinGenericScaleInvariance1991,metzgerGenericLongrangeCorrelations2026}.
The two-loop corrections of the NRCH model exhibited this feature, and led to expressions that were different from standard dimensional regularization integrals. We tackled this complexity by extending the method, as presented in \cref{app: complex,app: dim reg}. Other realizations of non-reciprocity have also been recently investigated \cite{klamserDirectedPercolationTransition2025,sezikCriticalDynamicsNonReciprocally2026,youngNonequilibriumFixedPoints2020,youngNonequilibriumUniversalityNonreciprocally2026}.

Our findings may have important implications on how biological systems can organize their dynamics and response despite the inherent slowness of the densely packed intracellular environments. The response time $\mathcal{T}_{\rm d} \sim \xiD^z$ can be shortened on demand by tuning $\az$ without the need to change $r$. The response appears to be extremely sensitive, due to the relatively large value of $z \simeq 4$, because of the conservation law. Therefore, the system can choose to respond relatively quickly by up-regulating metabolism while maintaining its structural features. We expect this mechanism to be robust with regards to the addition of more species.

\acknowledgements
We acknowledge support from the Max Planck School Matter to Life and the MaxSynBio Consortium which are jointly funded by the Federal Ministry of Education and Research (BMBF) of Germany and the Max Planck Society.

\appendix

\section{Response-field formalism}
\label{app: response field}

In this appendix, we summarize conventions and definition used through the paper, and give a more detaild definition of the response field path integral formulation.

\subsection{Conventions and definitions}

We use the following shorthand for the integrals throughout the text
\begin{align}
    \int_{\bm q} & \equiv \int \frac{\dd^d \bm q}{(2 \pi)^d} , &
    \int_\omega &\equiv \int \frac{\dd \omega}{2 \pi} , \\
    \int_{\bm x} & \equiv \int \dd^d \bm x , &
    \int_t &\equiv \int \dd t.
\end{align}

We also define the convenient shorthand 
\begin{align}
    \delta_{\bm q}\delta_{\omega} 
    = (2\pi)^{d+1} \delta^{d}(\bm q )\delta(\omega).
\end{align}
The geometric factor, where $\Omega_d$ is the solid angle in $d$ dimensions,  is 
\begin{align}
    \label{eq: geometric factor}
    A_d & \equiv \int \frac{ \dd \Omega_d  }{ (2\pi)^d } \, 1
    = \frac{ 2 }{ (4 \pi)^\frac{d}{2}  \Gamma(\tfrac{d}{2})  }
    = \frac{1}{8 \pi^2 } + \Oh(\epsilon),
\end{align}
where $\epsilon \equiv 4 - d$. 

\subsection{Correlation lengths}
\label{app: correlation lengths}
Consider the equal-time correlation function 
\begin{align}
    C_{\text{ET},ab}(\bm q) &\equiv  
    C_{ab}(\bm q, t = 0)
    = \int_\omega C_{ab}(\Q), 
\end{align}
and the total response function
\begin{align}
    \chi_{\text{T},ab}(\bm q)
    &\equiv \int_t  \chi_{ab}(\bm q, t)
    = \chi_{ab}(\bm q, \omega=0).
\end{align}
The former represents the correlation between fluctuations at different points at the same time. The latter is the linear response of the field to an external perturbation.
In (effective) equilibrium, they are related via the static FDT, $C_{\text{ET},ab}(\bm q) = k_{\rm B} T \chi_{\text{T},ab}(\bm q)$. 

It is instructive to calculate these functions at the linear level. Using \cref{eq: free corr}, we find
\begin{align}
    \label{eq: CET fourier space}
    C_{\text{ET},ab}(\bm q) =
    \int_\omega C_{ab}(\Q)
    =
    \frac{D}{\zeta K}
    \frac{\delta_{ab}}{q^2+\xiS^{-2}},
\end{align}
where $\xiS \equiv \sqrt{K / r}$.
Performing the integration, we find\footnote{Using identities 8.411.7 and 6.565.4 of Ref. \cite{gradshteynTableIntegralsSeries2014}.} 
\begin{align}
\label{eq: Bessel}
    \! \! \int_{\bm q} \frac{e^{i\bm q \cdot \bm x}}{q^2+\xi^{-2}}
    & = \Gamma(\tfrac{d}{2}) A_d \left( \frac{2}{x}\right)^{\frac{d-2}{2}}\!\!\!\!
    \int_0^\infty \!\! \dd q \,  \frac{q^{\frac{d}{2}} J_{\frac{d-2}{2}}(x q)}{q^2+\xi^{-2}},\\
    & = 
    \Gamma(\tfrac{d}{2}) A_d 
    \bigg( \frac{\sqrt{2}}{\xi}\bigg)^{\!{d-2}}\!\!
    \left( \frac{\xi}{x} \right)^{\!\frac{d-2}{2}} \!\!
    K_{\frac{d-2}{2}}(\tfrac{x}{\xi}),
\end{align}
for $\re \xi^{-2} \!>\! 0$ and $0\!<\!d\!<\!5$. The modified Bessel function of the second kind, $K_\alpha(z)$, has the limits
\begin{align}
    K_\alpha(z) = 
    \begin{cases}
        \sqrt{\frac{\pi}{2z}} e^{-z}, & z\gg 1, \\[2mm]
        \frac{\Gamma(\alpha)}{2}\left( \frac{2}{z} \right)^\alpha, & z\ll 1.
    \end{cases}
\end{align}
This gives us
\begin{align}
    C_{\text{ET},ab}(\bm x)
    & = 
    \Gamma(\tfrac{d}{2}) \frac{A_dD}{\zeta K} 
    \bigg( \frac{\sqrt{2}}{\xiS}\bigg)^{\!d-2}\!\! \left( \frac{\xiS}{x} \right)^{\!\frac{d-2}{2}}
    K_{\frac{d-2}{2}}(\tfrac{x}{\xiS})
    \delta_{ab}, 
    \\
    & \simeq x^{-\frac{d-1}{2}}e^{-x/\xiS} \delta_{ab}, 
    \quad x \gg \xiS.
\end{align}
Next, we can find the free response function $\chi$ from the free propagator $G$ [\cref{eq: free response}] using the relation $\chi = q^2 G$ [\cref{eq: response function}].
In position space, this can be written as
\begin{align}
    \label{eq: chi_T from complex}
    \chi_{\text{T},ab}(\bm x)
    & = \re \chi_\text{T}(\bm x) \delta_{ab} + \im \chi_\text{T}(\bm x) \varepsilon_{ab},
\end{align}
where we have defined
\begin{align}
    \chi_\text{T}(\bm x) 
    & = 
    \int_{\bm q} \frac{e^{i \bm q \cdot \bm x }}{(r + K q^2) + i(\az + \bz q^2)}, \\
    & = 
    \frac{\Gamma(\tfrac{d}{2}) A_d  }{K + i\bz}
    \bigg( \frac{\sqrt{2}}{\xiD}\bigg)^{d-2}\!\!
    \left( \frac{\xiD}{x} \right)^\frac{d-2}{2} \! K_{\frac{d-2}{2}}(\tfrac{x}{\xiD}),
\end{align}
where we have defined $\xiD \equiv \sqrt{(K + i\bz) / (r + i\az)}$. In \cref{app: complex}, where we present the complex formulation of NRCH, this form appears naturally. 
The above expression has the asymptotic form
\begin{align}
    \chi_\text{T}(\bm x)
    & \simeq x^{-\frac{d-1}{2}}e^{-x/\xiD}, 
    \quad x \gg \xiD.
\end{align}
For $r>0$ and $K > 0$, we find that $\re \xi_\text{d} > 0$.

\subsection{Generating functionals}
\label{app: generating functionals}

The stochastic field theory \cref{eq: EOM} can be represented as a path-integral using the response field formalism~\cite{martinStatisticalDynamicsClassical1973,dedominicisFieldtheoreticTechniquesCritical1975,janssenLagrangeanClassicalField1976}.
The expectation value of and operator $\Oh$ is expressed as
\begin{align}
    \label{eq: path integral app}
    \!\!\
    \E{\Oh[\varphi,\tilde\varphi]} 
    & = \int \D \varphi \D \tilde\varphi \,\, \Oh[\varphi,\tilde\varphi]\,\, e^{-\A[\varphi,\tilde\varphi]}.
\end{align}
Here, $\tilde \varphi_a$ is an auxiliary variable called the response field, and the action is
\begin{align}
    \label{eq: def A real space app}
    \!\!\A[\varphi,\tilde\varphi] 
    & =\! \int\limits_{\bm x, t} \! i \tilde \varphi_a\!
    \left[  \zeta\partial_t\varphi_a \!+\! \nabla^2(K_a \!+\! W_a \!+\! Di\tilde\varphi_a) \right]\!.\!\!
\end{align}
The model is fully described by the generating functional
\begin{align}
    Z[\j_a, \tilde J_a] & \equiv 
    \E{e^{\int_{\bm x, t}(\j_a\varphi_a + \tj_ai\tilde\varphi_a)}}.
\end{align}
Note that $Z[0,0] = 1$ by construction. The \emph{connected} generating functional is 
\begin{align}
    W[\j_a, \tj_a] &\equiv \ln Z[\j_a, \tilde J_a].
\end{align}
The (connected) two-point functions, defined in real space as
\begin{align}
    C_{ab}(\bm x, t, \bm x',t') 
    & \equiv \E{\varphi_a(\bm x,t) \varphi_b(\bm x',t')}_c, \\
    G_{ab}(\bm x, t, \bm x',t')
    & \equiv \E{\varphi_a(\bm x,t) i \tilde \varphi_b(\bm x',t')}_c,
\end{align}
can thus be written as functional derivatives:
\begin{align}
    C_{ab}(\bm x, t, \bm x',t') 
    & = \fdv{W[\j,\tj]}{{J_a(\bm x,t)},{J_b(\bm x',t')}}\bigg|_{\j,\tj=0}, \\
    G_{ab}(\bm x, t, \bm x',t')
    & = \fdv{W[\j,\tj]}{{J_a(\bm x,t)},{\tilde J_b(\bm x',t')}}\bigg|_{\j,\tj=0}.
\end{align}
The response propagator (or Green's function) $G$ is related to the linear response to an applied force $K_a\rightarrow K_a + h_a$ added to the equation of motion \cref{eq: EOM}.
In Fourier space, it is
\begin{align}
    \label{eq: response function}
    \chi_{ab}(\Q)\delta_{\bm q + \bm q'} \delta_{\omega + \omega'} 
     & \equiv 
     \fdv{\E{\varphi_a(\Q)}}{h_b(\Qp)}\bigg|_{h\rightarrow0}, \\
     \implies
     \chi_{ab}(\bm q, \omega) &= q^2 G_{ab}(\bm q, \omega).
\end{align}
Using a Legendre transformation, we obtain the One-Particle Irreducible (1PI) generating functional in terms of $\Phi \equiv \E{\varphi}$ and $\tilde\Phi =\E{\tilde \varphi}$:
\begin{align}
    &\Gamma[\tilde\Phi,\Phi] 
    \equiv - W[\tj, \j]
    + \int_{\bm x, t} (\j_a\Phi_a + \tj_a i\tilde\Phi_a),
\end{align}
where we have the following relations between the two formulations:
\begin{align}
    \!\!\Phi_a = \fdv{W}{\j_a}, \,\,
    i\tilde \Phi_a = \fdv{W}{\tj_a}, \quad
    \j_a = \fdv{\Gamma}{\Phi_a}, \,\,
    \tj_a = \fdv{\Gamma}{i\tilde\Phi_a}.\!\!
\end{align}
We obtain the vertex functions by taking derivatives, as follows
\begin{align}
    \label{eq: vector 1PI}
    \!
    \Gamma^{(\tilde N, N)}_{\tilde a_1,  ... \tilde a_{\tilde N}, a_1,  ...a_{N}}\!
    \equiv
    \frac{\delta^{\tilde N + N}\Gamma[\tilde\Phi, \Phi]}
    {\delta i\tilde\Phi_{\tilde a_1},...\delta i\tilde\Phi_{\tilde a_{\tilde N}},\delta \Phi_{a_{1}},...\delta \Phi_{a_{N}}}.\!\!
\end{align}
These are the physical observables of the field theory, which we will use to define renormalized coupling constants through renormalization conditions [see \cref{eq: renorm cond 1}].
One can readily find that the two-point vertex functions are indeed the two-point functions.
In particular,
\begin{align}
    G_{ab}^{-1}(\bm q,\omega) = \Gamma^{(1,1)}_{ab}(\bm q,\omega),
\end{align}
where $\Gamma^{(1,1)}_{ab}(\bm q,\omega)\delta_{\bm q+\bm q'}\delta_{\omega +\omega'} \equiv \Gamma^{(1,1)}_{ab}(\bm q,\bm q',\omega, \omega') $.

The 1PI generating functional is the effective action taking into account all fluctuations.
At the tree-level, i.e to zeroth order in perturbation-theory where the diagrammatic contributions are given by graphs without loops, the vertex is given by the microscopic action as described in \cref{sec: path integral}. Therefore, we have
\begin{align}
    \Gamma[\Phi,\tilde\Phi] = \A[\Phi, \tilde\Phi] - \Oh(\text{loops}).
\end{align}
In particular
\begin{align}
    \Gamma^{(1,3)}_{abcd} = 3! q^2U_{abcd} - \Oh(\text{loops}).
\end{align}


\section{The CS equation}
\label{app: CS}

In this Appendix, we give a more detailed exposition of the CS approach to RG~\cite{kleinertCriticalPropertiesPhi2001}.
We start by introducing renormalized fields, denoted by a bar:
\begin{align}
    Z^\frac{ 1 }{ 2 }\ren\varphi &= \varphi, &
    \tilde Z^\frac{ 1 }{ 2 } \ren{\tilde \varphi} &= {\tilde \varphi}.
\end{align}
Here, $Z$ and $\tilde Z$ are renormalization factors. The conservation law implies $\tilde Z = 1$ and $Z\in \mathbb{R}$, and therefore, the renormalized vertex-functions are given as $\smash{\ren\Gamma^{(\tilde N, N)} = Z^\frac{N}{2}\Gamma^{(\tilde N, N)}}$.
In addition, we define renormalized coupling constants, with relations such as $(Z_r/\sqrt{Z})\rr = r$. (see \cref{eq: bare constants} for a complete list). The renormalization factors will depend on the dimensionless couplings\footnote{It is a property of the minimal subtraction scheme that renormalization factors depend only on $g_i$, and not on $r$ or $\az$ directly.
This is why the flow function for $r$ is considered separately---only the flow of $g_i$ must vanish to reach a fixed-point.} 
\begin{align}
    \label{eq: def g CS}
    \!\!
    g_i =
    \left\{
        \l \equiv \ded u,\,\,
        \a \equiv \frac{\alpha_1}{u},\,\,
        \b \equiv \frac{\beta_0}{K},\,\,
        \c \equiv \frac{\alpha_0}{r}
    \right\}.\!
\end{align}
Their renormalized counterparts $\rg_i$ are defined in the same way, in terms of renormalized constants. For example, $\rl = \ded \ru$. For explicit calculation of the renormalization factors, see \cref{app: complex}.

\subsection{RG equation}

Regularization introduces an arbitrary scale, given by $\Lambda$. As discussed in \cref{sec: cs approach}, the CS equation is derived by enforcing the invariance upon changing $\Lambda$. 
We use $\ren\Gamma^{(\tilde N, N)}(\bm q, \omega; \ren g, \rt, \Lambda) = Z^\frac{N}{2}\Gamma^{(\tilde N, N)}$ to obtain
\begin{align}
    \label{eq: CS NN}
    \!\!\!\!
    0 & = 
    \Lambda \odv{  }{ \Lambda }\,\Gamma^{(\tilde N, N)}_{ab}  \Big|_{\rm bare}
    \\*
    & = 
    \left[ 
        \Lambda \pdv{  }{ \Lambda } 
        \!-\! \beta_i(\rg) \pdv{  }{ \ren g_i } 
        \!-\! \gamma_\t(\rg) \rt \pdv{  }{ \rt } 
        \!-\! N \eta(\rg) 
    \right]\! \ren\Gamma^{(\tilde N,N)}_{ab}\!\!,\!\!\!\!\!
\end{align}
where we have defined the $\beta$-functions
\begin{align}
    \label{eq: CS beta app}
    \beta_i(\rg) \equiv -\Lambda \odv{ \rg_i(\Lambda) }{ \Lambda }\Big|_{\rm bare},
\end{align}
and the flow-functions
\begin{align}
    \label{eq: CS flow}
    \gamma_\t(\rg) &\equiv - \Lambda \odv{  }{ \Lambda } \ln \frac{ \rt }{ \t }\Big|_{\rm bare}, \\
    \eta (\rg) & \equiv \Lambda \odv{  }{ \Lambda } \ln \sqrt{ Z }\big|_{\rm bare}.
\end{align}
An RG fixed-point is therefore defined by $\beta(g^*) = 0$. The CS-equation can be solved with the method of characteristics. This is done by introducing running couplings, as follows
\begin{align}
    \label{eq: def beta}
    \odv{ g_i(\ell) }{ \ell } & = \beta_i(g(\ell)), & g_i(0) &= g_i,\\
    \label{eq: def gamma}
    \frac{ 1 }{ \t(\ell) } \odv{ \t(\ell) }{ \ell } & = \gamma_\t(g(\ell)),&
    \t(0) &= \t,
\end{align}
and $\Lambda(\ell) = e^{-\ell}\Lambda$.
Using dimensional analysis, we write
\begin{align}
    e^{d_{N, \tilde N}\ell} \,&\ren\Gamma^{(\tilde N, N)}(\bm q, \omega; \rg, \rt, \Lambda) \nonumber \\
    =
    &\, \ren\Gamma^{(\tilde N, N)}(e^{\ell} \bm q, e^{4\ell}\omega; \rg, e^{2\ell} \rt, e^{\ell} \Lambda),
\end{align}
where $d_{N, \tilde N}$ is the physical (naive) dimension of $\Gamma^{(\tilde N, N)}$, namely
\begin{align}
    \hspace{-2mm}
    d_{N, \tilde N}
    & = 4 + (N-1) \left(1-\frac{d}{2}\right) - (\tilde  N-1) \left(1+\frac{d}{2}\right)
    \!.\hspace{-2mm}
\end{align}
Combining the above equations, we obtain the following form of the CS equation:
\begin{align}
    &0 = 
    \left( N\eta - d_{N, \tilde N} + \odv{  }{ \ell } \right)\!
    \\*
     & \hspace{5mm} \times
    \ren\Gamma^{(\tilde N,N)}(e^{\ell} \bm q, e^{4\ell}\omega; g(\ell), e^{2\ell}\t(\ell),e^{-\ell}\Lambda(\ell)).
\end{align}

\subsection{Fixed-point scaling form}

The CS equation is solved by
\begin{align}
    \label{eq: methods of ch}
    & \ren\Gamma_{ab}^{(\tilde N, N)}(\bm q, \omega; \rg_j, \rt, \Lambda)=
    e^{
    -\! \int_0^\ell\! \dd\ell'\left[d_{N, \tilde N} - N \eta(g_j(\ell')) \right] 
    }\nonumber \\*
    &
    \hspace{15mm}\times
    \ren\Gamma_{ab}^{(\tilde N, N)}(e^{\ell}\bm q, e^{4\ell}\omega; g_j(\ell), e^{2\ell} \t(\ell), \Lambda).
\end{align}
In a similar way, the flow-function \cref{eq: def gamma} gives
\begin{align}
    \label{eq: scaling tau}
    \t(\ell) =  \t e^{ - \int_0^\ell \dd \ell' \gamma_r(g(\ell')) } .
\end{align}
We now consider $\tilde N= N = 1$.
Close to the fixed-point, where the $\beta$-functions vanish, we can write
\begin{align}
    \label{eq: flow of g}
    g_i(\ell) \sim g_i^* + \left[e^{J\ell}\right]_{ij}  \delta g_j,
\end{align}
where $J_{ij} = \partial_i\beta_j(g^*)$.
As $\ell\rightarrow \infty$, all eigenvalues of $J$ (denoted $\omega_i$) must be negative for stability.
If, for some $i$ we have $\omega_{i} > 0$, then there is a finite scaling region for the following analysis, defined by $e^{\omega_{i}\ell} {\delta g_{i}} \ll g_{i}^*$ (no sum).
Beyond this, the system will cross over to a different regime.
With this assumption, the solution of the CS equation takes the form
\begin{align}
    \label{eq: scaling no matching}
    & \ren\Gamma^{(1,1)}(\bm q, \omega; \rg_j, \rt, \Lambda) \nonumber 
    \\
    & = 
    e^{(\es-4)\ell}
    \Gamma^{(1,1)}(e^{\ell} \bm q, e^{4\ell} \omega; g_j(\ell), e^{2\ell}\t(\ell), \Lambda),
\end{align}
where $\eta^* = \eta(g^*)$ and we used that by dimensional analysis, $d_{1, 1} = 4$.
The tuning parameter $r$ will scale as $\t(\ell) = \t e^{-\gamma_\t^*\ell}$, where $\gamma_\t^* = \gamma_\t(g^*)$.
We then set the matching condition by defining $\ell_\t$  so that 
\begin{align}
    e^{2\ell_\t} \t(\ell_\t) = \t e^{(2 - \gs)\ell_r}  = K \Lambda^2.
\end{align}
This corresponds to running the RG flow until $r(\ell)$ matches the renormalization scale $\Lambda^2$, up to the factor $K$.
From the form of \cref{eq: scaling no matching}, we identify
\begin{align}
    \!\! \xi_r\Lambda \equiv e^{\ell_r} \implies \xi_r=\Lambda^{-1}
    \left(\frac{r}{K \Lambda^2}\right)^{-\nu}\!\!,\,
    \nu \equiv \frac{ 1 }{ 2 - \gamma_r^* } .
\end{align}

These considerations and the rotational invariance of the model allow us to write down the scaling form of the vertex functions, to all relevant orders, as follows
\begin{align}
    & \ren\Gamma_{ab}^{(1,1)}(\bm q, \omega; \rg_i, \rt, \Lambda)\nonumber \\
    & = 
    - i \omega \ren\zeta
    \delta_{ab} +  \ren S (\bm q; \rg_i, \rt, \Lambda) \delta_{ab} + \ren A( \bm q; \rg_i, \rt, \Lambda) \varepsilon_{ab} ,
\end{align}
where the renormalization condition \cref{eq: renorm cond 1} is manifestly satisfied.
We now compare this to the scaling form given in \cref{eq: scaling no matching}.
Starting with the first term, we note that the conservation law implies $\zeta(\ell) = \zeta e^{-\eta\ell}$, which leads to the cancellation of the anomalous dimension of $\ren \Gamma$, leading to $\ren \zeta = \zeta$. Since all relevant parameters are tuned to their fixed-point values and the rest of the parameters flow to their fixed-point values, i.e. $g_i(\ell_r) \sim g_i^*$, we obtain the following scaling form for the vertex function
\begin{align}
    \!\!&
    \bar \Gamma_{ab}^{(1,1)}(\bm q, \omega; \rg_i, \rt, \Lambda) 
    = - i \ren\zeta\omega  \delta_{ab}
    \\* &\hspace{8mm}
    + 
    Kq^2(\Lambda \xi_\t)^{\eta-2}
    [ F_S(q \xi_\t) \delta_{ab} + F_A( q\xi_\t)\varepsilon_{ab} ].
\end{align}
where these scaling functions have the following asymptotic forms
\begin{align}
    F_{S,A}(s) 
    & \sim
    \begin{cases}
        1, & s\rightarrow 0, \\
        s^{2-\eta} , & s\rightarrow\infty, 
    \end{cases}.
\end{align}

\subsection{ NR scaling forms }

\label{sec: NR scaling}

Let now assume $\l = \l^*$, $\a = \a^*$ and $\b = \b^*$, while $\c = \c^* + \delta \c$, and denote $\a^* = \b^* = \c^* = \theta$. The scaling function will now, in general, depend on $\delta \c$.
Using the matching condition, we find that
\begin{align}
    \delta \c(\ell_r) =  \frac{\sigma}{K \Lambda^2} \left( \frac{r}{K \Lambda^2} \right)^{-\Delta}={\hat \sigma}/{\hat r}^{\Delta},
\end{align}
where $\sigma = \alpha_0 - \theta r,$, and we have defined $\Delta \equiv 1 + \omega_\c\nu = 1 + \frac{\epsilon}{5} + \Oh(\epsilon^2)$, $\hat r \equiv {r}/{(K \Lambda^2)} $, and $ \hat \sigma \equiv {\sigma}/{(K \Lambda^2)} $. With this, the scaling form \cref{eq: scaling no matching} can be written down as follows
\begin{align}
    \!\!&
    \bar \Gamma_{ab}^{(1,1)}(\bm q, \omega; \rg_i, \rt, \Lambda) 
    =     - i \ren\zeta\omega  \delta_{ab}
    + 
    q^2K\xi_r^{-2}(\Lambda \xi_\t)^{\eta} \nonumber \\
    &\hspace{8mm}
    \times
    [ F_S(q \xi_\t, \hat \sigma / \hat r^\Delta) \delta_{ab} + F_A(q\xi_\t,\hat \sigma / \hat r^\Delta)\varepsilon_{ab} ].
\end{align}

The renormalized response propagator is related to the vertex function we have found above by $\bar G_{ab}^{-1} = \bar \Gamma_{ab}^{(1,1)}$. We can write the scaling form of the renormalized correlation function as $\bar C_{ab} = 2q^2 D[\bar G \bar G^\dagger]_{ab}$, to all relevant orders.
The response function is related to the response propagator by $\ren\chi_{ab} = q^2 \bar G_{ab}$.
These can be expressed in terms of complex functions:
\begin{align}
    2\bar \chi_{ab}(\bm q, \omega)
    & = 
    [\bar\chi(\bm q, \omega) + \bar\chi^*(\bm q, -\omega)] \delta_{ab} \\
    &\quad
    - i[\bar\chi(\bm q, \omega) - \bar\chi^*(\bm q, -\omega)] \varepsilon_{ab}.
\end{align}
and
\begin{align}
    4 \bar C_{ab}(\bm q, \omega)
    & = 
    [\bar C(\bm q, \omega) + \bar C^*(\bm q, -\omega)] \delta_{ab} \\
    &\quad - i[\bar C(\bm q, \omega) - \bar C^*(\bm q, -\omega)] \varepsilon_{ab},
\end{align}
where we have defined
\begin{align}
    \label{eq: renorm complex G}
    \bar G(\bm q, \omega)
    & = 
    \frac{ 1 }{ -i (\bar\zeta \omega - \bar A) + \bar S},
    \\
    \label{eq: renorm complex C}
    \bar C(\bm q, \omega)
    &= 
    \frac{ 4 D }{ (\bar\zeta\omega - \bar A)^2 + \bar S^2 }
    = \frac{4 D q^2}{\bar S} \re \chi(\bm q, \omega).
\end{align}
With $\int_\omega \ren C(\bm q, \omega) = \frac{q^2 D }{\bar \zeta \bar S},$ we obtain
\begin{align}
    \bar C_{\text{ET},ab}(\bm q)
    & = 
    \frac{ D (\Lambda\xi_r)^{2-\eta\hspace{-2mm}} }{ \bar \zeta F_S(q \xi_r, \hat \sigma / \hat r^\Delta) } \delta_{ab}.
\end{align}
For the total response, we have
\begin{align}
    \ren \chi_\text{T}(\bm q) 
    & \equiv \bar \chi(\bm q, \omega = 0)\\
    & = \frac{(\xi_r^{2}/K)(\Lambda\xi_r)^{-\eta}}{ F_S(q \xi_r,\hat \sigma / \hat r^\Delta) + i F_A(q \xi_r,\hat \sigma / \hat r^\Delta) }.
\end{align}

The two different correlation lengths, as defined in \cref{sec: NRCH}, defined by the exponential decay of $\bar C_{\text{ET}}(\bm x)$ and $\ren \chi_\text{T}(\bm x)$ in position space.
This scale can be extracted from the poles of the $\bm q$-space two-point functions (or the zeros of the reciprocal functions) nearest to the real line, denoted as $q^*$~\cite{tarkoTheoryCriticalPoint1975}.
For example, at the linear level $C_\text{ET}^{-1}(\bm q) \propto (r + Kq^2)^{-1}$ has a pole at $q^* = i \sqrt{r / K}$, and thus $\xi_\text{s}^{-1} = - i q^{*}$, which yields $C_\text{ET}(\bm x\gg \xi_\text{s})\sim e^{-x/\xi_\text{s}} $. We now define
\begin{align}
    F_\text{d}(s, y)\equiv F_S(s, y) + i F_A(s, y).
\end{align}
The dynamical correlation length $\xiD$ is defined via the distance to the nearest pole in the complex $q$-plane $q^* = i \xiD^{-1}$, which implies $F_\text{d}(\xi_r q^*, y) = 0$.
We therefore define $X_\text{d}$ such that
\begin{align}
    F_\text{d}( iX^{-1}_\text{d}(y), y )  = 0.
\end{align}
This gives us the scaling form
\begin{align}
    \xiD \equiv X_\text{d}(y)\xi_r 
    &= \Lambda^{-1} \hat r^{-\nu} X_\text{d}\left( \frac{\hat \sigma}{\hat r^\Delta} \right), \nonumber \\
    & \equiv r^{-\nu} \G_{\sigma r}\! \left( \frac{\sigma}{r^\Delta} \right),
\end{align}
where we have defined $\G_{\sigma r}$ to absorb various factors of $\Lambda$ and $K$.
This matches our earlier scaling hypothesis, \cref{eq: scaling hypothesis}.
In principle, we must find $X_\text{s}(y)$ in the same way to obtain $\xiS$, which would give the same scaling hypothesis for the structural correlation length.
However, we obtain the asymptotes of $X_\text{d}$ and $X_\text{s}$ by considering the linear level, where $\xiD\sim {\alpha_0}^{-\frac{1}{2}}$ for $r\rightarrow0$, while $\xiS\sim r^{-\frac{1}{2}}$.
We therefore surmise that $X_\text{s} \sim \Oh(1)$, such that
\begin{align}
    \xiS\sim \xi_r = \Lambda^{-1} \hat r^{-\nu}.
\end{align}
At the linear level, we have
\begin{align}
    F_S(s, y) &= F_\text{s}(s, y) = 1 + s^2, \\
    F_A(s, y) &= \theta (1 + s^2) + y, \\
    F_\text{d}(s, y) &= [1 + i(\theta + y)][1 + (sX_\text{d}(y))^2].
\end{align}
and 
\begin{align}
    X_\text{s} = 1,  && X_\text{d}^2(y) = \frac{1 + i\theta}{1 + i(\theta + y)} = \frac{1 + i \b}{1 + i \c}. 
\end{align}
This gives us
\begin{align}
    \xiD^2 = \frac{K + i \bz}{r + i \az},
\end{align}
in agreement with the derivation given in \cref{app: correlation lengths}.

To simplify the presentation, we introduce new scaling functions
\begin{align}
    \G_\text{s}(s) &\equiv \Lambda^{\eta} F_\text{s}(s, y), \\
    \G_\text{d}(s) &\equiv \Lambda^{\eta} X_\text{d}^{2-\eta}(y) F_\text{d}(s X^{-1}_\text{d}(y), y),
\end{align}
and use \cref{eq: renorm complex G,eq: renorm complex C} to write the scaling forms of the full dynamic two-point functions as follows
\begin{align}
    &\bar \chi(\bm q, \omega) 
    = 
    \frac{1}{-i \bar \zeta \omega / q^2 + K \xiD^{\eta-2} \G_\text{d}(q \xiD)}, \\
     &\bar C(\bm q, \omega) = \frac{4 D \xiS^{2-\eta}}{\G_\text{s}(q\xiS)} \re\bar \chi(\bm q, \omega), \nonumber 
    \\
    &\hspace{-2mm}
    =
    \frac{4 D}{
    \!\big[ \frac{\bar\zeta\omega}{Kq^2}- \im \xiD^{\eta-2}\! \G_\text{d}(q \xiD)\big]^2
    \!+\!
    \big[\re\xiD^{\eta-2}\! \G_\text{d}(q \xiD)\big]^2\!
    }\,,\!
\end{align}
with $\re \xi_d^{\eta-2}\G_d(\bm q \xiD)  = \xi_s^{\eta-2} \G_s(\bm q \xiD)$. For the equal-time correlation function and the total response, we obtain
\begin{align}
    \bar C_\text{ET}(\bm q) &= \frac{D}{\bar \zeta K} \frac{ \xiS^{2 - \eta} }{\G_\text{s}(q\xiS)} \delta_{ab}, \\
    \bar \chi_{\text{T}, ab}(\bm q) & = \re \bar \chi_\text{T}(\bm q) \delta_{ab} + \im \bar \chi_\text{T}(\bm q)\varepsilon_{ab}, \\
    \bar \chi_\text{T}(\bm q) &= \frac{1}{K} \frac{ \xiD^{2 - \eta} }{\G_\text{d}(q\xiD)}.
\end{align}
With $\bar\zeta k_{\rm B} T = D$, we recover the FDT for $F_A = 0$.


\section{Complex formulation}
\label{app: complex}

We now define the complex path-integral formulation for the NRCH model, that complements the vector formulation in \cref{sec: path integral,app: response field}. By introducing the complex field
\begin{align}
    \varphi(\bm x, t) = \varphi_1(\bm x, t) + i \varphi_2(\bm x, t),
\end{align}
we can write the equation of motion of the NRCH, \cref{eq: EOM}, as follows
\begin{align}
    \label{eq: complex eom}
    &\hspace{15mm}
    E[\varphi] - \sqrt{ 2 D(-\nabla^2) }\,\xi=0,\\
    \label{eq: complex eom}
    & E[\varphi]  \equiv \zeta \partial_t \varphi+ \big[S+iA -\nabla^2(u + i \alpha_1) |\varphi|^2 \big]\varphi,
\end{align}
where $S$ and $A$ are the operators defined in \cref{eq: realspace S A}.
The complex noise, $\xi \equiv \xi_1 + i\xi_2$, has the correlations
\begin{align}
    \E{\xi(\bm x, t)\xi^*(\bm x', t')} = 2 \delta^d(\bm x - \bm x')\delta(t-t').
\end{align}
The probability-distribution in the response-field formalism can be found via
\begin{align}
&\E{ \delta\left(E[\varphi] \!-\! 
    \sqrt{\smash{2 D(-\nabla^2)}\phantom{b}\!\! } \xi \right) 
    \delta\left(E^*[\varphi]\!-\!\sqrt{\smash{2 D(-\nabla^2)}\phantom{b}\!\! } \xi^*\right) 
    }\nonumber  \\
    & =
    \int \D \tilde \varphi  \D \tilde \varphi^* 
    e^{- \A[\tilde\varphi,\varphi](\bm x, t)},
\end{align}
where we have defined the action density
\begin{align}
    \label{eq: complex action}
    \A = 
    \int\limits_{\bm x, t} \left(4 Di\tilde\varphi\nabla^2i\tilde\varphi^*\! + i\tilde \varphi^* E + i\tilde \varphi E^*\right).
\end{align}
The generating functional is defined as
\begin{align}
    Z[\Je] \equiv \E{e^{\int_{\bm x, t}(\Je^*_i\psi_i + \Je_i\psi^*_i)}},
\end{align}
where we have defined the combined physical and response variables
\begin{align}
    \psi_i \equiv 
    (i\tilde \varphi,\quad \varphi), &&
    \Je_i \equiv (\tj,\quad \j).
\end{align}
Beginning with the free theory, $U = 0$, we can calculate the generating functional exactly as a Gaussian integral.
In Fourier space, it is $\Z[\Je] = e^{ \Je^\dagger  M^{-1} \Je }$, where
\begin{align}
    M(q) & =
    \begin{pmatrix}
        - 4 D q^2 &  \!- i\omega \!+\! S(q)\!+\!iA(q) \\ i\omega \!+\! S(q)\!-\!iA(q) & 0
    \end{pmatrix},
\end{align}
This gives the free two-point functions
\begin{align}
    G_0(\Q,\Qp) &= 
    \E{ \varphi(\Q) i\tilde\varphi^*(\Qp) }_0,\nonumber \\
    & =
    \frac{ \delta_{\omega-\omega'}\delta_{\bm q -\bm q'} }{ \!-i(\omega - A) + S }
    ,\\
    C_0(\Q,\Qp)
    &= 
    \E{ \varphi(\Q) \varphi^*(\Qp) }_0,\nonumber \\
    & =
    \frac{ 4 D q^2\, \delta_{\omega-\omega'}\delta_{\bm q -\bm q'}  }{ (\omega - A)^2 + S^2 }
    .
\end{align}

The One-Particle Irreducible (1PI) generating functional is, with $\Psi = \E{\psi}$, the Legendre transform the logarithm of the generating functional,
\begin{align}
    \Gamma[\Psi] &\equiv - W[\Je] +\int_{\bm x, t} (\Je_i^*\Psi_i + \Je_i\Psi_i^*),\\
    W[\Je] & \equiv \ln Z[\Je], \quad
    \fdv{ \Gamma }{ \Psi_i^* } = \Je_i, \quad
    \fdv{ W }{ \Je_i^* } = \Psi_i.
\end{align}
Vertex functions are, 
in terms of $\Phi  = \E{\varphi}$ and $\tilde\Phi \equiv\E{\tilde \varphi}$, derivatives of the generating functional:
\begin{align}
    \Gamma_{(n,m)}^{[i_1, ... i_n, j_1, ... j_m]}
    \equiv
    \frac{\delta^{n+m} \Gamma[\Psi, \Psi^*] }{ \delta\Psi_{i_1}^*, ... \delta\Psi_{i_n}^*, \delta\Psi_{j_1}, ... \delta\Psi_{j_{m}} }.
\end{align}
at $\Je = 0$. We caution that the indices used here differ from those used in the vector formalism, \cref{eq: vector 1PI}. The upper indices in square brackets are $i = 1$ for response fields $i \tilde \Phi$ and $i = 2$ for physical fields $\Phi$, while $n$ and $m$ count the number of conjugated and non-conjugated fields, respectively. This can be related to the vector-formulation with the definition
\begin{align}
    \Gamma_{(n,m)}^{ (\tilde N, N) }
    \equiv
    \Gamma_{(n,m)}^{
    [
        \overbrace{\scriptscriptstyle{1...1}}^{\tilde N} 
        \overbrace{ \scriptscriptstyle{2...2}}^N
    ]
    }.
\end{align}
To compare, we have
\begin{align}
    \Gamma^{(1,3)}_{abcd}
    & =
    3! q^2 U_{abcd}
    - \Oh(\text{loops})
    , \\
    \Gamma^{(1,3)}_{(2,2)}
    & = 2 q^2(u + i\ao)
    - \Oh(\text{loops}),
\end{align}
where $U_{abcd}$ is defined in \cref{eq: def gabcd}.
For the two-point functions, we have
\begin{align}
    &G
    = \frac{1}{2} [ G_{11} + G_{22} - i(G_{12} - G_{21}) ], \\
    &C
    =  C_{11} + C_{22} - i(C_{12} - C_{21}),
\end{align}
in agreement with the total linear response we have calculated earlier [see \cref{eq: chi_T from complex}].

\subsection{Renormalization}

To go beyond the linear theory, we must consider the parameters that appear in the equation of motion, \cref{eq: complex eom} as \emph{bare} coupling constants.
These are related to the renormalized quantities through renormalization constants
\begin{gather}
    \label{eq: bare constants}
    \sqrt{ Z }\ren\varphi = {\varphi}, \quad\quad
    \frac{ Z_r }{ \sqrt{ Z } } \rr  = r, \quad\quad
    \frac{ Z_{u} }{ \sqrt{ Z^3 } }\ru = u, \\
    \frac{ Z_{\bz} }{ \sqrt{ Z } } \rbz = \bz, \quad
    \frac{ Z_{\az} }{ \sqrt{ Z } } \raz = \az, \quad
    \frac{ Z_{\ao} }{ \sqrt{ Z^3 } } \rao = \ao.
\end{gather}
A priori, we should also renormalize $\tilde \varphi$, $\zeta$, $D$, and $K$ as well by introducing $\tilde Z$ $Z_\zeta$, $Z_K$, and $Z_D$. Furthermore, $Z$ may be complex. We are, however, free to keep two quantities fixed, as we have two fields, hence we can choose $\bar K = K$ and $\bar D = D$.
Furthermore, the conservation law implies $\tilde Z = Z_\zeta = 1$, and $Z\in \mathbb{R}$.
This is not the case for the CGLE, which does not have the conservation law.
Indeed, this leads to the existence of another scaling exponent, $\eta'$ which is not present for the conserved dynamics~\cite{tauberCriticalDynamicsField2014,siebererDynamicalCriticalPhenomena2013}.
We have\footnote{The factor $2$ comes from an exchange-symmetry in two of the derivatives of $\smash{\Gamma^{(1,3)}_{(2,2)}}$.}
\begin{align}
    G^{-1}
    & = \Gamma^{(1,1)}_{(1,1)}
    = G_0^{-1} - \Oh(\text{loops}),\\
    \Gamma_U 
    & \equiv \frac{ 1 }{ 2 } \Gamma^{(1,3)}_{(2,2)}
    = q^2(u + i\ao) - \Oh(\text{loops}),
\end{align}
which we can use to write down the renormalization conditions
\begin{align}
    \label{eq: renorm cond 1}
    \partial_{\omega} G^{-1}\rgcond &= -i\zeta, \\ 
    \label{eq: renorm cond 2}
    \partial_{q^{2}} G^{-1}\rgcond &= r + i \alpha_0, \\
    \label{eq: renorm cond 3}
    \partial_{q^4} G^{-1}\rgcond &= K + i\beta_0,\\
    \partial_{q^2}\Gamma_U\rgcond & = u + i \alpha_1.
\end{align}
With these, we can find the connection between the bare vertex-functions, which we can calculate perturbatively, and the renormalization constants.
With the self-energy defined as $G^{-1} = G^{-1}_0 - \Sigma$, $\Gamma_U = q^2(u + i\ao) - \delta \Gamma$, and the corrections
\begin{align}
    \label{eq: corrections 1}
    \partial_{q^2} \Sigma\rgcond  &\equiv  -(\delta r + i \delta \az),\\
    \label{eq: corrections 2}
    \partial_{q^4} \Sigma\rgcond  &\equiv -(\delta K + i \delta \bz), \\
    \label{eq: corrections 3}
    \partial_{q^2} \delta\Gamma\rgcond &\equiv - (\delta u + i \delta \alpha_1),
\end{align}
the renormalization factors are
\begin{align}
    \label{eq: Z 1}
    Z^{-\frac{1}{2}}\! &= 1 \!+\! \delta K/K, &\!
    Z_{\beta_0}^{-1}\! &= 1 \!+\! \delta \beta_0/\beta_0, \\
    Z_r^{-1}\! &= 1 + \delta r / r, &
    \label{eq: Z 2}
    Z_{\alpha_0}^{-1}\! &= 1 + \delta \alpha_0/\alpha_0, 
    \\
    Z_u^{-1}\! &= 1 + \delta u / u, &\!
    Z_{\alpha_1}^{-1}\! &= 1 \!+\!\delta \alpha_1 / \alpha_1.
\end{align}
In terms of $g_i$, we can write $\ren g_{i} = {C g_i}/{(Z_i' \Lambda^{d_i})}$ (no sum), where $d_i$ is the physical dimension of the given term, $C$ is a constant, and
\begin{align}
    \label{eq: def Z'}
    Z'_i \equiv \begin{pmatrix}
    Z_u/ Z^{3/2}, & Z_{\alpha_1} / Z_u, & Z_{\beta_0}/Z^{1/2}, & Z_{\alpha_0} / Z_r 
    \end{pmatrix}.
\end{align}
The renormalized constants $g_i$ thus depend on $\Lambda$ through the $\Lambda^{d_i}$ factor directly, or indirectly through $Z_i'(g)$.
Applying the chain-rule to the definition of the CS $\beta$-function, \cref{eq: CS beta app}, gives
\begin{align}
    \label{eq: beta from Z}
    \beta_i(g)
        & =  g_{i} \left[ d_i + \beta_j \partial_j \ln  Z_i'(g) \right].
\end{align}
Similarly, from \cref{eq: CS flow}, the flow-functions are 
\begin{align}
    \gamma_\t = \beta_j\partial_j  \ln Z_\t'(g),
\end{align}
where $Z_r' = Z_r / \sqrt{ Z }$ and $Z_{\alpha_0}' = Z_{\alpha_0} / \sqrt{ Z }$.
All that is left now, is to calculate $\Sigma$ and $\Gamma^{(1,3)}$ to the desired order in perturbation theory.

\subsection{Diagramatics}

The perturbation series of vertex-functions is organized using Feynman diagrams. We read of the Feynman-rules from the action, \cref{eq: complex action}, as follows:
\begin{align}
    \Q \,
    \parbox{12mm}{
    \centering
    \begin{fmfgraph*}(12,5)
        \setval
        \fmfleft{i}
        \fmfright{o}
        \fmf{cc}{i,o}
    \end{fmfgraph*}
    } \,\Qp &= C_0(\Q)
    \delta_{\omega-\omega'}\delta_{\bm q-\bm q'},
    \\
    \Q\,
    \parbox{12mm}{
    \centering
    \begin{fmfgraph*}(12,5)
        \setval
        \fmfleft{i}
        \fmfright{o}
        \fmf{cg}{i,o}
    \end{fmfgraph*}
    }\,\Qp &=  G_0(\Q)
    \delta_{\omega-\omega'}\delta_{\bm q -\bm q'},\\
    \Q\,
    \parbox{12mm}{
    \centering
    \begin{fmfgraph*}(12,5)
        \setval
        \fmfleft{i}
        \fmfright{o}
        \fmf{cgc}{i,o}
    \end{fmfgraph*}
    }\, \Qp &= G^*_0(\Q)
    \delta_{\omega-\omega'}\delta_{\bm q -\bm q'},
\end{align}
and the interaction vertex is
\begin{align}
    \parbox{12mm}{
    \centering
    \begin{fmfgraph*}(12,10)
        \setval
        \fmfleft{i1,i2}
        \fmfright{o1,o2}
        \fmfv{label=$1$,label.angle=180,l.d=2pt}{i1}
        \fmfv{label=$2$,label.angle=180,l.d=2pt}{i2}
        \fmfv{label=$3$,label.angle=0,l.d=2pt}{o1}
        \fmfv{label=$4$,label.angle=0,l.d=2pt}{o2}
        \fmf{wigglyc}{i1,v}
        \fmf{plainc}{i2,v}
        \fmf{plain}{o1,v,o2}
    \end{fmfgraph*}
    }\,\,
    = -q^2_1 (u + i\ao) \; & \delta_{\omega_1+\omega_2 -( \omega_3+\omega_4)} \\[-4mm]
    & \hspace{-1mm}\times\! \delta_{\bm q_1+\bm q_2 -( \bm q_3+\bm q_4)},\nonumber
\end{align}
with its complex conjugate being
\begin{align}
    \parbox{12mm}{
    \begin{fmfgraph*}(12,10)
        \setval
        \fmfleft{i1,i2}
        \fmfright{o1,o2}
        \fmf{wiggly}{i1,v}
        \fmf{plain}{i2,v} 
        \fmf{plainc}{o1,v}
        \fmf{plainc}{o2,v}
    \end{fmfgraph*}
    }
    & =
    \left(
    \parbox{12mm}{
    \begin{fmfgraph*}(12,10)
        \setval
        \fmfleft{i1,i2}
        \fmfright{o1,o2}
        \fmf{wigglyc}{i1,v}
        \fmf{plainc}{i2,v} 
        \fmf{plain}{o1,v}
        \fmf{plain}{o2,v}
    \end{fmfgraph*}
    }
    \right)^{*}
    .
\end{align}
Here, a straight line represents $\varphi$, a wiggly line represents $i\tilde\varphi$, and the star represents the complex conjugate fields. One slight complication as compared to real-fields comes from the fact that $\varphi^*(\bm q, \omega)$ represents the Fourier-transform of $\varphi^*(\bm x, t)$, not the complex conjugate of $\varphi(\bm q, \omega)$.
Note the sign-difference in the Dirac delta, in contrast to the vector formulation. In general, ``momentum-conservation'' of vertices consists of equating to zero not the sum of all wave-vectors, but the sum of wave-vectors from $\varphi$ and $\tilde\varphi$ minus the sum of those from conjugate fields.

We now calculate the diagrams in \cref{eq: c diagrams 1,eq: c diagrams 2,eq: c diagrams 3}.
Following the Feynman-rules, we will write the integrals corresponding to these diagrams below.
The first, from \cref{eq: c diagrams 1}, is
\begin{align}
    \sigmaoneone & = - 2 q^2 (u + i\alpha_1)\!\int\limits_\K C_0(\K).
\end{align}
Next, from \cref{eq: c diagrams 2} we have
\begin{align}
    \hspace{-2mm}
    \gammaoneone\!\!
    & = 2 q^2 ( u \!+\! i\alpha_1)^2 \!\!\!\int\limits_\K\!\! k^2 G_0(\K) C_0(\Km),\\
    \hspace{-2mm}\gammaonetwo\!\!
    & = 4q^2 (u \!+\! i\alpha_1)^2 \!\!\!\int\limits_\K\!\! k^2 G_0(\K) C_0(\K),\\
    \hspace{-2mm}\gammaonethree\!\!
    & = 4 q^2 |u \!+\! i\alpha_1|^2  \!\!\!\int\limits_\K\!\! k^2 G^*_0(\K) C_0(\K).
\end{align}
Finally, from \cref{eq: c diagrams 3} we have
\begin{align}
    &\sigmatwoone
    = 2 q^2|u + i\alpha_1|^2 \nonumber \\
    &\times\int_{\bm k, \bm k', \omega, \omega'} k''^2 C_0(\K) C_0(\Kp)G_0^*(\Kppm), \\
    & \sigmatwotwo
    = 4 q^2 (u+i\alpha_1)^2 \nonumber
    \\ 
    &\times\int_{\bm k, \bm k', \omega, \omega'} k''^2 C_0(\K) C_0(\Kpm) G_0(\Kpp),
\end{align}
where we denote $\bm k'' = -\bm q - \bm k - \bm k'$ and $\varpi'' = - \omega - \varpi - \varpi'$.

We perform the $\varpi$-integrals with using contour integration. 
We define
\begin{align}
    \label{eq: def one loop integrals}
    I_n \equiv \int_{\bm k} \frac{ k^{2n} }{ S^n }, &&
    J_{(n,m)} \equiv \int_{\bm k} \frac{ (A/k^2)^n (S/k^2)^m }{ S(A^2 + S^2)/k^6 },
\end{align}
where $S(k)/k^2 = r + K k^2$ and $A(k)/k^2 = \ao + \bz k^2$.
The simplest case is
\begin{align}
    \int_{\bm k} C_0(K)
    & = \int_{\bm k} \int_\varpi \frac{ 4 D k^2 }{ (\gamma\varpi - A)^2 + S^2 } \\
    & = 2 \frac{ D }{ \zeta } \int_{\bm k} \frac{ k^2 }{ S } 
    = 2 \frac{ D }{ \zeta } I_2.
\end{align}
Next we have
\begin{align}
    \int_\K\!\! k^2 G_0(\K) C_0(\K)
    & = \frac{ D }{ \zeta }  [J_{(0,1)} - i J_{(1,0)}], \\
    \hspace{-2mm}
    \int_\K\!\! k^2 G_0(\K) C_0(\Km)
    &  = \frac{ D }{ \zeta } I_2 , \\
    \int_\K\!\! k^2 G_0^*(\K) C_0(\K)
    & =\frac{ D }{ \zeta } I_2 .
\end{align}
The two-loop integrals take the form
\begin{align}
    \hspace{-3mm}
    &\int_{\mathcal{K}} k''^2 C_0(\K) C_0(\Kp)G_0^*(\Kppm)
    \\\hspace{-3mm}
    &
    = 
    \frac{D^2}{\zeta^2} 
    \hspace{-2mm}\int\limits_{{\bm k},{\bm k'}} \hspace{-1mm}
    \frac{ 4 }{S S'}
    \frac{ k^2 k'^2 k''^2}{ S + S' + S'' + i(A + A' - A'') }, 
    \\\hspace{-3mm}
    & \int_{\mathcal{K}} k''^2 C_0(\K) C_0(\Kpm) G_0(\Kpp)
    \\\hspace{-3mm}
    & 
    =
    \frac{D^2}{\zeta^2} 
    \hspace{-2mm}\int\limits_{{\bm k},{\bm k'}} \hspace{-1mm}
    \frac{ 4 }{S S'}
    \frac{ k^2 k'^2 k''^2 }{ S + S' + S'' + i(-A + A' + A'') },
\end{align}
where $S = S(k)$, $S' = S(k')$, $S'' = S(k'')$, and similarly for $A$, $A'$ $A''$.
In total, the two-loop contribution to the self-energy is found as
\begin{align}
    &
    \hspace{-2mm}
    \Sigma^{(2)}\!
     =
    8 (u+i\alpha_1) \frac{ D^2 }{ \zeta^2 } q^2 \nonumber 
    \\ 
    &\hspace{-4mm}
    \times\!
     \!\!
    \int\limits_{{\bm k},{\bm k'}}\!
    \frac{ k^2 k'^2 k''^2 }{ S S' S'' }\!
    \left[
        \frac{ (u+i\alpha_1)^* S'' }{ \Sigma S + i \tilde\Sigma A'' }
        +
        \frac{ 2 (u+i\alpha_1) S'' }{ \Sigma S + i \tilde\Sigma A }
    \right] ,
\end{align}
where we have defined
\begin{align}
    \Sigma S &= S + S' + S'',\\
    \tilde\Sigma A & = - A + A' + A'', \\
    \tilde\Sigma A' & = A - A' + A'', \\
    \tilde\Sigma A'' & = A + A' - A''.
\end{align}
This integral can be written compactly by renaming integration variables:
\begin{align}
    \label{eq: H}
    & \hspace{-2mm}
    H\equiv K^3\!\!
    \int_{{\bm k},{\bm k'}}
    \frac{ k^2 k'^2 k''^2 }{ S S' S'' }
        \frac{\Sigma S+i ({\ao}/{u})  \tilde \Sigma S''}{ \Sigma S + i \tilde\Sigma A'' },
\end{align}
where $ \tilde \Sigma S'' = S + S'-S''$. In summary,
\begin{align}
    \Sigma^{(1)} & = - 4q^2 (u + i\alpha_1) \frac{ D }{ \zeta } I_1, \label{eq: Sigma 1}\\
    \Sigma^{(2)} &= 8q^2 (u + i\alpha_1) \frac{ D^2 u }{ \zeta^2 K^3 } H, \label{eq: Sigma 2}
\end{align}
and
\begin{align}
    \delta \Gamma
    & = 
    2 \frac{ D u }{ \zeta } \left[ u J_{(0,1)} + \alpha_1 J_{(1,0)} + 4 u I_2 \right]
       \nonumber  \\& \quad
    - 2 \frac{ D \alpha_1 }{ \zeta } \left[\alpha_1 J_{(0,1)} - u J_{(1,0)} \right]
       \nonumber \\& \quad
    + 2i \frac{ D \alpha_1 }{ \zeta } \left[u J_{(0,1)} + \alpha_1 J_{(1,0)} + 4 u I_2 \right]
    \nonumber \\& \quad
    + 2i \frac{ D u }{ \zeta } \left[ \alpha_1 J_{(0,1)} - u J_{(1,0)} \right].
\end{align}
We now need to evaluate the integrals $I_n$, $J_{(n,m)}$ and $H$ to order $\epsilon^{-1}$, as they alone determine the renormalization factors. The details of these calculations are presented in \cref{app: dim reg}. In particular, in \cref{eq: I1,eq: I2,eq: J01,eq: J10} for the one-loop results, and \cref{eq: H,eq: H1,eq: H2} for the two-loop results. Combining these with \cref{eq: corrections 1,eq: corrections 2,eq: corrections 3}, we find the following corrections at one-loop
\begin{align}
    \label{eq: Z-factors}
    \frac{\delta r}{r} & = - 4 \l \frac{ 1 }{ \epsilon }, \\
    \frac{ \delta \alpha_0}{\alpha_0} & = - 4 \l \frac{ \a }{ \c } \frac{ 1 }{ \epsilon }, \\
    \frac{\delta u}{u} & = - 2 \l \frac{ 5 + 4 \b^2 - \a^2 + 2 \a\b }{ 1 + \b^2 }\frac{ 1 }{ \epsilon }, \\
    \frac{\delta \alpha_1}{\alpha_1} & = - 2 \l \frac{ 1 }{ \a } \frac{ \b(\a^2 - 1) + \a(6 + 4 \b^2) }{ 1 + \b^2 }\frac{ 1 }{ \epsilon }.
\end{align}
Here, we have written the results in terms of the dimensionless couplings, which in the CS approach are defined by \cref{eq: def g CS}.
At two-loop order, we obtain
\begin{align}
    \label{eq: correction K}
    \hspace{-2mm}
    \frac{\delta K}{K} 
    & = \l^2
    \left\{
        1 + (\b - \a)[ \im \g(\b) + \a \re \g(\b)]
    \right\}\frac{ 1 }{ \epsilon }, \\
    \label{eq: correction b}
    \hspace{-2mm}
    \frac{\delta \beta_0}{\beta_0} 
    & = \frac{ \l^2 }{ \b }
    \left\{
        \a + (\b - \a)[\a \im \g(\b) - \re \g(\b)]
    \right\}\frac{ 1 }{ \epsilon },\!
\end{align}
for $K$ and $\beta_0$, where $\g(\b)$ is defined in \cref{app: two loop}.
We can now insert the renormalized quantities, such as $\rl$, instead of their bare counterparts, as this is consistent within our perturbation calculation, to the leading order. The renormalization factors [using \cref{eq: Z 1,eq: Z 2}] are found as
\begin{align}
    \label{eq: Z r}
    \hspace{-2mm}
    Z_r &= 1 + 4 \rl\, \frac{ 1 }{ \epsilon }, \\
    \label{eq: Z a0}
    \hspace{-2mm}
    Z_{\alpha_0} &= 1 + 4 \rl \,\frac{ \ra }{ \rc }\, \frac{ 1 }{ \epsilon }, \\
    \label{eq: Z}
    \hspace{-2mm}
    \sqrt{\!Z} &= 1 - \rl^2\{ 1 + (\rb - \ra)[\im \g(\rb) + \ra\re \g(\rb)] \} \frac{ 1 }{ \epsilon }.
\end{align}
For the $\beta$-functions, we need the combinations defined in \cref{eq: def Z'}, which we find to be
\begin{align}
    \label{eq: Z' l}
    Z_\l' &= 1 + 2 \rl \,\frac{ 5 + 4 \rb^2 - \ra^2 + 2 \ra\rb }{ 1 + \rb^2 } \, \frac{ 1 }{ \epsilon }, \\
    \label{eq: Z' a}
    Z_\a' 
    &= 1 + 2 \rl \,\frac{ \ra - \rb }{ \ra } \frac{ 1 + \ra^2 }{ 1 + \rb^2 } \,\frac{ 1 }{ \epsilon }, \\
    \label{eq: Z' b}
    Z_\b' 
    & = 1 + 2 \rl^2 \,\frac{ \rb - \ra }{ \rb }[1 + \h(\ra, \rb)] \, \frac{ 1 }{ \epsilon }, \\
    \label{eq: Z' c}
    Z_\c' &= 1 - 4\rl \,\frac{ \rc - \ra }{ \rc } \, \frac{ 1 }{ \epsilon },
\end{align}
where we have defined the function
\begin{align}
    \label{eq: h}
    & \h(\a, \b) \equiv (1 + \a\b)\re \g(\b) +  (\b - \a)\im \g(\b),
\end{align}
which is plotted in \cref{fig: h}.
With \cref{eq: beta from Z}, we obtain the result of the main text, \cref{eq: beta l,eq: beta a,eq: beta b,eq: beta c}.


\section{Dimensional Regularization}
\label{app: dim reg}

In this appendix, we evaluate the integrals derived above using dimensional regularization.

\subsection{Identities}
\label{app: identities}

The Gamma function
\begin{align}
    \label{eq: gamma function}
    \Gamma(z) = \int_0^\infty \dd t\, t^{z - 1} e^{-t}, && \re( z )> 0,
\end{align}
obeys (for integer $n > 0$)
\begin{align}
    \label{eq: gamma div}
    \Gamma(\delta - n) &= \frac{ (-1)^n }{ n! } \left( \frac{ 1 }{ \delta } - \gamma + \sum_{m=1}^n \frac{ 1 }{ m } \right) + \Oh(\delta),
\end{align}
This is related to the Beta function (not to be confused with the RG $\beta$-functions) by
\begin{align}
    \label{eq: beta function}
    B(a, b) 
    \equiv \int_0^1 \dd t \, t^{a - 1} (1 - t)^{b - 1}
    = \frac{ \Gamma(a) \Gamma(b) }{ \Gamma(a + b) }.
\end{align}
Feynman-parametrization uses the identity
\begin{align}
    \label{eq: feynman para}
    \frac{ 1 }{ A^n B^m }
    & = \frac{ \Gamma(n + m) }{ \Gamma(n) \Gamma(m) } \int_0^1 \dd x \frac{ x^{n-1} (1-x)^{m-1} }{ [x A + (1-x)B]^{n+m} }.
\end{align}
We observe that the NRCH field theory requires a generalization of the standard dimensional regularization integrals. This is in contrast with model B, for example, for which any non-zero Feynman-diagrams are the same as for model A, as the conservation law does not affect the integrals. We define
\begin{align}
    \label{eq: Ia def}
    I^{(a)}_{(n,m)}(r) 
    &\equiv 
    \int_{\bm p} \frac{ (|\bm p|^a)^n }{ (r + |\bm p|^a)^m }.
\end{align}
With the substitution $x = 1 / (1 + |\bm p|^{a}/ r)$, we can rewrite this in terms of the $\Gamma$-functions using \cref{eq: beta function}:
\begin{align}
    \label{eq: Ia}
    I^{(a)}_{(n,m)}(r)
    &
    = 
    \frac{ A_d }{ a } \frac{ \Gamma(n \!+\! \tfrac{d}{a}) }{ \Gamma(m) } 
    \Gamma(m\! -\! n\! -\! \tfrac{d}{a}) r^{\tfrac{d}{a}-n-m}.
\end{align}
By a change of variables, we also find
\begin{align}
    \label{eq: I'}
    I_m'(r,q)
    & \equiv
    \int_{\bm k} \frac{ 1 }{ (k^2 + 2 \bm k \cdot \bm q + r)^m } \\
    & =
    \frac{ A_d }{ 2 } \frac{ \Gamma(m - \frac{ d }{ 2 }) }{  \Gamma(m) } (r - q^2)^{\frac{ d }{ 2 } - m}.
\end{align}
%


\subsection{One-loop integrals}
\label{app: one loop}

Here, we evaluate the integrals defined in \cref{eq: def one loop integrals}.
Using \cref{eq: Ia}, we identify
\begin{align}
    I_n = \int_{\bm k} \frac{ 1 }{ (r + K k^2)^n } = \frac{1}{K^n} I_{(0,n)}^{(2)}(r/K).
\end{align}
Using \cref{eq: Ia}, we have in particular
\begin{align}
    \label{eq: I1}
    I_1 & = - \frac{1}{8 \pi^2 \epsilon } \frac{r^{\frac{ d }{ 2 } - 1}}{ K^{\frac{ d }{ 2 }} }+ \Oh(\epsilon^0), \\
    \label{eq: I2}
    I_2 & = \frac{1}{8 \pi^2 \epsilon } \frac{r^{\frac{ d }{ 2 } - 2}}{ K^{\frac{ d }{ 2 }} }+ \Oh(\epsilon^0).
\end{align}
Next, the integrals
\begin{align}
            J_{(n,m)} \equiv \int_{\bm k} \frac{ (A/k^2)^n (S/k^2)^m }{ S(A^2 + S^2)/k^6 },
\end{align}
where $S/k^2  = r + K k^2$ and $A/k^2 = \ao + \bz k^2$,
can be evaluated using Feynman parametrization.
This procedure utilizes the identity \cref{eq: feynman para} to rewrite the momentum-integral in terms of $I^{(a)}_{(n,m)}$, which we evaluated in \cref{eq: Ia def}.
As we use minimal subtraction, and we are only seeking the leading order corrections, it suffices for us to calculate the $\epsilon^{-1}$ divergence. Beginning with $J_{(0,1)}$, we write 
\begin{align}
    J_{(0,1)} 
    = 
    \frac{ 1 }{K^2 + \bz^2\! }\! \int_{\bm k} \frac{ 1 }{ (k^2 - R_+)(k^2 - R_-) }.
\end{align}
Here, $R_\pm$ are the roots of a polynomial of the form $k^4 + C_1 k^2 + C_2$. We then introduce Feynman parameters and calculate
\begin{align}
    J_{(0,1)}&=  \frac{ 1 }{K^2 + \bz^2\! }\!\int_{\bm k} \int_0^1 \! \frac{ \dd x }{ [x(k^2 \!-\! R_+) + (1\!-\!x)(k^2 \!-\! R_-)]^2} ,\nonumber  \\
    & = \frac{ 1 }{K^2 + \bz^2\!}\! \int_0^1\dd x\, \int_{\bm k}  \frac{ 1 }{ [k^2 + R_1(x)]^2}, \nonumber \\
    & = \frac{ 1 }{ K^2 + \bz^2 } \frac{ \Gamma(2 - \frac{ d }{ 2 }) }{ (4\pi)^d \Gamma(2) } \int_0^1 \dd x\,  R_1(x)^{\frac{ d }{ 2 } - 2},
\end{align}
where $R_1(x) \equiv - x R_+ - (1-x)R_-$.
We now use $R_1(x)^{\frac{ d }{ 2 } - 2} = e^{\epsilon/2 \ln R_1(x)} = 1 + \Oh(\epsilon)$ to extract the divergence.
With this, we can expand and evaluate the $\Gamma$-functions using \cref{eq: gamma div} to obtain
\begin{align}
    \label{eq: J01}
    J_{(0,1)}= \frac{1}{8 \pi^2  \epsilon }\frac{1}{ K^2 + \bz^2 } + \Oh(\epsilon^0).
\end{align}
Next, in a similar fashion, we calculate
\begin{align}
    J_{(1,0)} = \frac{ 1 }{ K(\bz^2 + K^2) }
    \int_{\bm k} 
    \frac{ \alpha_0 + \bz k^2 }{ (k^2 + \frac{r}{K}) (k^2 - R_+)(k^2 - R_-) }.
\end{align}
Using the definition $R_2(x, y) = - x R_+ - y R_- + (1 - x - y) \tfrac{r}{K}$, we find
\begin{align}
    J_{(1,0)}& = \int_0^1 \dd x \, \int_0^{1-x} \dd y \,
    \frac{\alpha_0 + \bz k^2}{ [k^2 + R_2(x, y)]^3 },\nonumber \\
    & = 
    \az \int_0^1 \dd x \, \int_0^{1-x} \dd y \,
    I_{(0,3)}^{(2)}(R_2(x, y)) \nonumber \\
    &+ \bz \int_0^1 \dd x \, \int_0^{1-x} \dd y \, I_{(1,3)}^{(2)}(R_2(x, y)).
\end{align}

Extracting the divergence yields
\begin{align}
    \label{eq: J10}
    J_{(1,0)} &= \frac{1}{8 \pi^2 \epsilon } \frac{ 1 }{ \bz^2 + K^2 } \frac{ \bz }{ K } + \Oh(\epsilon^0).
\end{align}
%


\begin{widetext}
\subsection{Two-loop integrals}
\label{app: two loop}

Two loop self energy contribution is $\Sigma^{(2)}=8 (u + i\az) \tfrac{ D^2 u }{ \zeta^2 K^3 } q^2 H $ [\cref{eq: Sigma 2}], where the integral $H$ is given as follows
\begin{align}
    \hspace{-3mm}
    H & = K^3
    \hspace{-3mm}\int\limits_{\bm k,\bm k'\!,\bm k''}\hspace{-2mm}
    \frac{ \delta_{\bm k + \bm k' + \bm k'' + \bm q} }{ (r+K k^2) (r+K k'^2)(r+K k''^2) } 
    \\
    &
    \hspace{12mm}
    \times 
    \left[
        1 \!+\!
    \frac{
        i r \left(\a - \c \right)(k^2 + k'^2 - k''^2)
        +
        i K \left(\a - \b \right)(k^4 + k'^4 - k''^4)
    }{ 
        r (1 + i \c)(k^2 + k'^2)  
        + r (1 - i \c) k''^2 
        + K (1 + i\b)(k^4 + k'^4) 
        + K (1 - i\b)k''^4
     }
    \right].
\end{align}
We only need the leading order of $Z$ and $Z_\bz$, and can therefore take the the limits $r, \az \rightarrow 0$. 
This yields
\begin{align}
    \label{eq: H}
    H & = 
    H_1 + i \left(\a - \b\right) H_2, 
    &
    H_1 &\equiv
    \hspace{-3mm}\int\limits_{\bm k,\bm k'\!,\bm k''}\hspace{-3mm}
    \frac{  \delta_{\bm k + \bm k' + \bm k'' + \bm q} }{ k^2\, k'^2\, k''^2 } , &
    H_2& \equiv
    \frac{ 1 }{ 1 + i \b }
    \hspace{-3mm}\int\limits_{\bm k,\bm k'\!,\bm k''}\hspace{-3mm}
    \frac{  \delta_{\bm k + \bm k' + \bm k'' + \bm q}}{ k^2\, k'^2\, k''^2 } \frac{k^4 + k'^4 - k''^4}{ (k^4 + k'^4) + B k''^4},
\end{align}
where $B \equiv \dfrac{1 - i\b}{1 + i\b}$. In the small wavevector limit, we obtain
\begin{align}
    q^2 \partial_{q^2}H_1|_{q = 0}
    = 
    - q^2
    \left( 1 - \frac{ 4 }{ d } \right) H', &&
    H' \equiv
    \int_{\bm k_1} \frac{ 1 }{ k_1^2 } 
    \int_{\bm k_2} \frac{ 1 }{ k_2^2 } \frac{ 1 }{ (\bm k_1 + \bm k_2)^4 }.
\end{align}
We next employ Feynman-parametrization and \cref{eq: I'}, to find
\begin{align}
    \label{eq: sol H'}
    H' 
    = 
    \Gamma(3) \int_0^1 \dd x\, x
    \int\limits_{\bm k_1} \frac{ 1 }{ k_1^2 } 
    \int\limits_{\bm k_2}
    \frac{ 1 }{ [ k_2^2 + 2 x \bm k_1 \cdot \bm k_2 + xk_2^2 ]^3 }
    & =
    \Gamma(3) \int_0^1 \dd x\, x\!
    \int\limits_{\bm k_1} \frac{ 1 }{ k_1^2 } 
    I'_3(x(1-x)k_1^2) 
    = \frac{1}{128 \pi^4 \epsilon^2} + \Oh(\epsilon^{-1}).
\end{align}
This gives,
\begin{align}
    \label{eq: H1}
    H_1 \sim 
    - \frac{q^2 }{512 \pi^4 \epsilon } 
    + \mathrm{const.} + \Oh(q^4, \epsilon^0)
    .
\end{align}
To calculate $H_2$, we expand to order $q^2$, and obtain
\begin{align}
    \label{eq: H' expansion}
    & \hspace{-4mm} H_2=  
    \mathrm{const.} + \Oh(q^4, \epsilon^0) + 
    \frac{ q^2\! }{ d }
    \frac{ 1 }{ 1 + i \b }\nonumber
    \\*
    & \hspace{-6mm} 
    \times\!\!
    \left[
    \epsilon K_{(0,0)}
    \!- d K_{(1,0)}
    \!+ 2 \left( 6 + d  \right) K_{(2,0)}
    \!- 16 K_{(3,0)}
    \!- d K_{(0,1)}
    \!+ 3 \left( 4 + d \right) K_{(1,1)}
    \!- 2 \left( 14 + d \right)  K_{(2,1)}
    \!+ 16 K_{(3,1)}
    \right]\!.\hspace{-2mm}
\end{align}
Here, we have defined
\begin{align}
    K_{(n,m)}
    & \equiv
    B^n
    \int\limits_{{\bm k},{\bm k'}}
    \frac{ 1 }{ k^2 k'^2 } \frac{ 1 }{ (\bm k + \bm k')^{4} } 
    \left[
        \frac{  (\bm k + \bm k')^{4} }{k^4 + k'^4 + B (\bm k + \bm k')^{4} }
    \right]^{n}
    \left[
        \frac{ (\bm k + \bm k')^{4} }{ k^4 + k'^4 }
    \right]^{m}, &&
    B = \frac{ 1 - i \b }{ 1 + i \b }.
    \label{eq: Hnm}
\end{align}
We observe that most of these integrals can be derived from a master integral upon differentiation. 
To see this, note that for any function $G(\bm k, \bm k')$, we have
\begin{align}
    F(B)
    & \equiv
    \int\limits_{{\bm k},{\bm k'}} \frac{ G(\bm k, \bm k') }{ k^2 k'^2 }
    \frac{ 1 }{ k^4 + k'^4 + B (\bm k + \bm k')^4 }, &&
    \implies &
    \odv[l]{ F(B) }{ B }
    & = 
    \int\limits_{{\bm k},{\bm k'}} \frac{ G(\bm k, \bm k') }{ k^2 k'^2 }
    \frac{ (-1)^l \Gamma(l + 1) (\bm k + \bm k')^{4 l} }{ [k^4 + k'^4 + B (\bm k + \bm k')^4]^{l + 1} }.
\end{align}
Hence, we can write \cref{eq: Hnm} as follows
\begin{align}
    \label{eq: H(nm) derivative}
    K_{(n,m)} = 
    - 
    \frac{ (-B)^n }{ \Gamma(n) }
    \odv[n-1]{  }{ B }
    \left[\frac{ K_{(1,m)} }{ B }\right].
\end{align}
We therefore only need a few of the integrals, which we calculate next.

We note that $K_{(0,0)}$ is given by $H'$, which we have already calculated in \cref{eq: sol H'}:
\begin{align}
    \label{eq: K00}
    K_{(0,0)} 
    =  \int_{{\bm k},{\bm k'}}\frac{ 1 }{ k^2 k'^2 (\bm k + \bm k')^4 }
    = -\frac{1}{128 \pi^4 \epsilon^2 }+ \Oh(\epsilon^{-1}).
\end{align}
Then, for $K_{(0, 1)}$ we have
\begin{align}
    K_{(0, 1)} 
    = \int_{{\bm k},{\bm k'}} \frac{ 1 }{ k^2 k'^2 } \frac{ 1 }{ k^4 + k'^4 }
    = \int_{{\bm k},{\bm k'}} \frac{ 1 }{ k^2 k'^4 } \frac{ k'^2 }{ k^4 + k'^4 }. 
\end{align}
With Feynman parametrization, and using \cref{eq: beta function} and \cref{eq: Ia}, we get
\begin{align}
    \label{eq: K01}
    K_{(0, 1)} & = 
    \int_0^1 \dd x \, \int_{\bm k} \frac{ 1 }{ k^2 } 
    \int_{{\bm k}'} \frac{ 1 }{ [ k'^4 + x k^4 ]^2 } 
    = 
    \int_0^1 \dd x \, \int_{\bm k} \frac{ 1 }{ k^2 } I^{(4)}_{(1/2, 2)}(x k^4) 
    = \frac{1}{512 \pi^3 \epsilon }+ \Oh(\epsilon^0).
\end{align}
For the next integral, we can rewrite \cref{eq: Hnm} as follows
\begin{align}
    K_{(1, 1)} 
    & = 
    \int_{{\bm k},{\bm k'}} \frac{ 1 }{ k^2 k'^2 }
    \frac{ (k^4 + k'^4) - (k^4 + k'^4) + B (\bm k + \bm k')^4 }{ k^4 + k'^4 + B (\bm k + \bm k')^4 }
    \frac{  1 }{ k^4 + k'^4 }, \nonumber \\
    & = 
    \int_{{\bm k},{\bm k'}} \frac{ 1 }{ k^2 k'^2 }
    \left( \frac{  1 }{ k^4 + k'^4 }  - \frac{ 1 }{ k^4 + k'^4 + B (\bm k + \bm k')^4 } \right),\nonumber
    \\
    & = K_{(0,1)}
    - \frac{ 1 }{ B }  K_{(1,0)}.
    \label{eq: K11}
\end{align}
The last integral we need is
\begin{align}
    K_{(1,0)} 
    & = \int_{{\bm k},{\bm k'}} \frac{ 1 }{ k^2 k'^2 } \frac{ B }{ k'^4 + k^4 + B(\bm k' + \bm k)^4 },\nonumber \\
    & = 
    B A_d \int \frac{ \dd \Omega(\theta) }{ (2\pi)^d }
    \int \dd k \int \dd k' 
    \frac{ (k k')^{d - 3} }{ k'^4 + k^4 + B(k^2 + k'^2 + 2 k k' \cos\theta)^2 }.
\end{align}
Using a change of coordinates $k = R \cos\phi$, $k' = R \sin\phi$, (with a Jacobian $R$) and $ \int \frac{ \dd \Omega(\theta) }{ (2\pi)^d }\, f(\theta) = A_d \frac{ S_{d-2} }{ S_{d-1} } \int_0^{\pi} \dd \theta \, \sin^{d-2}\!\theta f(\theta)$, we find
\begin{align}
    K_{(1,0)}
    = 
    A_d^2 \frac{ S_{d-2} }{ S_{d-1} } B
    \int \dd R \, R^{2(d-3) + 1 - 4}
    \int_0^\pi \dd \theta \int_0^\frac{\pi}{2}\dd \phi
    \frac{ \sin^{d-2}\theta \cos^{d-3}\phi \sin^{d-3}\phi }{ \cos^4\phi + \sin^4\phi + B[1 + 2\cos\phi\sin\phi\cos\theta]^2 }.
\end{align}
Using dimensional regularization to extract the divergence, we find
\begin{align}
    \int \dd R \, R^{2(d-3) + 1 - 4} = \int \dd R \, R^{-2\epsilon - 1}=\frac{ 1 }{ 2\epsilon } + \Oh(\epsilon^0).
\end{align}
\begin{figure}[!t]
    \centering
    \includegraphics[width=.63\textwidth]{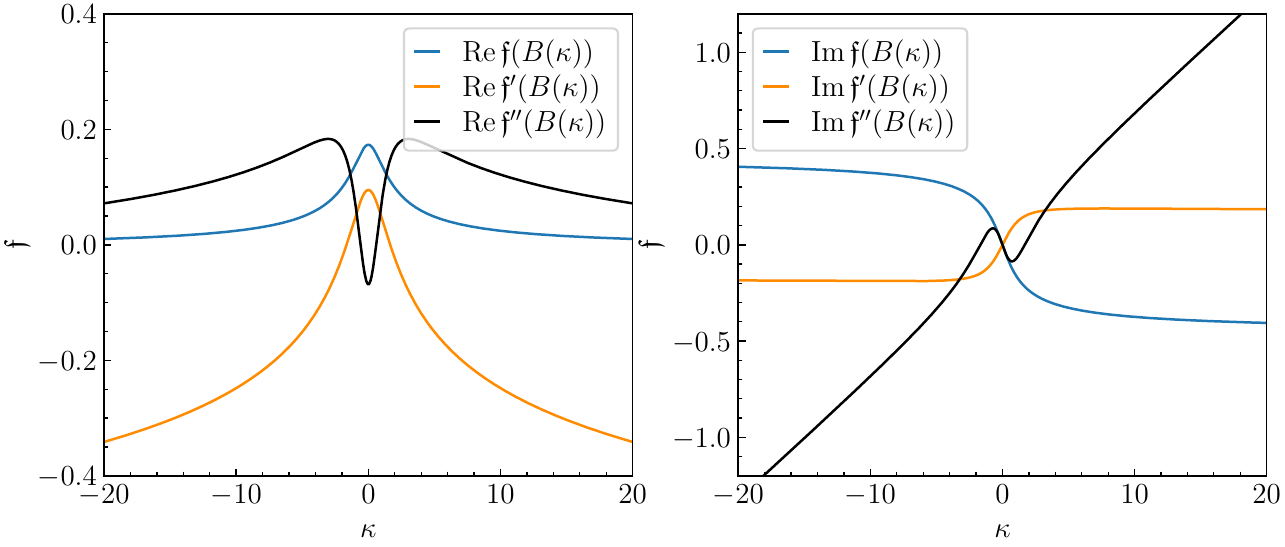}
    \includegraphics[width=.34\textwidth]{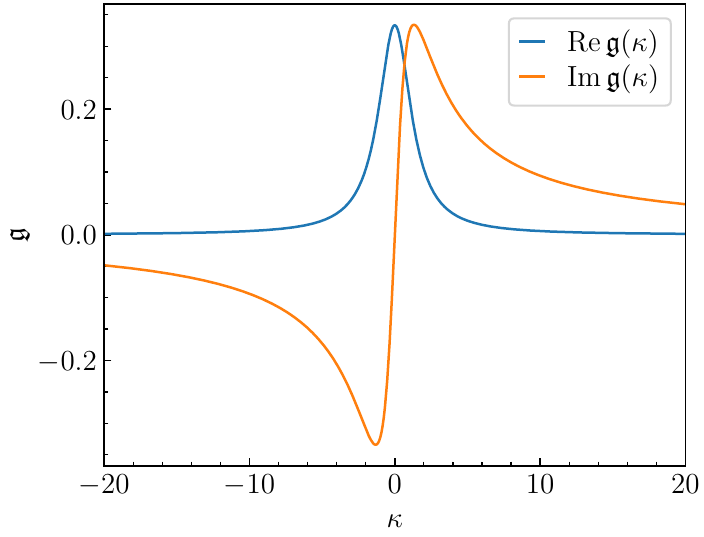}
    \caption{The functions $\f(\b)$ and $\g(\b)$.}
    \label{fig: fg}
\end{figure}
Together with $\frac{ S_{d-2} }{ S_{d-1} } =\frac{ 2 }{ \pi }+ \Oh(\epsilon^0)$, we obtain
\begin{align}
    \label{eq: K10}
    K_{(1,0)} = \frac{1}{64 \pi^4 \epsilon  } \,\f(B)+ \Oh(\epsilon^0),
\end{align}
where we have defined the function
\begin{align}
    \f(B) \equiv
    \int\limits_0^\pi \!\! \int\limits_0^\frac{\pi}{2}
    \frac{ \dd\theta \dd\phi }{ 2 \pi }
    \frac{ B \sin^{2}\!\theta \sin2\phi }
    { \cos^4\!\phi + \sin^4\!\phi + \!B[1 + \sin2\phi\cos\theta]^2 }.
\end{align}
Combining \cref{eq: K00,eq: K01,eq: K11,eq: K10} with \cref{eq: H(nm) derivative} to find all the terms in \cref{eq: H' expansion}, we get
\begin{align}
    \label{eq: H2}
    H_2 &= 
    - \frac{q^2 }{512 \pi^4 \epsilon } \, \g(\b)
    + \mathrm{const.} + \Oh(q^4, \epsilon^0),
\end{align}
where we have defined
\begin{align}
    \label{eq: g}
    \g(\b) & \equiv
    \frac{ 1  }{ 1 + i \b }
    \left\{ 1 + 
    \frac{16}{1+i\b}
        \left[ 
            \f'\Big(\frac{1-i\b}{1+i\b}\Big)
            + 2 \,\frac{1-i\b}{1+i\b} \,
            \f''\Big(\frac{1-i\b}{1+i\b}\Big)
        \right]
    \right\}.
\end{align}
The functions $\f$ and $\g$ are illustrated in \cref{fig: fg}.

\end{widetext}

\end{fmffile}
\bibliography{ref,manual,Golestanian}
\end{document}